\documentclass[fleqn,usenatbib]{mnras}

\usepackage{newtxtext,newtxmath}
\usepackage[mathscr]{euscript}

\usepackage[T1]{fontenc}
\usepackage{ae,aecompl}
\usepackage[utf8]{inputenc}
\usepackage{ulem}
\DeclareRobustCommand{\VAN}[3]{#2}
\let\VANthebibliography\thebibliography
\def\thebibliography{\DeclareRobustCommand{\VAN}[3]{##3}\VANthebibliography}

\usepackage{graphicx}	
\usepackage{amsmath}	

\usepackage{amssymb}	
\usepackage{xcolor}

\usepackage[displaymath, mathlines]{lineno}
\newcommand*\patchAmsMathEnvironmentForLineno[1]{%
 \expandafter\let\csname old#1\expandafter\endcsname\csname #1\endcsname
 \expandafter\let\csname oldend#1\expandafter\endcsname\csname end#1\endcsname
 \renewenvironment{#1}%
    {\linenomath\csname old#1\endcsname}%
    {\csname oldend#1\endcsname\endlinenomath}}%
\newcommand*\patchBothAmsMathEnvironmentsForLineno[1]{%
 \patchAmsMathEnvironmentForLineno{#1}%
 \patchAmsMathEnvironmentForLineno{#1*}}%
\AtBeginDocument{%
\patchBothAmsMathEnvironmentsForLineno{equation}%
\patchBothAmsMathEnvironmentsForLineno{align}%
\patchBothAmsMathEnvironmentsForLineno{flalign}%
\patchBothAmsMathEnvironmentsForLineno{alignat}%
\patchBothAmsMathEnvironmentsForLineno{gather}%
\patchBothAmsMathEnvironmentsForLineno{multline}%
}

\newcommand{\dd}{{\rm d}}

\title[Lennard-Jones strangeon stars]{Rotation and deformation of strangeon
stars in the Lennard-Jones model}

\author[Yong Gao et al.]{
Yong Gao,$^{1,2}$
Xiao-Yu Lai,$^{3,4}$
Lijing Shao$^{2,5}$\thanks{Corresponding author: lshao@pku.edu.cn}
and Ren-Xin Xu$^{1,2}$\thanks{Corresponding author: r.x.xu@pku.edu.cn }
\\
$^{1}$Department of Astronomy, School of Physics,
Peking University, Beijing 100871, China\\
$^{2}$Kavli Institute for Astronomy and
Astrophysics, Peking University, Beijing 100871, China\\
$^{3}$Department of Physics and Astronomy, Hubei
University of Education, Wuhan 430205, China\\
$^{4}$Research Center for Astronomy, Hubei
University of Education, Wuhan 430205, China\\
$^{5}$National Astronomical Observatories,
Chinese Academy of Sciences, Beijing 100012, China
}

\date{Accepted XXX. Received YYY; in original form ZZZ}

\pubyear{2021}

\begin{document}
\label{firstpage}
\pagerange{\pageref{firstpage}--\pageref{lastpage}}
\maketitle

\begin{abstract}
The strong interactions at low energy scales determine the state of the
supranuclear matter in the pulsar-like compact objects. It is proposed that
the bulk strong matter could be composed of strangeons, which are quark
clusters with a nearly equal number of three light-flavor quarks.  In this
work, to characterize the strong-repulsive interactions at short distances
and the non-relativistic nature of the strangeons, the Lennard-Jones model
is used to describe the equation of state (EoS) of strangeon stars (SSs). 
We investigate the static, the slowly rotating, and the tidally deformed SSs
in detail. The corrections resulted from the finite surface densities are
considered crucially in the perturbative approaches. We also study the
universal relations between the moments of inertia, the tidal
deformabilities, and the quadrupole moments. Those results are ready to be
used for various purposes in astrophysics, and possible constraints from
contemporary observations on the parameter space of the Lennard-Jones model 
are discussed. Future observations of the pulsars' radio signals, the X-ray
emissions from the hot spots on the surface of the stars, and the
gravitational waves (GWs) from binary mergers can give tighter constraints
or even verify or falsify the existence of SSs.
\end{abstract}

\begin{keywords}
dense matter -- pulsars: general -- methods: numerical
\end{keywords}



\section{Introduction}
\label{sec:ontroduction}

The equation of state (EoS) of the supranuclear matter in pulsar-like compact
stars remains unknown.  Traditionally, people believe that neutron stars (NSs)
are formed in the supernova explosions.  However, there is another possibility
that the energetic electrons in the collapsed stellar core are eliminated by
transforming the $u/d$ quarks to the $s$ quark. \citet{Witten:1984rs}
conjectured that the pulsar-like compact objects should be quark stars (QSs),
which contain strange quark matter (SQM) with nearly equal number of $u$, $d$,
$s$ quarks, and a small fraction of electrons to keep the neutrality. Some
efforts have been made to understand the state of pulsar-like compact stars in
the framework of conventional QSs, including the MIT bag model with
almost free quarks~\citep{Alcock1986} and the color-superconducting state
model~\citep{Alford:2007xm}.

Realistic densities inside pulsar-like compact stars ($\rho\sim 2-10 \, \rho_0$
where $\rho_0$ is the saturation density of nuclear matter) are not high enough
to justify the validity of perturbative quantum chromodynamics (QCD).  The
perturbative QCD, based on asymptotic freedom, works well only at high energy
scales, $E_{\rm scale} \gtrsim 1$ GeV.  However, the state of pressure-free
strong matter at supranuclear density pertains to non-perturbative QCD because
$E_{\rm scale} \lesssim 1$ GeV, analogous to the case of normal atomic nuclei. 
In addition, the difficulties to obtain the relativistic EoS of cold quark
matter at several nuclear densities also come from the fact that the vast
assemblies of interacting particles face the complicated quantum many-body
problem~\citep{Xu_2009,Xu:2018zdo,Alford:2007xm,Baym_2018}. The uncertainty in
EoS introduces additional uncertainty to the strong-field gravity tests as
well~\citep{Shao:2017gwu, Shao:2019gjj}.

It is phenomenologically conjectured that, in cold matter at supranuclear
density, the constituent units could be strange quark
clusters~\citep{Xu:2003xe}, since the strong interaction may render quarks 
grouped in quark-clusters. Each quark-cluster is composed of several quarks
(including $u$, $d$ and $s$ quarks) condensing in position space rather than in
momentum space. A strange quark cluster is named ``strangeon,'' being coined to
``strange nucleon''~\citep{XG2016,LX2017}.  The astrophysically compressed
baryonic matter, therefore, should be in a state of strangeon matter, and
pulsar-like compact stars could be strangeon stars (SSs). Moreover, at realistic
baryon densities of compact stars, the residual interaction between strangeons
could be stronger than their kinetic energy, so strangeons would be trapped in
the potential well and the bulk of the dense matter in the compact stars are
crystallized into a solid state~\citep{Xu:2003xe,Xu_2009}.

Although the state of bulk strong matter is essentially a non-perturbative QCD
problem and is difficult to answer from first principles, the astrophysical
point of view could give some hints that bulk strong matter could be in the form
of strangeon matter. Strangeon matter may constitute the true ground state of 
strongly-interacting matter rather than $^{56}$Fe and neutron matter, and this
could be seen as a {\it generalized Witten's conjecture}, while  the traditional
Witten's conjecture focuses on the matter composed of almost free $u$, $d$ and
$s$ quarks.

Strangeon matter, similar to strange quark matter, is composed of nearly equal
numbers of $u$, $d$ and $s$ quarks; however, different from that in strange
quark matter, quarks in strangeon matter are localized inside strangeons due to
the strong coupling between quarks. There are differences and similarities
between SSs and NSs/QSs.  On the one hand, quarks are thought to be localized in
strangeons in SSs, like neutrons in NSs, but a strangeon has three flavors and
may contain more than three valence quarks. The matter at the surface of SSs is
still strangeon matter, i.e., SSs are self-bound by strong force, like QSs.

Eventually the theoretical models, including NSs, QSs and SSs, need to be tested
by the astrophysical observations.  The model of SSs has been found to be
helpful to understand different manifestations of pulsar-like compact stars.
SSs had been found to be massive~\citep{Lai_2009128,Lai:2009cn} before the
discovery of pulsars with $M>2M_\odot$~\citep{Demorest2010}.  The surface of SSs
could naturally explain the pulsar magnetospheric activity~\citep{Xu1999ApJL} as
well as the subpulse-drifting of radio pulsars~\citep{Lu2019}. Starquakes of
solid SSs could induce glitches~\citep{Zhou2004,Zhou2014,Lai2018MN} and explain
the glitch activity of normal radio pulsars~\citep{WangWH2020}.  The plasma
atmosphere of SSs can reproduce the Optical/UV excess observed in X-ray dim
isolated NSs~\citep{Wang2017APJ}. The tidal deformability~\citep{LZX2019} as
well as the light curve~\citep{Lai2018RAA,Lai2021} of merging binary SSs are
consistent with the results of gravitational wave (GW) event
GW170817~\citep{LIGOScientific:2017vwq} and its multiwavelength electromagnetic
counterparts~\citep{Kasliwal2017,Kasen2017}.

Some phenomenological models have been applied to investigate the EoS of
SSs~\citep{Lai_2009128,Lai:2009cn}, which may indicate some properties of QCD at
low energy scales and have implications on possible astronomical observations
that can give constraints on such models. The strangeons are colorless, just
like the molecules are neutral in the bulk of inert gas. \citet{Lai:2009cn}
utilized the Lennard-Jones model~\citep{jones1924determination} that describes
inert gas to characterize the interactions between strangeons. This model well
characterizes the non-relativistic nature and the strong-repulsive interactions
at short distances. In this work we use this model as the EoS and span a wide
range of parameter space to investigate the global properties of SSs. We model
the static SSs and compare the structures with that of Tolman IV and Tolman VII
solutions~\citep{Tolman:1939jz,Lattimer:2004sa}.

Astrophysically, pulsars have  spins and the rotation will affect the structures
of the stars and the spacetime itself.  \citet{Hartle:1967he} and
\citet{Hartle:1968si} gave a perturbative approach to describe slowly rotating
relativistic stars to the second order of angular frequency $\Omega$. Later,
\citet{Hartle:1973zza} developed this formalism to the third order of $\Omega$
and calculated the variations of moments of inertia for distorted NSs. One can
evaluate the change of physical quantities at each order of $\Omega$ and
investigate various properties of the relativistic
stars~\citep{Chandra1974,Weber1991,Lattimer:2000nx,Benhar:2005gi,Urbanec:2013fs,Yagi:2013awa,Yagi:2013bca}.
Remarkably, It has been shown that this perturbative approach can be applied
with great accuracy for most observed NSs, even for most millisecond
pulsars~\citep{Berti:2003nb,Berti:2004ny,Benhar:2005gi}. In some models of NSs,
the relative errors compared to the results obtained by numerical relativity for
most quantities are less than $2\%$ if the spin frequency is less than $\sim 600
\,\rm Hz$~ \citep{Benhar:2005gi,Berti:2003nb,Berti:2004ny}. 

In this work, we use the Hartle-Thorne approximation to study the rotating SSs
to the third order of $\Omega$.  The match conditions at the boundary of the
star which are resulted from the finite surface density are crucially
considered.  We present systematic results for the moments of inertia, the
quadrupole moments, the eccentricities, changes in the gravitational and
baryonic masses, and universal relations between some of these quantities. The
measurement of the moment of inertia from Lense-Thirring precession can give us
constraints on the parameter space of SSs. The other physical quantities and the
universal relations are ready to be used to interpret astrophysical
observations, such as the light curves of X-ray hot spots on the surface of SSs
and GWs from binary SS mergers.

For coalescing binary compact stars, the finite size of the stars at the end of
inspiral can not be ignored.  Each star is deformed in the tidal field of the
companion. The energy goes to deform the star and the tidal induced quadrupole
moments will contribute to the GW
phasing~\citep{Hinderer:2007mb,Damour:2009vw,Dietrich:2020eud}.  The phase
evolution of the GWs will be faster compared to non-spinning stars with the same
component masses.  The GW170817 event gives the first constraint on the tidal
deformability of
NSs~\citep{LIGOScientific:2017vwq,LIGOScientific:2018cki,LIGOScientific:2018hze}.
In this paper, we study the tidal properties of SSs based on the work
in~\citet{LZX2019} and use the posterior from LIGO/Virgo collaborations to
put constraints on the parameter space of the Lennard-Jones model.

\citet{Yagi:2013awa,Yagi:2013bca} found a remarkable universal relation between
the moments of inertia, the tidal deformabilities, and the spin induced
quadrupole moments, also known as I-Love-Q relation.  The relation is nearly EoS
independent and the relative errors can be less than $1\%$ for various EoS
models, including NSs and QSs. Based on our calculations of rotation and tidal
deformation, we find that SSs in the Lennard-Jones model also satisfy
the I-Love-Q relation.

The organization of the paper is as follows. In Sec.~\ref{sec:eos}, we introduce
the Lennard-Jones model of SSs.  The structures of static SSs are discussed in
Sec.~\ref{sec:background}. Based on the background solutions, in
Sec.~\ref{sec:rotation}, we investigate the global properties of rotating SSs in
the Hartle-Thorne approximation.  In Sec.~\ref{sec:tidal}, we calculate the
tidal deformabilities of SSs and discuss the constraints on the parameter space
in our model.  Then we study the universal relations between the moment of
inertia, the tidal deformability, and the quadrupole moment for SSs in
Sec.~\ref{sec:universal}. Finally, we summarize our work in
Sec.~\ref{sec:discussion}. The ordinary differential equations that determine
the structures of slowly rotating relativistic stars are given in
Sec.~\ref{append:A}.

Throughout this paper, we use the geometric units where $G=c=1$. The convention
of the metric is $(-,+,+,+)$.

\section{Strangeon stars in the Lennard-Jones model}
\label{sec:eos}

In the Lennar-Jones model~\citep{jones1924determination}, the potential between
two strangeons is \citep{Lai:2009cn},
\begin{equation}
    \label{eqn:eos}
    u(r)=4 \epsilon\left[\left(\frac{\sigma}{r}\right)^{12}-\left(\frac{\sigma}{r}\right)^{6}\right]\,, 
\end{equation} 
where $\epsilon$ is the depth of the potential well, $r$ is the distance between
two strangeons, and $\sigma$ is the distance when $u(r)=0$. Though simple, this
model has the properties of long-range attraction and short-range repulsion.
Note that the $\sigma$-$\omega$ model, which is commonly used to describe the
interactions between nucleons, can also be well described by long-range
attraction and short-range repulsion~\citep{WALECKA1974491}.  The lattice
simulations of QCD indicate a short-distance
repulsion~\citep{Ishii:2006ec,Wilczek2007}.  Moreover, the repulsive hardcore is
essential to generate a stiff EoS for dense matter constituted of strangeons and
plays a fundamental role in determining the structures of SSs. 

We take simple cubic lattice structures and ignore the surface tension. The
potential energy density is \citep{Lai:2009cn}
\begin{equation}
    \rho_{\rm P}=2 \epsilon\left(A_{12} \sigma^{12} n^{5}-A_{6} \sigma^{6} n^{3}\right)\,,
\end{equation}
where $A_{12}=6.2$, $A_{6}=8.4$, and $n$ is the number density of strangeons.
The lattice of strangeons may form other structures in reality, but the
differences should be small and do not affect the structures of SSs
significantly. The energy density should also include the rest energy of
strangeons, $\rho_{\rm r}$, the kinetic energy from lattice vibrations, 
$\rho_{\rm L}$, and the kinetic energy from electrons, $\rho_{\rm e}$. 

The rest energy of a strangeon is $N_{\rm q}m_{\rm q}c^{2}$, where $m_{\rm q}$
is the mass of a quark and $N_{\rm q}$ is the number of quarks in a strangeon.
We take the mass of quarks to be $m_{\rm q}=300\,\rm MeV$, which is about one
third of the nucleon's mass. The parameter $N_{\rm q}$ is not known and we leave
it as a free parameter. In the calculations, we take $N_{\rm q}=18$ and $N_{\rm
q}=9$. A strangeon with 18 quarks is called quark-alpha~\citep{Michel:1991pk},
which is completely symmetric in spin, flavor, and color spaces.

The vibration energy can be obtained by using the Debye approximation of lattice. The vibrations of 
the cubic lattice can be decomposed into $3N$ independent oscillation modes~\citep{Lai:2009cn} and 
the vibration energy density $\rho_{\mathrm{L}}=9(6 \pi^{2})^{1/3} \hbar v n^{4/3}/8$, where $v$ is 
the speed of sound waves. Compared to the sum of potential energy and rest energy, the lattice vibration
energy is small even if one assumes $v$ is equal to the speed of light~\citep{Lai:2009cn}. For this reason, 
we ignore the contribution of lattice vibration energy in our calculations.

A small fraction of electrons is needed to keep the equilibrium of chemical
potential~\citep{Witten:1984rs,Alcock1986}.  In the MIT bag model of SQM,
electrons per baryon are determined by the mass of strange quarks $m_{\rm s}$
and the coupling constant $\alpha_{\rm s}$ between quarks~\citep{Farhi:1984qu}.
If one fixes the mass $m_{\rm s}$ of the strange quarks, a larger coupling
constant $\alpha_{\rm s}$ will lead to a larger fraction of electrons per
baryon.  When $\alpha_{\rm s}$ equals 0.3, the number of electrons per baryon is
smaller than $10^{-4}$~\citep{Farhi:1984qu,Lai_2009128,Lai:2009cn}.  In the case
of strangeons, we have applied the strong interactions between strangeons. The
number density of electrons should be small despite the concrete number is still
not clear. Even if we take $10^{-4}$, the Fermi energy of electrons still can be
ignored compared to the contribution of potential energy and rest-mass energy of
strangeons~\citep{Lai:2009cn,LZX2019}.

The total energy density of the zero temperature dense matter composed of
strangeons reads
\begin{equation}
    \label{eqn:energy_density}
    \rho=2 \epsilon\left(A_{12} \sigma^{12} n^{5}-A_{6} \sigma^{6} n^{3}\right) + nN_{\rm q}m_{\rm q}c^{2}\,.
\end{equation}
From the first law of thermodynamics, one derives the pressure
\begin{equation}
    \label{eqn:pressure}
    P=n^{2} \frac{\mathrm{d}(\rho / n)}{\mathrm{d} n}=4 \epsilon\left(2 A_{12} \sigma^{12} n^{5}-A_{6} \sigma^{6} n^{3}\right)\,.
\end{equation} 
At the surface of the SSs, the pressure becomes zero and we obtain the surface
number density of strangeons as $\big(A_{6}/(2A_{12}\sigma^{6})\big)^{1/2}$. For
convenience, we transform it to the 
number density of baryons 
\begin{equation}
    \label{eqn:surface}
    n_{\rm s}=\left(\frac{A_{6}}{2A_{12}}\right)^{1/2}\frac{N_{\rm q}}{3\sigma^{3}}\,.
\end{equation}

For a given number of quarks $N_{\rm q}$ in a strangeon, the EoS of SSs is
completely determined by the depth of the potential $\epsilon$ and the number
density of baryons $n_{\rm s}$ at the surface of the star. 
The nucleon-nucleon scattering data indicate that the internucleon potential well lies in the 
range of $\sim$ $50$--$120$$\,\rm MeV$ for the $^{1}S_{0}$ (spin singlet and $s$-wave) channel
~\citep{Wiringa:1994wb,Stoks:1994wp,Machleidt:2000ge}; see Fig.~1 of \citet{Ishii:2006ec}. Since the strong interactions are not sensitive to 
the flavor of quarks, we choose $\epsilon$ spanning in the range of $20$--$100\,\rm MeV$, 
which is similar to the internucleon potential and enough to trap the strangeons in 
the potential well. The surface baryonic density $n_{\rm s}$ should be in the same order as 
the nuclear saturation density, $n_{0}=0.16\,\rm fm^{-3}$.
But unlike proton or neutron with 3 quarks, we take $N_{\rm q}$ to be 9 or 18 for strangeons. 
The interactions may group the quarks more compactly compared to 
nuclei with the same number of quarks. 
Therefore, we let $n_{\rm s}$ lie in the range of $0.24\,\rm fm^{-3}$ to $0.36\,\rm fm^{-3}$, 
which corresponds to $1.5\,n_0$ to $2.25$$\,n_{0}$.
\begin{figure}
    \centering
    \includegraphics[width=8cm]{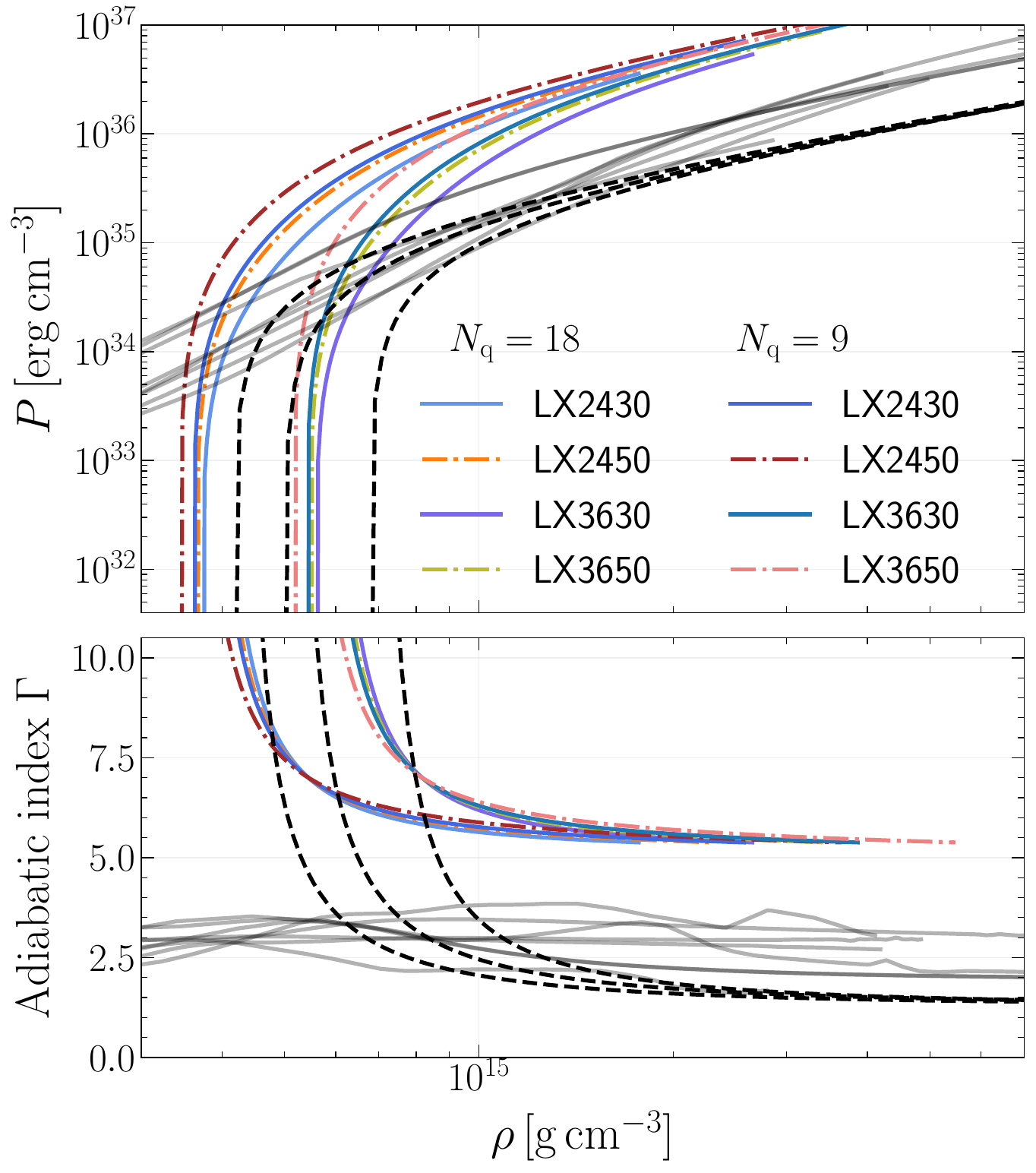}
    \caption{The upper panel shows the relation between mass--energy density
    $\rho$ and pressure $P$.  The Lennard-Jones models of SSs with different
    choices of the potential depth $\epsilon$, the surface number density of
    baryons $n_{\rm s}$, and the number of quarks in a strangeon $N_{\rm q}$ are
    displayed.  For comparison, we also show the EoS models of normal NSs (grey)
    and QSs (black). Correspondingly, the lower panel shows the adiabatic index
    $\Gamma$ for SSs (colored), normal NSs (grey), and QSs (black). We name EoS
    of SSs according to different combinations of $n_{\rm s}$ and $\epsilon$.
    For example, ``LX2450'' means the surface baryon number density $n_{\rm
    s}=0.24\,\rm fm^{-3}$ and the potential depth $\epsilon=50\,\rm {MeV}$.}
    \label{fig:eos}
\end{figure}

In the upper panel of Fig.~\ref{fig:eos}, we show the EoS of SSs for $N_{\rm
q}=18$ and $N_{\rm q}=9$.  Different surface densities and potential depths are
chosen. We also present the EoS of normal NSs and
QSs~\citep{Lattimer:2000nx,Lackey:2005tk}, including the models for normal NSs
and SQMs for QSs. Compared to normal NS models, the surface densities of SSs and
QSs are not zero and in the order of nuclear density, which originate from the
fact that SSs and QSs are self-bound systems while normal NSs are
gravitational-bound systems.

One can surely neglect the entropy gradient for the zero temperature EoS of SSs.
Therefore, the increases in pressure and density toward the center of the star
are adiabatic~\citep{Shapiro1983}.  The adiabatic index $\Gamma$ is defined as
\begin{equation}
   \Gamma=\frac{n}{P}\frac{{\rm d} P}{{\rm d} n}=\frac{\rho+P}{P}\frac{\partial P}{\partial \rho}\,,
\end{equation}
which determines the change of pressure $P$ associated with the variations of
the local density $n$ of particles \citep{Shapiro1983}. For relativistic stars
in equilibrium, the adiabatic index $\Gamma$ measures the stiffness of the EoS.
In the lower panel of Fig.~\ref{fig:eos}, we show the relation between the
adiabatic index $\Gamma$ and the mass-energy density $\rho$. One notices that
the adiabatic index for SSs is larger than that of NSs, which indicates that SSs
are stiffer than normal NSs.  The adiabatic index for QSs has the same trend as
the SSs. But it becomes much smaller than that of SSs at high densities. The EoS
of QSs is taken as $P=(\rho-4B)/3$, where $B$ is the bag constant. At low
densities, $\Gamma$ goes up since the bag constant is very large. But as the
densities increase, the free quarks in QSs become relativistic and the EoS is
softened. For SSs, the repulsive hardcore and the non-relativistic nature of the
particles make the EoS always be stiff.

One main concern of the Lennard-Jones model for SSs is that the adiabatic sound
speed $c_{\rm s}=(\partial P/ \partial \rho)^{1/2}$ turns into superluminal in
high pressure. The possibility that the adiabatic sound speed $c_{\rm s}$ in
ultra-dense matter exceeds the speed of light has been discussed in some
literature~\citep{Bludman:1968zz,Ruderman:1968zz,Caporaso:1979uk,Caporaso:1983hwn,Ellis:2007ic}.
The central question is: what does $\partial P/ \partial \rho$ mean for a
relativistic fluid? The expression $\partial P/ \partial \rho$ is borrowed from
Newtonian hydrodynamics and comes from a static calculation of the EoS
$P=P(\rho)$ by ignoring the dynamics of the medium~\citep{Lai:2009cn}. One
assumes the infinite speed of interactions and finite temperature. The static
sound speed $(\partial P/ \partial \rho)^{1/2}$ agrees with the dynamical one.
However, for relativistic ultra-dense matter, if one assumes zero temperature
and finite speed of the interactions between particles, the adiabatic sound
speed is no longer a dynamically meaningful speed of the disturbances, but only
a measurement of local stiffness~\citep{Caporaso:1983hwn,Lai:2009cn}.  From the
underlying microscopic picture, \citet{Bludman:1968zz} and
\citet{Caporaso:1979uk} gave dynamical calculations that particles in dense
matter interacting with one another by retarded fields. They found that the
propagating sound waves always have a speed less than or equal to the speed of
light although $\partial P/ \partial \rho>1$. 

For SSs, the bulk of the stars are in a solid state and strangeons form lattice.
Inspired by \citet{Bludman:1968zz} and \citet{Caporaso:1979uk},
\citet{Lu:2018kls} carried out a 1$-$dimensional chain model to calculate the
dynamical speed of the sound waves in SSs. They found that the causality
condition is always satisfied although $(\partial P/ \partial \rho)^{1/2}$ can
be larger than the speed of light. The ultra stiffness and the violation of
commonly used causality limit can lead to many interesting global properties of
SSs. Interested readers are referred to the above literature for details.


\section{Equilibrium background of spherical and static stars}
\label{sec:background}

The line element of an isolated and non-spinning relativistic star can be
written as
\begin{equation}
    \label{eqn:static}
       {\rm d} s^{2}=g_{\alpha\beta}\dd x^{\alpha}\dd x^{\beta}= -e^{\nu} \dd t^{2}+e^{\lambda} 
       \dd r^{2}+r^{2}\left(\dd \theta^{2}+\sin ^{2} \theta \dd \phi^{2}\right) \,,
\end{equation}
where $\nu$ and $\lambda$ are functions of $r$. Since the star is static, we
take the four-velocity $u^{\alpha}$ as 
\begin{equation}
    u^{t}=e^{-\nu/2},\quad u^{i}=0\,, \quad\quad \left(i=r\,,\theta\,,\phi\right)\,.
\end{equation} 
We approximate the stress-energy tensor of SSs as perfect fluid, 
\begin{equation}
    \label{eqn:fluid}
    T_{\alpha \beta}=(\rho+P) u_{\alpha} u_{\beta}+P g_{\alpha \beta}\,,
\end{equation}
where $\rho$ and $P$ are the energy density and the pressure. By taking
$e^{\lambda}=1/(1-2m/r)$ and substituting Eq.~(\ref{eqn:static}) and
Eq.~(\ref{eqn:fluid}) in the Einstein equations, one obtains the
Tolman-Oppenheimmer-Volkoff (TOV) equations for spherical and static
relativistic stars, 
\begin{align}
    \label{eqn:dmdr}
    & \frac{\dd m}{\dd r}=4 \pi r^{2} \rho \,,\\
    &\frac{\dd \nu}{\dd r}=\frac{2( m+4 \pi  r^{3} P)}{r(r-2 m)}\,,\\
    \label{eqn:tov}
    &\frac{\dd P}{\dd r}=-\frac{(\rho+P)\left( m+4 \pi  r^{3} P\right)}{r(r-2  m)}\,,
\end{align}
where $m$ is the gravitational mass enclosed in radius $r$. Integrating
Eqs.~(\ref{eqn:dmdr}--\ref{eqn:tov}) with the boundary conditions at the center
of the star,
\begin{equation}
    m (r) \mid_{r=0} = 0\,, \quad \nu (r) \mid_{r=0} = \nu_{\rm c}\,, \quad P(r) \mid_{r=0} = {P}_{\rm c}\,,
\end{equation}
and the EoS of SSs in Eqs.~(\ref{eqn:energy_density}--\ref{eqn:pressure}), one
obtains the stellar structures of isolated and non-spinning SSs. Here $P_{\rm
c}$ is the pressure at the center of the star.  The constant $\nu_{\rm c}$ can
be determined by matching the interior and exterior solutions at the boundary of
the star.

\begin{figure}
    \centering
    \includegraphics[width=8cm]{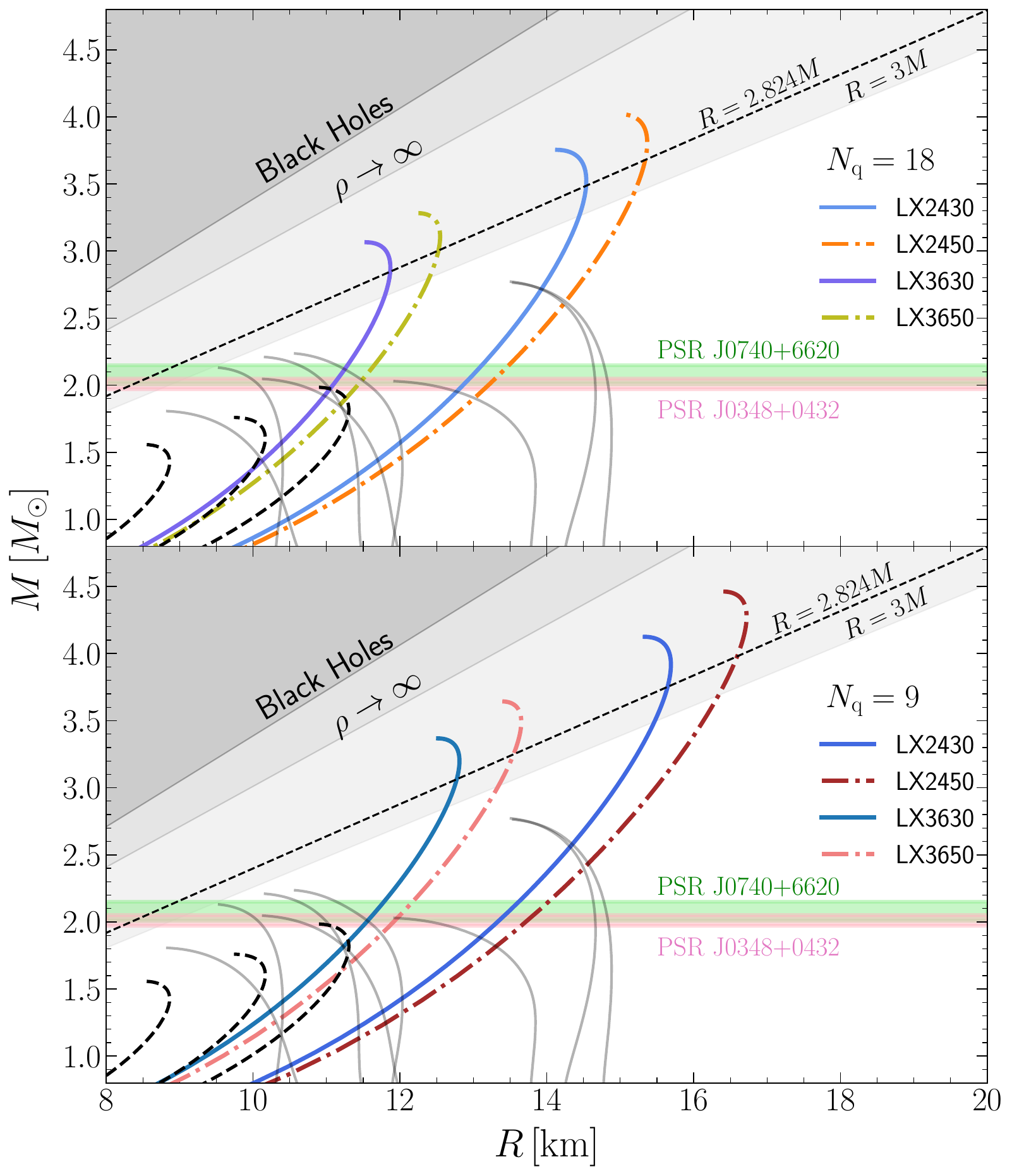}
    \caption{Mass radius relations of SSs for $N_{\rm q}=18$ ({\it upper panel})
    and $N_{\rm q}=9$ ({\it lower panel}) with different combinations of the
    surface baryonic density $n_{\rm s}$ and the potential depth $\epsilon$. 
    For comparison, we also show the mass radius relations for normal NSs (grey)
    and QSs (black).  The black hole limit ($R=2M$), the limit for the central
    density to be infinite ($R=9M/4$), the limit for $\dd P/\dd \rho=1$
    ($R=2.824M$), the photon spheres ($R=3M$), and the $1$-$\sigma$ region of the
    mass measurements of PSR~J0348+0432 \citep{Antoniadis1233232} and
    PSR~J0740$+$6620 \citep{Fonseca:2021wxt} are illustrated.}
    \label{fig:mr}
\end{figure}

In Fig.~\ref{fig:mr}, we display the mass-radius relations of SSs with different
choices of $N_{\rm q}$, $n_{\rm s}$, and $\epsilon$. Because the EoS of SSs in
the Lennard-Jones model is quite stiff and the maximal mass of SSs can be over
$3M_{\odot}$ for large parameter space~\citep{Xu:2018zdo,Lu:2018kls}. The
discoveries of massive pulsars, PSR~J0348+0432~\citep{Antoniadis1233232} and
PSR~J0740$+$6620~\citep{Fonseca:2021wxt}, at $\sim2 \,M_{\odot}$ via pulsar
timing support the stiffness property and more massive ones (e.g., $\geq 2.5 \,
M_{\odot}$) are expected for our model. 

SSs are self-bounded by strong interactions. The trend of the mass-radius
relation is basically the same as that of QSs. We can understand it with the
help of the adiabatic index $\Gamma$ shown in the lower panel of
Fig.~\ref{fig:eos}. At low densities, $\Gamma \rightarrow \infty$ and the
gravitational field is weak, which leads to $R \propto  M^{1 / 3}$. As the
central densities become larger, the GR effect becomes dominant and it results
in the formation of a maximal mass. For QSs, the quarks are nearly 
free and the interactions are added in a perturbative way. As the central density 
increases, the quarks become more and more relativistic and the EoS is softened with 
$\Gamma \rightarrow 4/3$. Therefore, the maximal mass can hardly reach $2\,M_{\odot}$~\citep{Lattimer:2000nx}. 
While for SSs, the quarks are grouped into non-relativistic clusters and the interactions 
between strangeons are non-perturbative. We conjecture that the hardcore exists for strangeons just like that 
for nuclei \citep{Ishii:2006ec,Wilczek2007}, and use the Lennard-Jones model to characterize this important feature. 
The hardcore will make the EoS very stiff. The adiabatic index $\Gamma$ is much larger than 
QSs at high densities (see Fig.~\ref{fig:eos}). Therefore, the maximal mass $\gtrsim 3\,M_{\odot}$ is possible.

The detailed behaviors of the mass-radius relations for different choices of
parameters can be basically understood as follows. For given $n_{\rm s}$, a
smaller $N_{\rm q}$ means that the distance $\sigma$ should be smaller according
to Eq.~(\ref{eqn:surface}). Compared to the case of $N_{\rm q}=18$, the
adiabatic index $\Gamma$ for $N_{\rm q}=9$ is larger and the EoS is stiffer.
Consequently, for given $n_{\rm s}$ and $\epsilon$, the radius at given mass and
the maximal mass are larger for $N_{\rm q}=9$.  If $N_{\rm q}$ is fixed, the
parameters $n_{\rm s}$ and $\epsilon$ completely determine the EoS.  Both  the
increase of the potential depth $\epsilon$ and the decrease of surface density
$n_{s}$ make a stiffer EoS since the repulsive force is amplified. For example,
the LX2430 is softer than LX2450, but stiffer than LX3630.

\begin{figure}
    \centering
    \includegraphics[width=8cm]{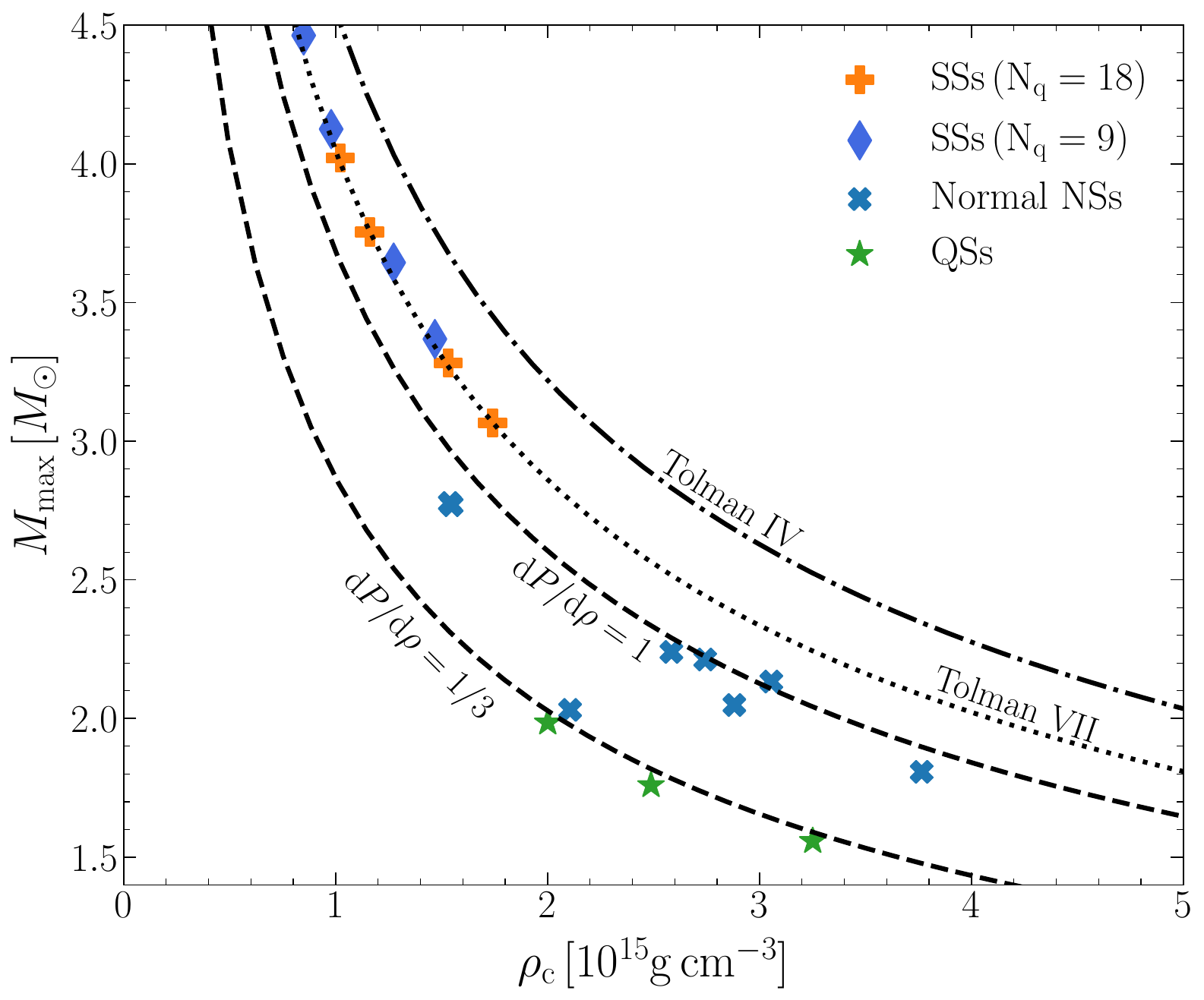}
    \caption{The relation between the maximal mass and the central mass--energy
    density for SSs, normal NSs, and QSs.  The $M_{\rm max}$-$\rho_{\rm c}$
    relation for $\dd P/\dd \rho=1$ and $\dd P/\dd \rho=1/3$ are showed. The
    former bounds the central densities of NSs while SSs can pass over this
    line.  The latter effectively bounds the existence of deconfined quarks in
    stars.  The Tolman IV and Tolman VII solutions coupled with $\dd P/\dd
    \rho=1$ are also illustrated \citep{Lattimer:2004sa,Lattimer:2012nd}.}
    \label{fig:m_rhoc}
\end{figure}

Another important feature is that the mass-radius relations of SSs can invade
into the commonly used causality limit $R=2.824M$ \citep{Lattimer:2012nd}. This
limit assumes that the EoS has a single parameter $\rho_{0}$ and satisfies the
relation $P=0$ for $\rho<\rho_{0}$ and $P=\rho-\rho_{0}$ for $\rho\geq\rho_{0}$.
Integrating TOV equations with this EoS, one obtains the maximal mass of NSs to
be $M_{\rm max }\simeq 4.09 (\rho_{\rm s}/\rho_{0})^{1/2}
M_{\odot}$~\citep{Lattimer:2012nd}, where $\rho_{\rm s}$ is the nuclear
saturation density. If one knowns the EoS approximately up to $2\rho_{\rm s}$,
the maximal mass of NSs should be about $3 \, M_{\odot}$ and the central density
is constrained to  $\rho_{\rm c} \leq 50.8 \rho_{\rm s}\left(M_{\odot} / M_{\rm
max }\right)^{2}$~\citep{Rhoades:1974fn,Lattimer:2004sa,Lattimer:2012nd}, which
is denoted as $\dd P/\dd \rho=1$ in Fig.~\ref{fig:m_rhoc}. As we suggested
before, this limit just represents the fact that the EoS is very stiff but the
causality condition is still satisfied at least for SSs.  Therefore, the maximal
mass can be easily larger than $3 \, M_{\odot}$ and the $M_{\rm max}$-$\rho_{\rm
c}$ relation for SSs can surpass this limit. 

The MIT bag model of QSs satisfies $P=(\rho-4B)/3$ and $\dd P/\dd \rho$ equals
1/3 all over the star.  The central density is constrained to $\rho_{\rm c} \leq
30.8 \rho_{\rm s}\left(M_{\odot} / M_{\rm max }\right)^{2}$, which is a factor
of 0.607 lower than hadronic NSs
\citep{Rhoades:1974fn,Lattimer:2004sa,Lattimer:2012nd}.  The $M_{\rm
max}$-$\rho_{\rm c}$ relation in this condition is also presented in
Fig.~\ref{fig:m_rhoc}.  As mentioned by~\citet{Lattimer:2012nd}, this relation
not only constrains the central densities of pure SQM but also bounds a
significant deconfined quarks that are mixed in normal NSs. This feature is
vital to distinguish QSs and SSs. If we believe that the pulsar-like compact
objects have significant number of $s$ quarks, the quarks may not be deconfined
and SSs are supported in some way once much more massive pulsars (e.g., $\geq2.5
\, M_{\odot}$) are found.

To compare the analytical solutions of the Einstein equations with numerical
solutions with modelled EoS can give us some insights to the nature of
pulsar-like compact objects. In Fig.~\ref{fig:m_rhoc}, we also illustrated the
Tolman VII solution for gravitational bound systems and Tolman IV solution for
self-bound systems \citep{Tolman:1939jz,Lattimer:2004sa} that coupled with $\dd
P/\dd \rho=1$. For normal NSs and QSs, the Tolman VII solution sets a stringent
upper limit to the central density of the maximal mass~\citep{Lattimer:2004sa}. 
However, the SSs in the parameter space that we used can surpass the Tolman VII
solution.  The Tolman IV solution still has larger maximal masses than the SSs.

In the following sections, we will investigate the slow rotation and the tidal
deformation of SSs as perturbations on the background solution of spherical and
static stars that are given above.  
 
\section{Slowly rotating strangeon stars}
\label{sec:rotation}

To the order $\Omega^{3}$, the line element of a slowly rotating relativistic star 
is \citep{Hartle:1967he,Hartle:1968si,Hartle:1973zza,Benhar:2005gi}
\begin{align}
    \label{eqn:hartle_thorne}
    &\dd s^{2}=-e^{\nu(r)}[1+2 h(r, \theta)] \dd t^{2}+e^{\lambda(r)}\left
    [1+\frac{2 m^{*}(r, \theta)}{r-2 m(r)}\right] \dd r^{2} \nonumber\\
    &+r^{2}[1+2 k_{2}(r, \theta)]\left\{\dd \theta^{2}+\sin ^{2} \theta[\dd \phi-w(r, \theta) 
    \dd t]^{2}\right\} +O(\Omega^{4})\,.
\end{align}
The metric is invariant under the combined transformations of $t\rightarrow -t$
and $\phi \rightarrow -\phi$.  Therefore, the function $w(r,\theta)$ only
contains odd orders of $\Omega$ while the functions $h(r,\theta)$, $m^{*}(r,
\theta)$, and $k_{2}(r, \theta)$ only include the even orders of $\Omega$.  We
can expand these perturbative corrections with the spin-weighted spherical
harmonics \citep{Hartle:1967he,Hartle:1973zza}. 

The functions $h(r,\theta)$, $m^{*}(r, \theta)$, and $k_{2}(r, \theta)$ are of
order $\Omega^{2}$ and only contain $l=0$ and $l=2$ terms, 
\begin{align}
    h(r, \theta)=&h_{0}(r)+h_{2}(r) P_{2}(\cos\theta)+O\left(\Omega^{4}\right)\,,\\
    m^{*}(r, \theta)=&m_{0}(r)+m_{2}(r) P_{2}(\cos\theta)+O\left(\Omega^{4}\right)\,,\\
    k_{2}(r, \theta)=&\left[v_{2}(r)-h_{2}(r)\right] P_{2}(\cos\theta)+O\left(\Omega^{4}\right)\,,
\end{align}
where $P_{2}(\cos\theta)$ is the Legendre polynomial with $l=2$ and the function
$v_{2}(r)$ is introduced for simplicity. Note that the $l=0$ contribution to
$k_{2}$ has been eliminated by a coordinate transformation.  The function
$w(r,\theta)$ can be expanded as~\citep{Hartle:1973zza,Benhar:2005gi}
\begin{equation}
    w(r, \theta)=\omega(r)+w_{1}(r)-w_{3}(r) \frac{1}{\sin \theta} 
    \frac{\dd P_{3}(\cos\theta)}{\dd \theta}+O\left(\Omega^{5}\right)\,.
\end{equation}
Here $P_{3}(\cos\theta)$ is the Legendre polynomial with $l=3$. The function
$\omega(r)$ is the $l=1$ term in the first order of $\Omega$. The functions
$w_{1}(r)$ and $w_{3}(r)$ are of order $\Omega^{3}$, which represent $l=1$ and
$l=3$ terms respectively. 

In the Hartle-Thorne coordinate, the four-velocity of the fluid can be
represented as 
\begin{align}
    u^{t}=&\left[-\left(g_{t t}+2 \Omega g_{t \varphi}+\Omega^{2} g_{\varphi \varphi}\right)\right]^{-1 / 2}\nonumber\\
    =&e^{-\nu / 2}\left[1+\frac{1}{2} r^{2} \sin ^{2} \theta \bar{\omega}^{2} e^{-\nu}-h_{0}-h_{2} P_{2}\right]\,,\nonumber\\
    u^{r}=&u^{\theta}=0\,, \quad u^{\varphi}=\Omega u^{t}\,,
\end{align}
where the quantity, 
\begin{equation}
    \bar{\omega}(r)=\Omega-\omega(r)\,,
\end{equation}
represents the angular velocity of the fluid element relative to the local
inertial frame to order $\Omega$.  It plays an important role in determining
stellar structures.

For rotating stars, the fluid elements are displaced. To guarantee the self
consistency of the perturbation theory, \citet{Hartle:1967he} used a special
coordinate system that maps the isodensity surface which lies at coordinate 
$(\tilde R,\tilde \Theta)$ in an unperturbed star to 
\begin{equation}
    r=\tilde R+\xi_{0}(\tilde R)+\xi_{2}(\tilde R) P_{2}(\cos \tilde \Theta)+ O(\Omega^{4})\,,\quad \theta=\tilde \Theta \,,
\end{equation}
for the rotating star. The displacements $\xi_{0}$ and $\xi_{2}$ describe the
spherical and quadrupole deformations of the star separately. In this way, the
pressure and density are known functions for both of the non-rotating and
rotating configurations. It is formally equivalent to work in the original ($r,
\theta$) coordinate with the variations of pressure $P$, energy density $\rho$,
and baryonic density $\bar \rho$ to be~\citep{Hartle:1967he,Hartle:1968si}
\begin{align}
    \delta P=&(\rho+P)\left[p_{0}+p_{2}P_{2}(\cos\theta)\right]\,,\\
    \delta \rho=&(\rho+P)\frac{{\rm d} \rho}{{\rm d} P}\left[p_{0}+p_{2}P_{2}(\cos\theta)\right]\,,\\
    \delta \bar \rho=&(\rho+P)\frac{{\rm d} \bar \rho}{{\rm d} P}\left[p_{0}+p_{2}P_{2}(\cos\theta)\right]\,.
\end{align}
The dimensionless quantities $p_{0}$ and $p_{2}$ are defined as
\begin{equation}
    \label{eqn:p0p2}
    p_{0}(r)=-\xi_{0}(r)\left[\frac{1}{\rho+P} \frac{\dd P}{\dd r}\right]\,, 
    \quad p_{2}(r)=-\xi_{2}(r)\left[\frac{1}{\rho+P} \frac{\dd P}{\dd r}\right]\,,
\end{equation}
which are functions of $r$ and evaluate the pressure perturbation.  Then, the
stress-energy tensor of the slowly rotating star reads 
\begin{equation}
    \label{eqn:stress_energy}
    T_{\alpha}^{\ \beta}=-(\rho+\delta \rho+P+\delta P) u_{\alpha} u^{\beta}+(P+\delta P) \delta_{\alpha}{ }^{\beta}\,.
\end{equation}

The structures of slowly rotating relativistic stars are determined by the
perturbative functions $\omega$, $h_{0}$, $m_{0}$, $p_{0}$, $h_{2}$, $m_{2}$,
$v_{2}$, $p_{2}$, $w_{1}$, and $w_{3}$, which can be calculated from systematic
differential equations with appropriate boundary conditions.  The types of stars
that we want to solve determine the boundary conditions. A rigidly rotating star
is specified by two parameters which can be diversely taken as the central
density $\rho_{\rm c}$ and the angular velocity $\Omega$, or the baryonic mass
$\bar{M}$ and the angular momentum $J$, or other
combinations~\citep{Hartle:1973zza,Stergioulas:2003yp}. 

The constant central density sequence is commonly used and can be proceeded as follows.
    (I) Choose a central density $\rho_{\rm c}$ and integrate the TOV equation
    with a given EoS.  The structure of the static and spherical background is
    determined. One obtains the gravitational mass $M$, the baryonic mass
    $\bar{M}$, and the radius of star $R$.
    (II) Keep the central density fixed and give a rigid angular frequency
    $\Omega$ to the star.  Then, one calculates the corrections to the first,
    the second, and the third order of $\Omega$. 
This procedure is first formulated in \citet{Hartle:1967he} to the second order
of $\Omega$ and has been used in numerous literature to discuss slowly rotating
relativistic stars
\citep{Hartle:1968si,Chandra1974,Weber1991,Weber1992,Berti:2004ny,Urbanec:2013fs,Yagi:2013awa,Yagi:2013bca}.

An advantage of this method is that all the configurations with different
angular velocities but the same central density can be obtained by rescaling a
single case with a specific angular velocity. We take the angular velocity
$\Omega_{*}=(M/R^{3})^{1/2}$ as the reference angular frequency in our
calculations. For normal NSs, this angular frequency basically represents the limit when the matters on the
equator of the star are shed. The mass $M$ and radius $R$ are that of
non-rotating configuration.  In practice, we first calculate a physical quantity
$F^{*}_{n}$ at order $n$ for a given central density and the angular frequency
$\Omega_{*}$. Then the quantity $F_{n}$ for a smaller frequency $\Omega$, where
the slow rotation approximation is satisfied, can easily be obtained by 
\begin{equation}
    F_{n}=(\Omega/\Omega_{*})^{n}F^{*}_{n}\,,\quad n=1, 2, 3\,.
\end{equation}

We present detailed differential equations and boundary conditions for $\omega$,
$h_{0}$, $m_{0}$, $p_{0}$, $h_{2}$, $m_{2}$, $v_{2}$, $p_{2}$, $w_{1}$, and
$w_{3}$ in the Appendix~\ref{append:A}. In the following sections, we will
discuss the physical quantities related to the rotation at each order. Since the
change in moment of inertia for a given baryonic mass is important in some
physical processes, such as pulsar glitches and spin evolutions, we will also
study the constant baryonic mass sequence in the third order of $\Omega$ and the
corrections to the moment of inertia as well.

\subsection{First order: Angular momentum, moment of inertia, and the dragging of the local inertial frame}

The axial symmetry of the system leads to the existence of a conserved angular
momentum current  
\begin{equation}
    {J^{\alpha}_{\rm tot}}=T^{\alpha \beta}\eta_{\beta}\,,
\end{equation}
where $\eta_{\beta}$ is the Killing vector corresponding to the rotation
symmetry. A conserved total angular momentum $ J_{\rm tot}$ can be defined by
integrating $J_{\rm tot}^{\alpha}$ over any space-like hypersurface
\citep{Hartle_Sharp1967,Hartle:1973zza,Misner:1974qy}. We can naturally choose
the $t=\rm constant$ hypersurface and the total angular momentum is 
\begin{equation}
    \label{eqn:J_total}
    J_{\rm tot}=\int J^{0} \dd V=\int \sqrt{-g} T_{\ \phi}^{t} \dd^{3} x= 
    \frac{1}{8\pi} \int \sqrt{-g} R_{\ \phi}^{t} \dd^{3} x \,,
\end{equation}
where $g$ is the determinant of the 4-dimensional metric and $\dd V=\sqrt{-g}\dd
^{3}x$ is the proper volume element.  To the first order of $\Omega$, the star
remains to be spherical and the angular momentum $J$ is of order $\Omega$, 
\begin{equation}
    J=\frac{1}{6}\left[r^{4} j(r) \frac{\dd \bar{\omega} }{\dd r}\right] {\bigg|_{r=R}}\,,
\end{equation}
where we have introduced $j(r)=e^{-(\nu+\lambda)/2}$. Then the moment of inertia
can be calculated with $I=J/\Omega$, which is a zeroth-order quantity and only
depends on the structure of spherical and static background solution.  

\begin{figure}
    \centering
    \includegraphics[width=8cm]{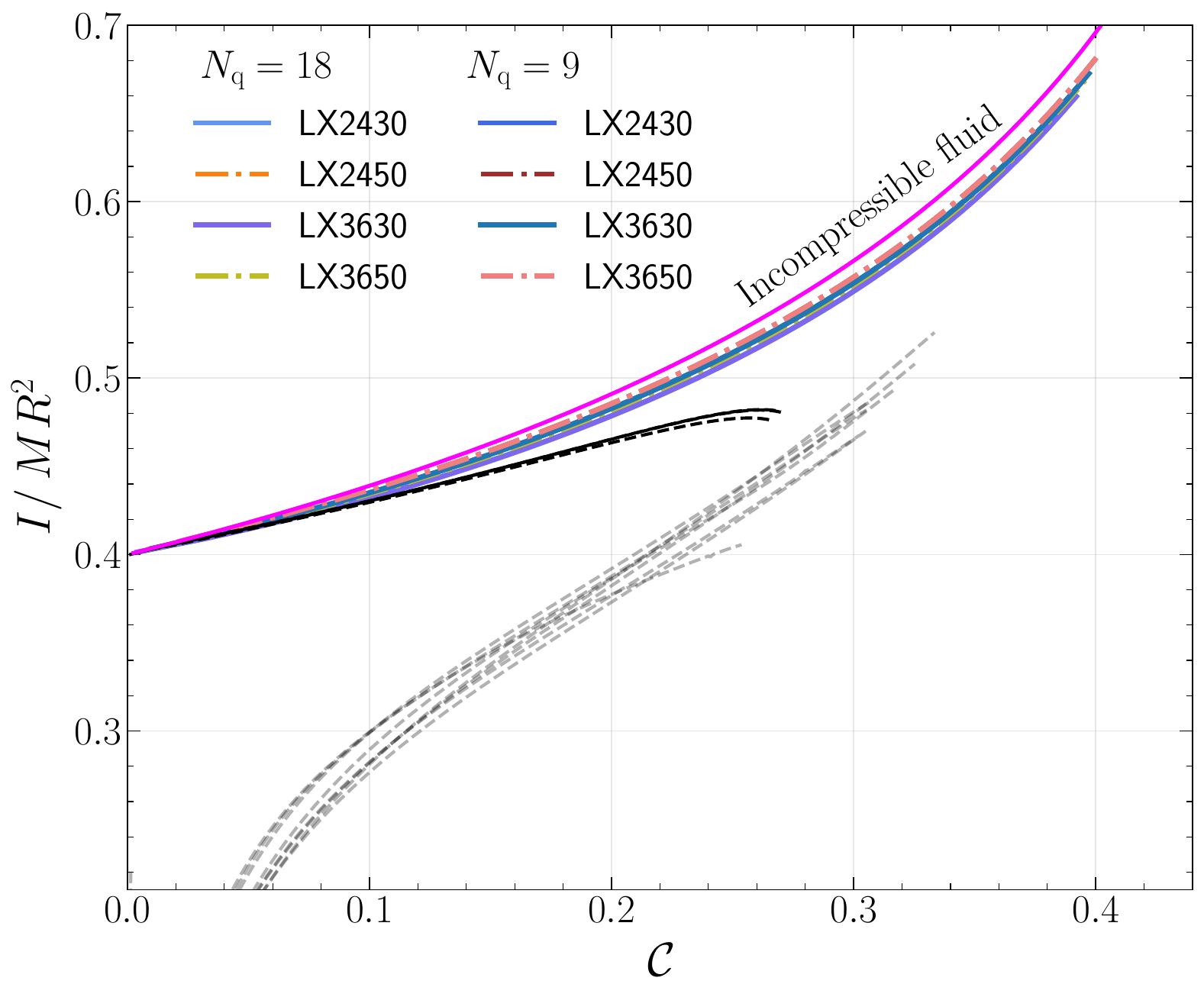}
    \caption{The relation between dimensionless moment of inertia $I/MR^{2}$ and
    the compactness $\mathcal{C}$ for $N_{\rm q}=18$ and $N_{\rm q}=9$. For
    comparison, we also plot the relations for normal NSs (grey), QSs (black),
    and incompressible fluid (magenta).}
    \label{fig:I_C}
\end{figure}

For normal NSs and QSs, it has been shown by~\citet{Lattimer:2000nx}
and~\citet{Bejger:2002ty} that the dimensionless quantity $I/MR^{2}$ and the
compactness $\mathcal{C}\equiv M/R$ satisfy two distinct EoS-insensitive
quasi-universal relations. In Fig.~\ref{fig:I_C}, we plot the relation between
$I/MR^{2}$ and $M/R$ for normal NSs and QSs. For SSs, we find that different
parameters also indicate a nearly EoS-insensitive curve, which is distinct from
normal NSs and also deviates from QSs in the condition of large compactness. As
the compactness $M/R \rightarrow 0$, The quantity $I/MR^{2}$ for SSs and QSs
tends to be the value for incompressible fluid in Newtonian gravity, namely 0.4.
Moreover, as the compactness increases, QSs deviate from the incompressible
fluid limit while SSs are still  very close to this limit, which results from
the fact that SSs have hardcore at short distances and the EoS is much stiffer
than QSs. 

A spinning pulsar in binary system will drag the frame and introduce
relativistic spin-orbit couplings~\citep{Barker:1975ae}. This coupling between
the orbit angular momentum and the spin angular momenta, also known as the
Lense-Thirring precession, is related to the moment of inertia of the spinning
pulsar. It produces two observable effects which could be observed with pulsar
timing \citep{Damour:1988mr,Lattimer:2004nj}. First, the spin angular momenta of
the pulsars will precess around the total angular momentum of the binary system
if the spin and the orbit angular momenta are not aligned. Since the total
angular momentum is conserved,\footnote{The losses of the angular momentum due
to the GW radiation are higher-order contributions, which can be neglected in
this problem.} the precession of the spins will induce a compensating
precession of the orbit angular momentum and the orbit inclination angle will
change correspondingly. Second, the spin-orbit coupling makes an advance of the
periastron of the orbit \citep{Hu:2020ubl}.

The spin of one star (component $\rm A$) usually spins much faster than the
companion (component $\rm B$).  The contribution to the Lense-Thirring effect of
star $\rm B$ can usually be neglected. The precession of the orbital plane
causes a periodic deviation of the time-of-arrival of pulses from pulsar $\rm
A$.  The period departure is proportional to $I_{\rm A}\Omega_{A}\cos i$
\citep{Lattimer:2004nj}, where $I_{\rm A}$ and $\Omega_{\rm A}$ are the moment
of inertia and spin angular velocity of pulsar A, and $i$ is the inclination
angle of the orbit plane. The periastron advance due to Lense-Thirring effect is
proportional to $I_{\rm A}\Omega_{A}$, which is a tiny effect compared to the
first post-Newtonian (PN) term and is opposite to the direction of orbital
motion. One can notice that the moment of inertia will enter into those
observational effects. Thus the search of Lense-Thirring effect of pulsars will 
tell us information on the moment of inertia.  

The most promising candidate of detecting Lense-Thirring precession is the
double pulsar system PSR J0737$-$3039A/B.  This relativistic system (orbital
period $\simeq 2.45\,\rm h$) contains two pulsars with spin period $P_{\rm
A}\simeq 22\,\rm ms$ and $P_{\rm B}$ about 122 times larger than A
\citep{Burgay:2003jj}. One can neglect the contribution to Lense-Thirring effect
of pulsar B. The angle between the spin angular momentum of pulsar A and the
orbital angular momentum is very small, which makes the periodic modulations due
to the precession of the orbital plane hard to measure since $\cos i\simeq 0$.
However, it is found that the advance of the periastron is possible to be
measured to a accuracy of $\sim 10 \%$ in several years of
timing~\citep{Kramer_2009}.  This measurement can put important constraints on
the EoS of NSs~\citep{Lattimer:2004nj}. Recently, \citet{Hu:2020ubl} 
investigated the prospects for constraining the moment of inertia of pulsar A in
details by simulating the timing observations with the MeerKAT and the SKA
\citep{Shao:2014wja, Weltman:2018zrl}. The results suggest a measurement of
moment of inertia $I_{\rm A}$ to an accuracy of $11\%$ at 68$\%$ confidence
level. 

\begin{figure}
    \centering
    \includegraphics[width=8cm]{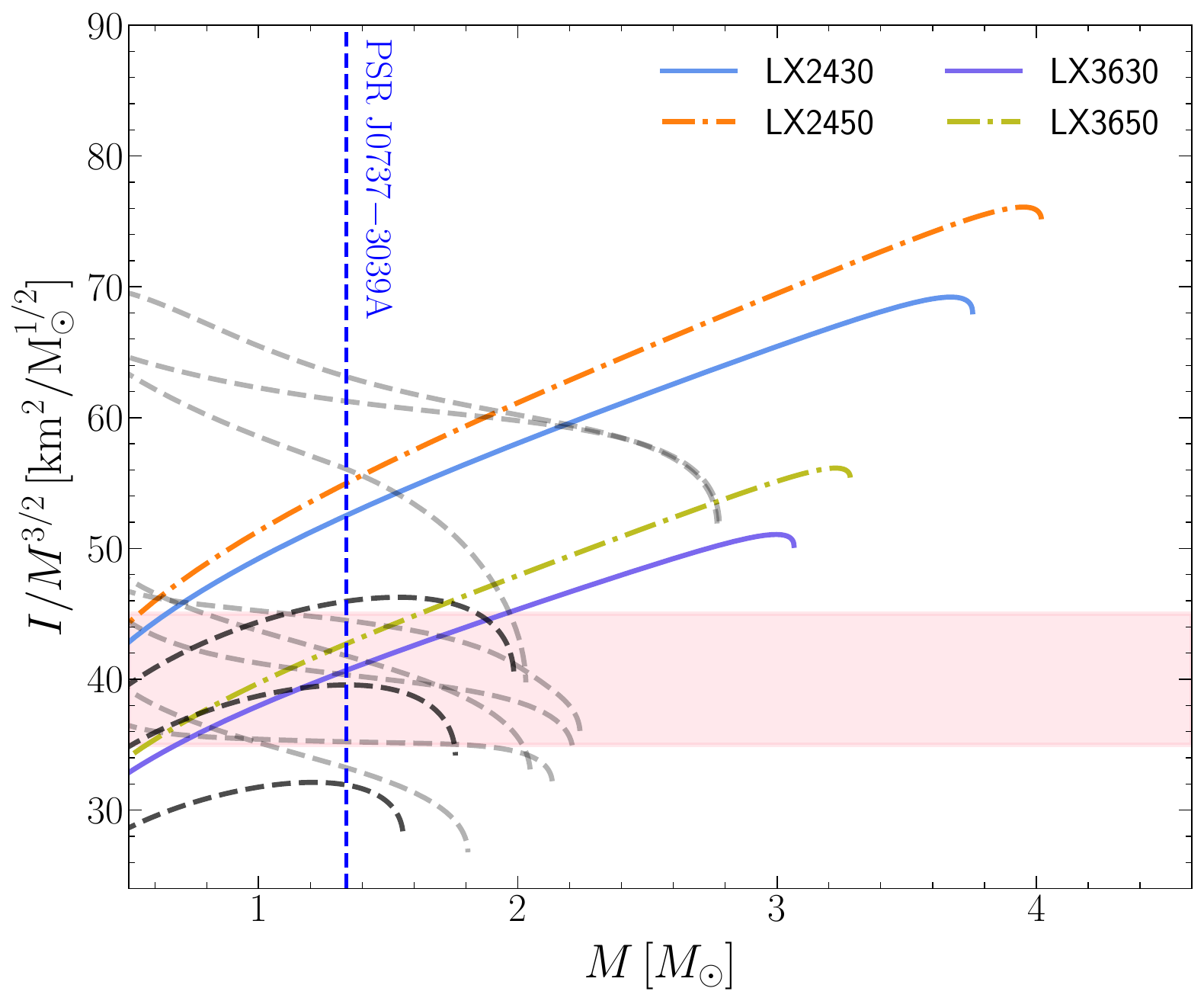}
    \caption{The relation between the rescaled moment of inertia $I/M^{3/2}$ and
    mass of stars $M$. The relations for normal NSs (grey dashed lines) and QSs
    (black dashed lines) are also displayed. The vertical line represents the
    mass of PSR 0737$-$3039A, $M=1.338M_{\odot}$.  The pink shaded region
    denotes a hypothetical $10\%$ measurement of the moment of inertia for PSR
    0737$-$3039A with its center value $40\,{\rm km^{2}}/M^{1/2}_{\odot}$.}
    \label{fig:nq18moi}
\end{figure}

In Fig.~\ref{fig:nq18moi}, we plot the relations between the rescaled moment of
inertia $I/M^{3/2}$ (to reduce the range of the coordinate) and mass of stars
$M$ for SSs with $N_{\rm q}=18$, normal NSs, and QSs.  For normal NSs, the
rescaled moment of inertia is nearly monotonically decreasing with respect to
the increase of mass.  While the relations for QSs and SSs are inverse except
for very large masses. We also plot a hypothetical $10\%$ measurement of the
quantity $I/M^{3/2}$ for PSR 0737$-$3039A \citep{Hu:2020ubl} with the central
value to be $40\,{\rm km^{2}}/M^{1/2}_{\odot}$.  If this is the case, some stiff
EoSs of normal NSs, as well as EoSs LX2430 and LX2450 of SSs with $N_{\rm
q}=18$, will be excluded.

\begin{figure}
    \centering
    \includegraphics[width=7.5cm]{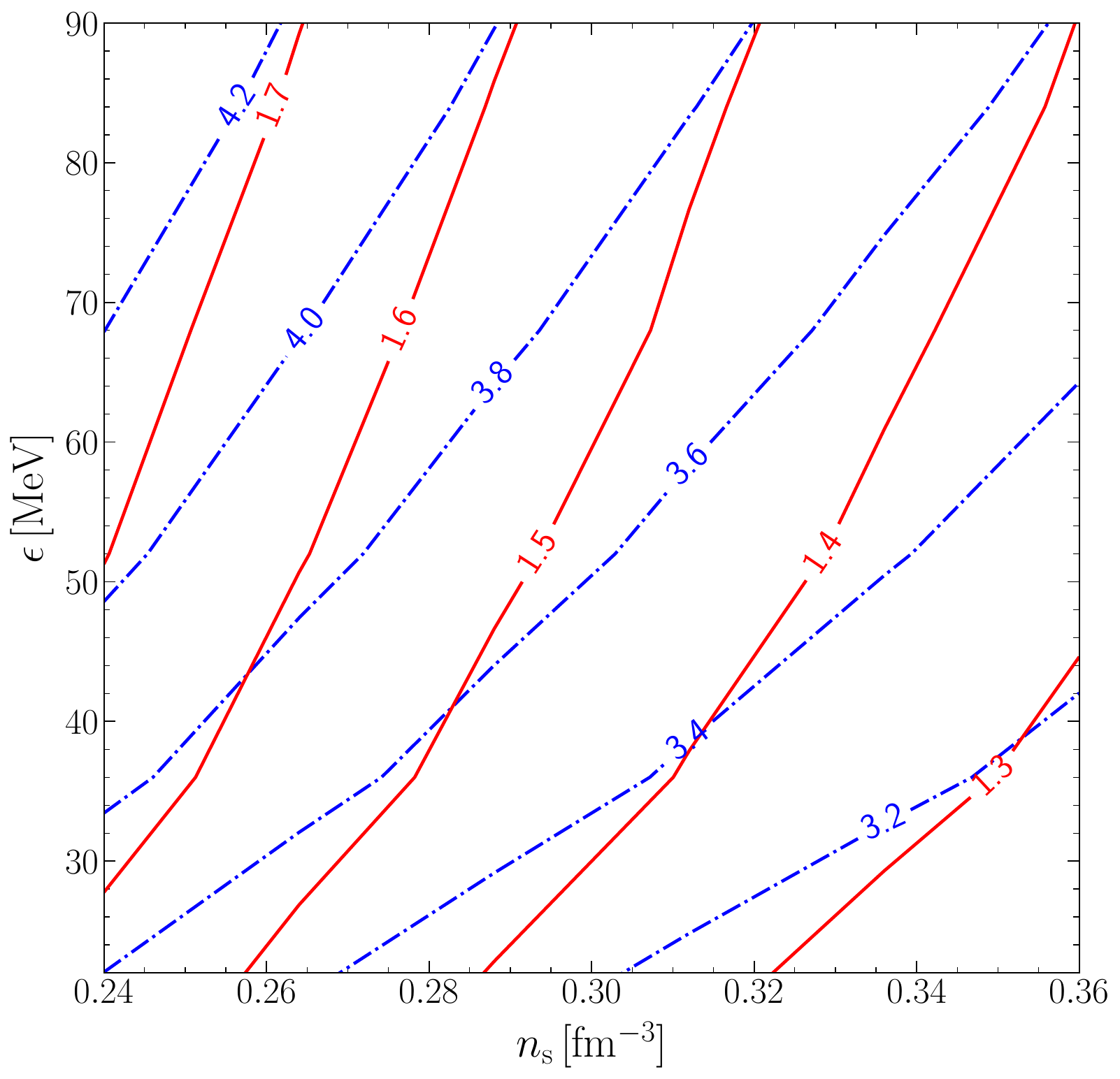}
    \caption{Contours of maximal masses $M_{\rm max}$ (blue) and moment of
    inertia $I_{45}$ ($I/10^{45}\,\rm g\,cm^{2}$) of PSR 0737$-$3039A (red) for
    SSs with $N_{\rm q}=18$.}
    \label{fig:contourI_nq18}
\end{figure}

\begin{figure}
    \centering
    \includegraphics[width=7.5cm]{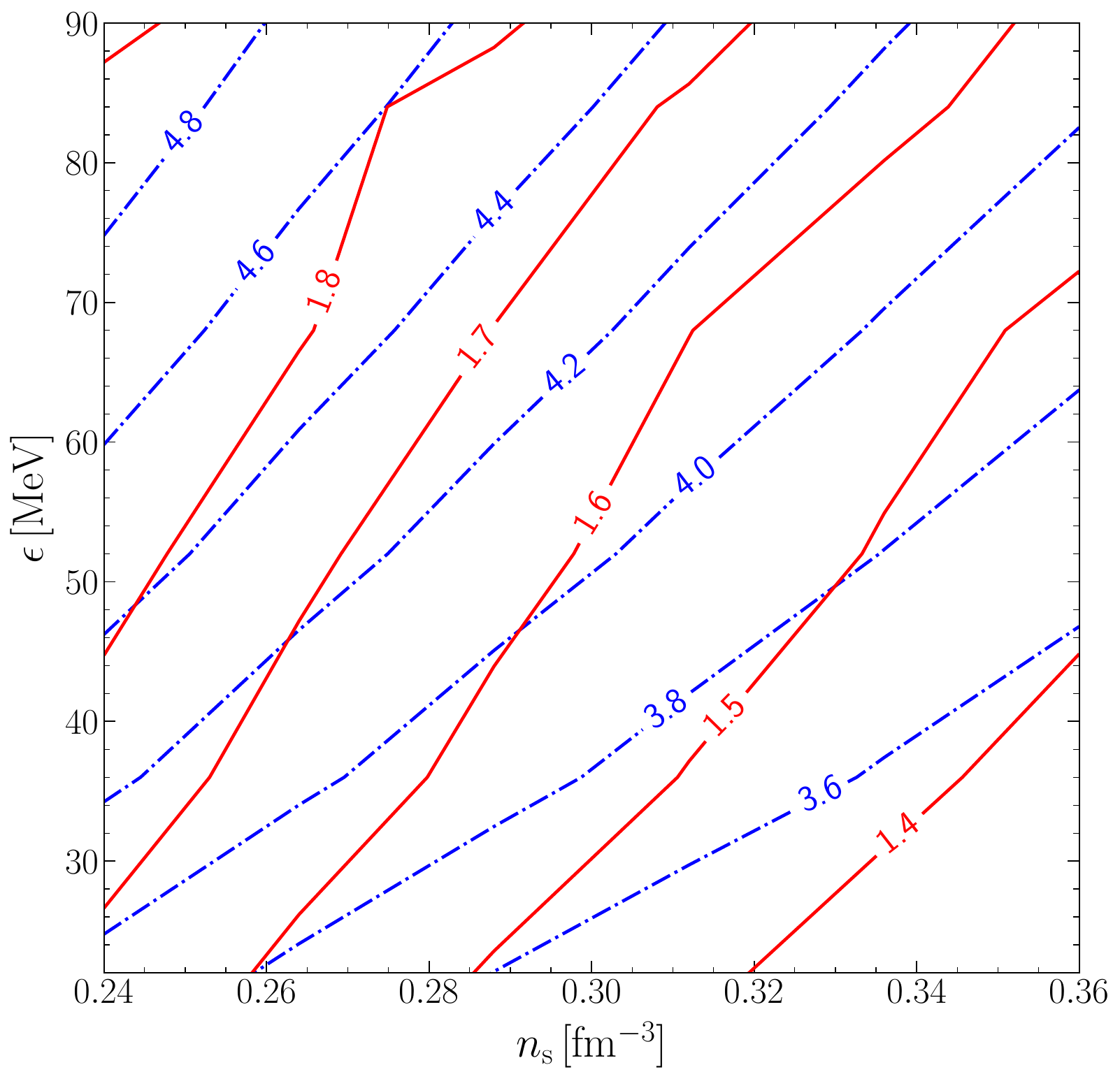}
    \caption{Same as Fig.~\ref{fig:contourI_nq18}, for $N_{\rm q}=9$.}
    \label{fig:contourI_nq9}
\end{figure}

For given $N_{\rm q}$, the EoS of SSs in the Lennard Jones model is determined
by two parameters, the potential depth $\epsilon$ and the surface baryonic
number density $n_{\rm s}$. Each pair of $\epsilon$ and $n_{\rm s}$ also 
corresponds to a specific maximal mass of SSs and a unique moment of inertia for
PSR 0737$-$3039A. Therefore, in Fig.~\ref{fig:contourI_nq18} and
Fig.~\ref{fig:contourI_nq9}, we plot the contour lines for maximal mass and 
moment of inertia spanning across the parameter space of $\epsilon$ and $n_{\rm
s}$ for $N_{\rm q}=18$ and $N_{\rm q}=9$ respectively.  If the moment of inertia
of PSR 0737$-$3039A is measured in the future, one can put constraints on the 
parameter space of $\epsilon$ and $n_{\rm s}$, and the maximal mass of SSs. 

\subsection{Second order: Spherical and quadrupole deformations}

As we mentioned before, to the second order of $\Omega$, the star will be
deformed and the isodensity surface at radial coordinate $r$ in the non-rotating
star will be displaced to $r+\xi_{0}(r)+\xi_{2}(r)P_{2}(\cos \theta)$ in the
rotating configuration \citep{Hartle:1968si}.  According to the definition of
the ``pressure perturbation factor'' $p_{0}$ and $p_{2}$ in
Eq.~(\ref{eqn:p0p2}), the displacements $\xi_{0}$ and $\xi_{2}$ can be
represented as,  
\begin{align}
    \xi_{0}(r)=&-p_{0}(\rho+P)/(\dd P/\dd r)=p_{0}\,r(r-2m)/(m+4\pi r^{3}P)\,,\\
    \xi_{2}(r)=&-p_{2}(\rho+P)/(\dd P/\dd r)=p_{2}\,r(r-2m)/(m+4\pi r^{3}P)\,,
\end{align}
which are second order of $\Omega$ and are related to the spherical and
quadrupole deformations respectively.  Correspondingly, the perturbative
functions of order $\Omega^{2}$ can be divided into two classes: (i) the $l=0$ 
functions, $m_{0}$, $h_{0}$, and $p_{0}$, that describe the spherical stretching
of the star; (ii) the $l=2$ functions, $h_2$, $v_{2}$, $m_{2}$, and $p_{2}$, that
describe the quadrupole deformation of the star. 

One may directly use $\xi_{0}(r)$ and $\xi_{2}(r)$ to define the mean radius and
the eccentricity of the isodensity surface in the Hartle-Thorne coordinate
\citep{Hartle:1967he,Hartle:1968si},
\begin{align}
    \bar{r}_{\rm  HT}=& \,r+\xi_{0}(r)\,,\\
    e(r)_{\rm HT}=&\left[\text { (radius at equator) }^{2} /(\text { radius at pole })^{2}-1\right]^{1 / 2}\nonumber \\
    =& \left[-3\xi_{2}(r)/r\right]^{1/2}\,,
\end{align}
which are not invariant under the transformation of coordinate system. To give
an invariant parametrization of the isodensity surface, one needs to embed the
geometry into a three-dimensional flat space (denoted with polar coordinates
$r^{*}$, $\theta^{*}$, and $\phi^{*}$) and search for the surface that has the
same intrinsic geometry as the isodensity surface of the star
\citep{Hartle:1968si,Chandra1974}. 

To the second order of $\Omega$, the desired surface in flat space is a spheroid
with the equation \citep{Hartle:1968si}
\begin{equation}
    r^{*}\left(\theta^{*}\right)=r+\xi_{0}(r)+\Big\{\xi_{2}(r)+
    r\left[v_{2}(r)-h_{2}(r)\right]\Big\} P_{2}\left(\cos \theta^{*}\right)\,.
\end{equation}
The mean radius of the spheroid is 
\begin{equation}
    \bar{r}^{*}=r+\xi_{0}(r)\,,
\end{equation}
and the eccentricity can be defined as 
\begin{align}
    \label{eqn:eccentricity}
    e(r) &=\left[\text { (radius at equator) }^{2} /(\text { radius at pole })^{2}-1\right]^{1 / 2}\nonumber \\
    &=\left[-3\left(v_{2}-h_{2}+\xi_{2} / r\right)\right]^{1 / 2}\,.
\end{align}
The mean radius of the star, $\bar R$, and the eccentricity of the surface of
the star, $e_{\rm s}$, can be obtained by setting $r=R$. 

Since the fluid elements are displaced and the star comes to a new equilibrium
state, the baryonic mass, the gravitational mass, and the quadrupole moment also
change. To obtain the deformation of the star and changes in various physical
quantities, we need to give the solutions of the $l=0$ and $l=2$ perturbative
functions.  Numerically, One integrates the differential equations of those
perturbative functions in Appendix~\ref{append:A} with appropriate boundary
conditions at the center of the star and at infinity.  Fortunately, the
analytical solutions exist outside of the star with some undetermined constants.
One therefore can integrate the differential equations for perturbative
functions to the radius $R$ and match the results with the exterior solutions to
ascertain the undetermined constants.

A technical problem needs to be stressed. For SSs or QSs, the surface density
drops from nuclear densities to zero and some thermodynamical quantities such as
pressure do not admit regular Taylor expansions in $(r-R)$ when $r\rightarrow R$
\citep{Damour:2009vw}. For example, the differential equations involving the
terms  $(\rho+P) \mathrm{d} \bar{\rho}/\mathrm{d} P$ or $(\rho+P) \mathrm{d}
\rho/\mathrm{d} P$ are singular across the surface of the star. To solve the
issue, one can treat the baryonic density $\bar{\rho}(r)$ and the mass-energy
density ${\rho}(r)$ as inverted step functions across the surface of the star.
Then the terms $(\rho+P) \mathrm{d} \bar{\rho}/\mathrm{d} P$ and $(\rho+P)
\mathrm{d} \rho/\mathrm{d} P$ at the boundary of the star can be represented as
\begin{align}
    \frac{\mathrm{d} \bar{\rho}}{\mathrm{d} P}(\rho+P) =&\bar{\rho} \delta(r-R) r(r-2m)/(m+4\pi r^{3}P)\,,\\
    \frac{\mathrm{d} {\rho}}{\mathrm{d} P}(\rho+P) =&{\rho} \delta(r-R) r(r-2m)/(m+4\pi r^{3}P)\,,
\end{align}
where we have used the expression of $\dd P/\dd r$ in Eq.~(\ref{eqn:tov}).
Numerically, we first integrate the differential equations in the open interval
$0<r<R$ and obtain the value $y(R_{-})$ just inside of the star, where $y$ is a
physical quantity depending on the radial coordinate $r$.  Second, we add the
contributions from the integration of $\delta$ function at the surface of the
star and get the value just outside of the star $y(R_{+})$. Then, the
undetermined constants are obtained by matching $y(R_{+})$ with the exterior
solutions. Physically, it means that one must consider the match conditions to
guarantee the continuity of spacetime. The physical quantities in the interior
of the star $\Sigma^{-}$ and the exterior vacuum region $\Sigma^{+}$ are matched
on a common boundary $\Sigma_{0}=\Sigma^{\pm}(r=R_{\pm})$.  For convenience, we
define $[y]:=y^{+}\big|_{\Sigma_{0}}-y^{-}
\big|_{\Sigma_{0}}$~\citep{Reina:2015jia,Reina:2017mbi}.

\subsubsection{Spherical deformations: change in the gravitational mass and the baryonic mass}

\begin{figure}
    \centering 
    \includegraphics[width=8cm]{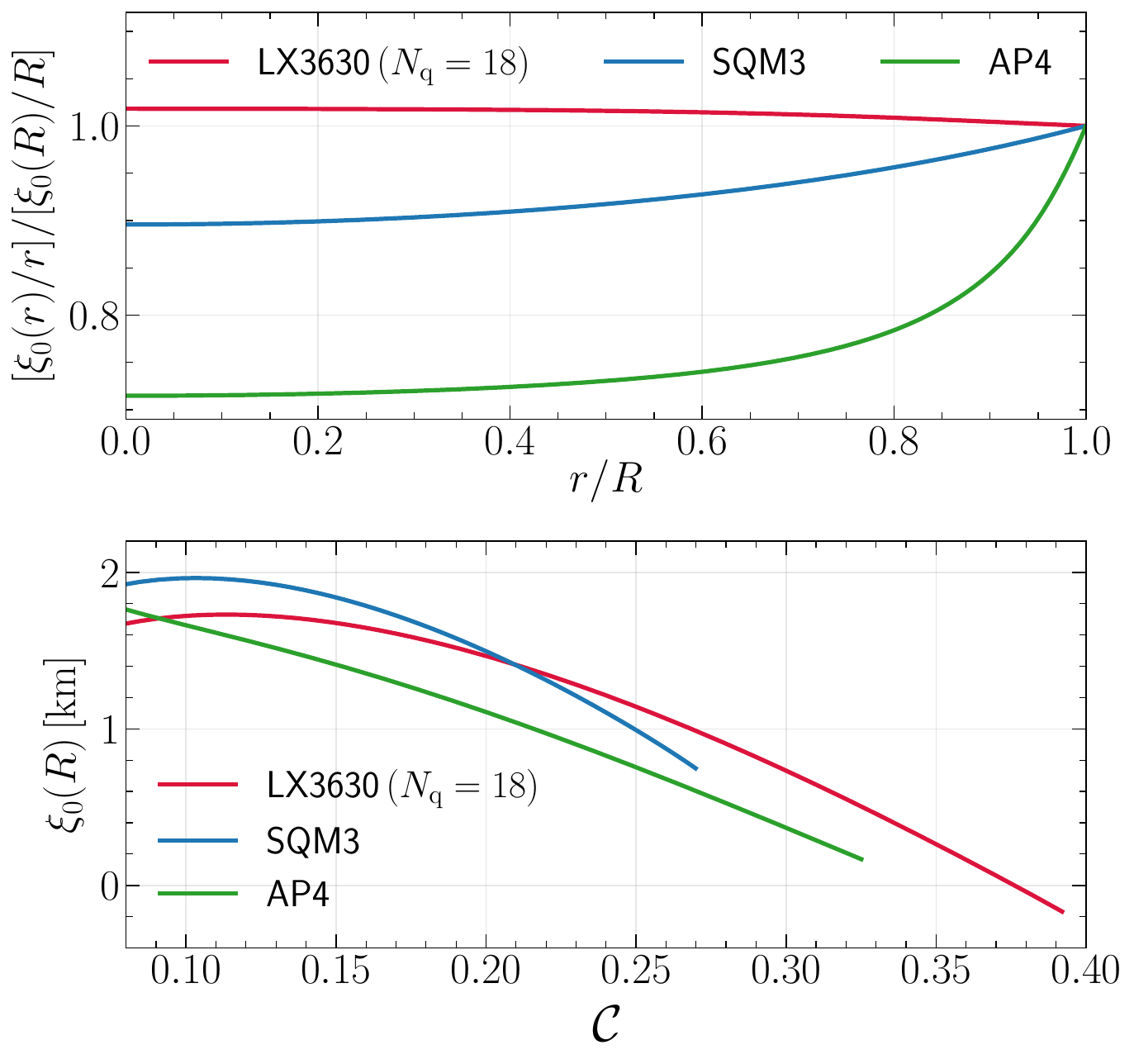}
    \caption{The spherical stretching due to rotation. For clarity, we only
    shows the results for EoS LX3630 ($N_{\rm q}=18$), AP4, and SQM3. The upper
    panel shows the fractional change $\xi_{0}(r)/r$ of the isodensity surface
    at the radial coordinate $r$, normalized by the fractional change
    $\xi_{0}(R)/R$ at the surface of the star, where we take $M=1.4\,M_{\odot}$
    for the non-rotating configuration. In the lower panel, we display the
    relation between the spherical displacement at the surface of the star
    $\xi_{0}(R)$ and the compactness $\mathcal C$. We take the angular frequency
    to be $\Omega_{*}$.  For a fixed central density, the stretching at the
    surface $\xi_{0}(R)$ for a smaller angular velocity $\Omega$ can be obtained
    by multiplying the factor $(\Omega/\Omega_{*})^{2}$.}
    \label{fig:stretching}
\end{figure}

In the upper panel of Fig.~\ref{fig:stretching}, we show the fractional change
$\xi_{0}(r)/r$ of the isodensity surface at the radial coordinate $r$ normalized
by the fractional change $\xi_{0}(R)/R$ at the surface of the star. The
behaviors of $(\xi_{0}(r)/r)/(\xi_{0}(R)/R)$ reveal the density distribution
inside of the star \citep{Hartle:1968si}.  For an EoS of normal NSs, AP4, the star
has a dense core and an envelope, and the density distribution is more diffusive
than the EoS of SSs and QSs. As a consequence, the spherical stretching is small
in the core and increases in the envelope of the star. For SSs and QSs, the
densities inside of the star are more homogeneous than normal NSs, and the
variations of the fractional stretching are smaller. This tendency is just what
we expect from simple Newtonian intuition. Particularly, the fractional change
of the stretching of the SSs is nearly a constant through the star, which
indicates that the EoS of SSs is close to the incompressible fluid.  Besides,
for SSs, the fractional change $\xi_{0}(r)/r$ decreases monotonically from the
center to the boundary of the star, which is different from QSs and normal NSs. 

In the lower panel of Fig.~\ref{fig:stretching}, we plot the spherical
stretching $\xi_{0}(R)$ versus the compactness of the star. The spherical
stretching decreases with the increase of the compactness in the case of
$M/R\geq 0.1$.  Another feature is that the spherical stretching can be smaller
than zero near the maximal mass for SSs. The ``pressure perturbation factor''
$p_{0}$ is negative and the rotation makes the star contract. 
\citet{Chandra1974} also showed this feature for incompressible fluid when
$R/R_{\rm s}\rightarrow 9/8$, where $R_{\rm s}$ is the Schwarzschild radius
(see the first three columns of Table I and Figure 3 therein).

The baryonic mass and the gravitational mass are rotational invariant quantities
and do not change under parity transformation ($\Omega \rightarrow -\Omega$).
Thus, perturbations are only determined by the  $l=0$ functions, namely $h_{0}$,
$m_{0}$, and $p_{0}$, and the non-rotating background solutions.  In practice,
we numerically integrate $m_{0}$ and $p_{0}$ inside of the star. Outside of the
star, $p_{0}$ vanishes and $m_{0}$ is
\begin{equation}
    m_{0}=\ \delta M-J^{2} / r^{3}\,,
\end{equation}
where $\delta M$ is a constant and $J$ is the angular momentum. The interior and
the exterior solutions are matched at $r=R$ with the match condition
\begin{equation}
    [m_{0}]=4 \pi R^{3} \rho\left(R_{-}\right)(R-2 M) p_{0}(R) / M \,, 
\end{equation}
where $\rho\left(R_{-}\right)$ represents the energy density just inside of the
star.  The function $h_{0}$ can be obtained by algebraic relations
\begin{align}
    h_{0}=&-p_{0}+\frac{1}{3}r^{2}e^{-\nu}\bar{\omega}^{2}+h_{0\rm c} \,,\quad(\rm  inside\ of\ the \ star),\\
    \label{eqn:h0_exterior}
    h_{0}=&-\frac{\delta M}{r-2 M}+\frac{J^{2}}{r^{3}(r-2 M)}\,, \quad(\rm  outside\ of\ the \ star).
\end{align}
The constant $h_{0\rm c}$ is the value of $h_{0}$ at the center of the star,
which can be obtained by matching the interior and exterior solutions.

\begin{figure*}
    \centering 
    \includegraphics[width=12cm]{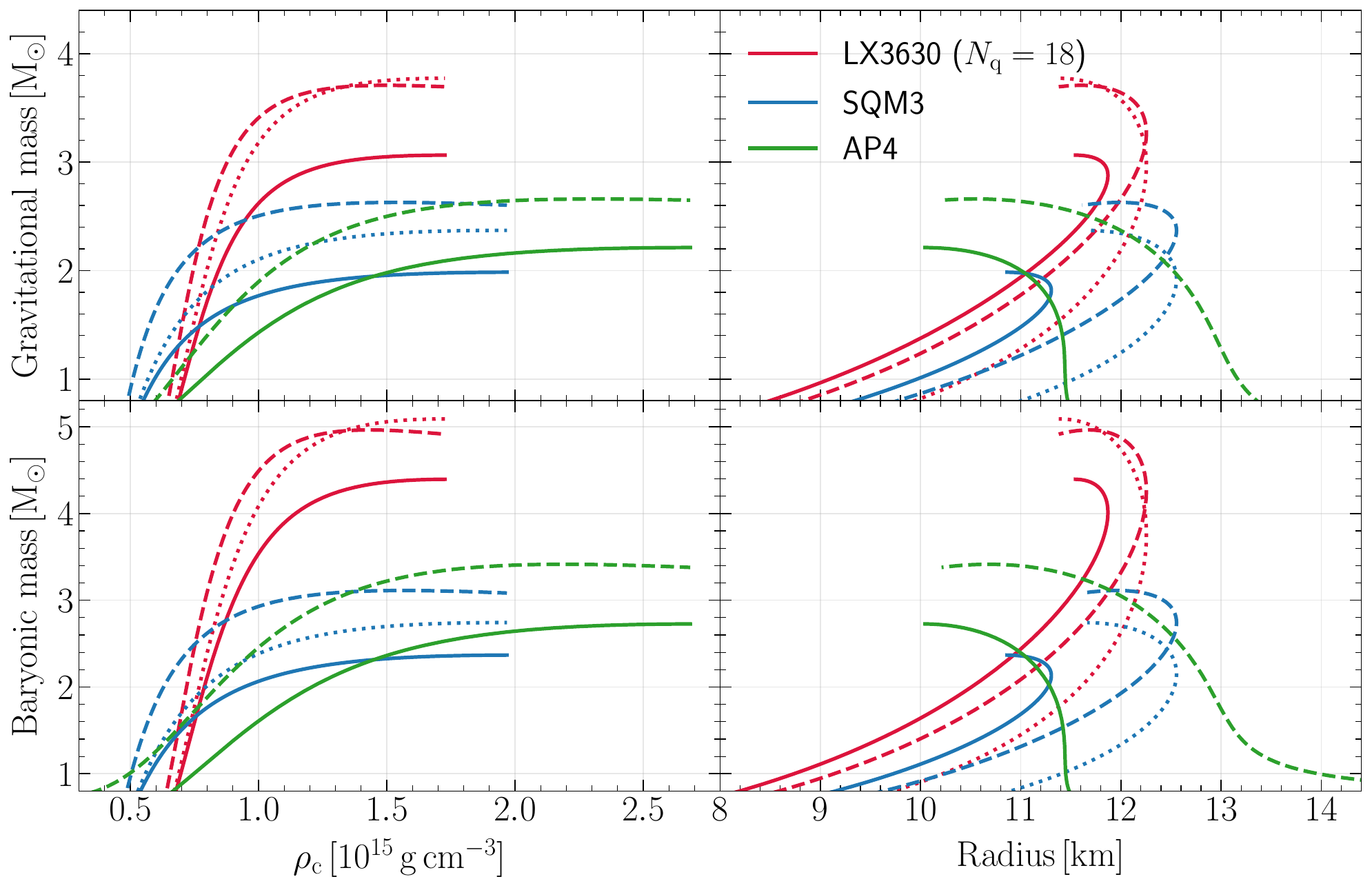}
    \caption{The first column shows the the gravitational mass and the baryonic
    mass versus the central mass-energy density for representative EoSs $\rm
    LX3630\,(N_{\rm q}=18)$, SQM3, and AP4. The solid lines represent
    non-rotating configurations. The dashed lines are the ones for rotating
    configurations with the angular frequency $\Omega_{*}$. The gravitational
    mass and the baryonic mass for a smaller frequency $\Omega$, where the slow
    rotation approximation is valid, can be easily obtained by an upward
    displacement from the solid curves by the fraction $(\Omega/\Omega_{*})^{2}$
    of the distance to the dashed curves (see text for more details). The second
    column shows the relation between the gravitational mass and the mean
    radius. The solid curves are plots of the mass $M$ versus the radius $R$ for
    non-rotating stars. The dashed curves represent the relation between mass
    $\mathcal{M}=M+\delta M$ and mean radius $\bar{R}=R+\xi_{0}(R)$ for rotating
    configurations with angular frequency $\Omega_{*}$. The mass radius
    relations for angular velocities that satisfies the slow rotation
    approximation can be obtained by multiplying the rescaling factor
    $(\Omega/\Omega_{*})^{2}$ along a constant central density sequence. For
    comparison, we also display the relations for $\rm LX3630\,(N_{\rm q}=18)$,
    SQM3 without the corrections induced by the match conditions with dotted
    lines.}
    \label{fig:mass_correction}
\end{figure*}

In general relativity, the mass $\mathcal{M}$ of a rotating relativistic star is
determined by the spherical part of the metric $g_{tt}$ at large distances 
\begin{equation}
    -\frac{1+g_{tt}}{2}\Big|_{\rm spherical\  part}=-\frac{\mathcal M}{r} \,,\quad r\rightarrow \infty\,.
\end{equation}
Combining the metric of a slowly rotating relativistic star in
Eq.~(\ref{eqn:hartle_thorne}) and the exterior solution of $h_{0}$ in
Eq.~(\ref{eqn:h0_exterior}), one obtains the total gravitational mass 
\begin{equation}
    \mathcal{M}=M+\delta M\,,
\end{equation}
where $M$ is the background contribution and $\delta M$ is the second order
correction which appears as an integration constant in the exterior solution of
$h_{0}$ and $m_{0}$.  After matching the interior and the exterior solutions of
$m_{0}$, one obtains
\begin{equation}
    \delta M= m_{0}(R_{-})+4 \pi R^{3} \rho(R_{-})(R-2 M) p_{0}(R)/M+J^{2} / R^{3}\,.
\end{equation}
The baryonic mass is a conserved quantity. Integrating the differential form of
the baryonic mass conservation law, $\big[\sqrt{-g}\bar{\rho} u^{\mu}\big]_{,
\mu}=0$, at a $t=\rm const$ hypersurface, one obtains the baryonic mass of the
star~\citep{Hartle:1967he,Misner:1974qy}
\begin{equation}
    \bar{\mathcal M}=\int \sqrt{-g}\bar{\rho}u^{t}\dd^{3}x\,.
\end{equation}
Note that the integration extends to the whole region of the deformed star. To
second order of $\Omega$, the expansion of the baryonic mass can be represented
as $\bar{\mathcal M}=\bar{M}+\delta \bar{M}+O(\Omega^{4})$.  The baryonic mass
of the non-rotating star is
\begin{equation}
    \bar{M}=\int^{R}_{0}4\pi r^{2}\left(1-\frac{2m}{r}\right)^{-1/2}\bar{\rho}\dd r\,.
\end{equation}
The correction at the second order of $\Omega$ is 
\begin{align}
    \label{eqn:baryonic_correction}
    \delta \bar M&= \int_{0}^{R} 4 \pi r^{2} \left(1-\frac{2 m}{r}\right)^{-1 / 2}\nonumber\\
    &\times \bigg\{\Big[\frac{m_{0}}{r-2 m} +\frac{1}{3} r^{2}\bar{\omega}^{2} 
    e^{-\nu(r)}\Big] \bar{\rho} +\frac{\dd \bar{\rho}}{\dd P}(\rho+P) p_{0}\bigg\}\dd r\,,
\end{align}
which can be represented as 
\begin{equation}
    \delta \bar M= \delta \bar {M}(R_{-})
    +4\pi R^{3}\left(1-\frac{2M}{R}\right)^{-1/2} \frac{\bar{\rho}(R_{-})(R-2M)p_{0}(R)}{M}\,,
\end{equation}
where we have considered the corrections from the matching condition at the
surface of the star.

In Fig.~\ref{fig:mass_correction}, we show the gravitational mass and the
baryonic mass at frequency $\Omega_{*}$ for EoS LX2430 ($N_{\rm q}=18$), SQM3,
and AP4. The slow-rotation approximation breaks down at this frequency. But the
mass and the radius for a given central density and a smaller angular frequency
$\Omega$ can be simply obtained by multiplying the factor
$(\Omega/\Omega_{*})^{2}$. The maximal mass at the rotating frequency
$\Omega_{*}$ increases by $\sim 20\%$.  We also plot the lines without the
corrections from the match conditions at the surface of the star. It is obvious 
that the corrections are crucial and cannot be ignored.

\subsubsection{Quadrupole deformations: the production of quadrupole moments}

\begin{figure}
    \centering 
    \includegraphics[width=8cm]{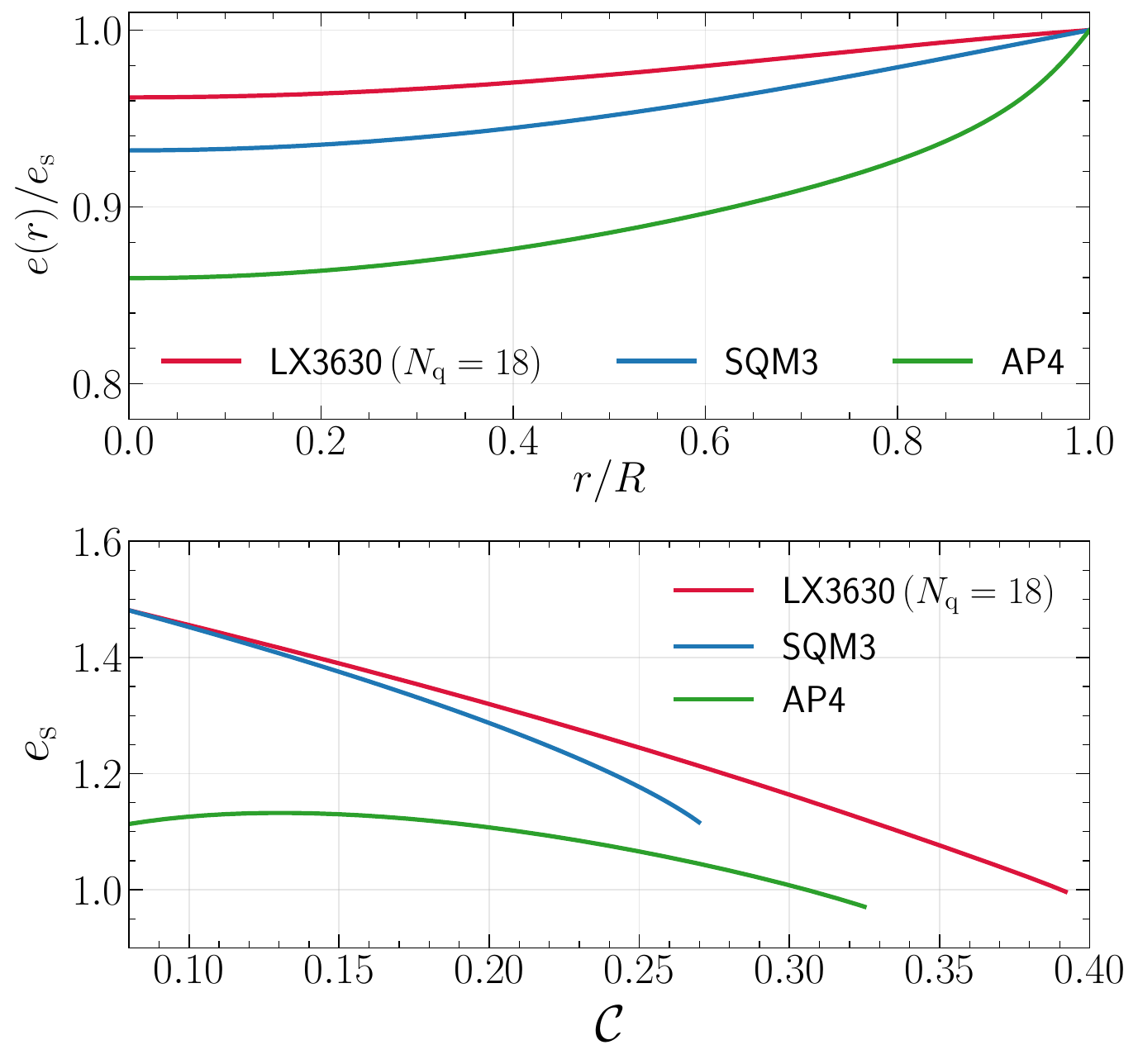}
    \caption{The upper panel shows the eccentricity $e(r)$ of isodensity surface
    at radial coordinate $r$, normalized by the eccentricity $e_{\rm s}$ at the
    surface of the star, where the mass for the non-rotating configuration is
    taken as $M=1.4\,M_{\odot}$. The lower panel shows the relation between the
    surface eccentricity $e_{\rm s}$ and the compactness $\mathcal C$ with the
    angular frequency chosen to be $\Omega_{*}$. For a smaller frequency
    $\Omega$ where the slow-rotation approximation is valid, the surface
    eccentricity $e_{\rm s}$ can be obtained by multiplying the factor
    $\Omega/\Omega_{*}$ for a fixed central density. }
    \label{fig:eccentricity}
\end{figure}

The quadrupole deformation can be described by the eccentricity $e(r)$ of the
isodensity surface at radial coordinate $r$. In the upper panel of
Fig.~\ref{fig:eccentricity}, we show the fractional change of the eccentricity
$e(r)/e_{\rm s}$ at different radial coordinates inside of the star.  The
relations also reveal the internal density distribution, which has the same
tendencies as the fractional change of the radial stretching in the upper panel
of Fig.~\ref{fig:stretching}.  The relation between the surface eccentricity
$e_{\rm s}$ and the compactness $\mathcal C$ is shown in the lower panel of
Fig.~\ref{fig:eccentricity}, where we set the angular frequency to be
$\Omega_{*}$.  One can notice that the surface eccentricity for LX2430 and SQM3
are larger than that of AP4. The difference between SSs and normal NSs can be as
large as $\sim 20\%$ for some choices of the compactness. 

The quadrupole moment depends on the $l=2$ perturbative functions: $h_{2}$,
$v_{2}$, $p_{2}$, and $m_{2}$.  In practice, we integrate the differential
equations to obtain the interior solutions of $h_{2}$ and $v_{2}$.  The exterior
solutions of $h_{2}$ and $v_{2}$ are 
\begin{align}
    \label{eqn:h2_exterior}
    h_{2}=&J^{2}\left(\frac{1}{M r^{3}}+\frac{1}{r^{4}}\right)+K Q_{2}^{2}\left(\frac{r}{M}-1\right) \,,\\
    v_{2}=&-\frac{J^{2}}{r^{4}}+K \frac{2 M}{[r(r-2 M)]^{1 / 2}} Q_{2}^{1}\left(\frac{r}{M}-1\right)\,,
\end{align}
where $Q_{2}^{1}$ and $Q_{2}^{2}$ are the associated Legendre functions of the
second kind.  The constant $K$ is determined by matching the exterior and
interior solutions at $r=R$ with the match conditions
\begin{equation}
    [h_{2}]=0\,,\quad [v_{2}]=0\,.
\end{equation}
The functions $p_{2}$ and $m_{2}$ can be obtained from the algebraic relations
\citep{Hartle:1968si},
\begin{align}
    m_{2}=&\ (r-2 m)\left[-h_{2}-\frac{1}{3} r^{3}\left(\dd j^{2} / \dd r\right) 
    \bar{\omega}^{2}+\frac{1}{6} r^{4} j^{2}(\dd \bar{\omega} / \dd r)^{2}\right]\,,\\
    p_{2}=&-h_{2}-\frac{1}{3}r^{2}e^{-\nu}\bar{\omega}^{2}\,,
\end{align}
which come from the first integrals of the Einstein field equations.

The quadrupole moment can be read out from the coefficient of
$P_{2}(\cos\theta)/r^{3}$ in the Newtonian potential
\citep{Hartle:1968si,Thorne:1984mz}. As the radial coordinate $r$ goes to
infinite asymptotically, the quadrupole part of the Newtonian potential is
\begin{equation}
    \label{eqn:rotation_expansion}
    -\frac{1+g_{tt}}{2}\Big|_{\rm quadrupole\ part}= -\frac{Q_{\rm r}}{r^3}P_{2}(\cos\theta) \,,\quad r\rightarrow \infty\,,
\end{equation}
where $Q_{\rm r}$ is the rotation-induced quadrupole moment. Calculating the
effective potential $-(1+g_{tt})/2$ with the expansion of $h_{2}$ in
Eq.~(\ref{eqn:h2_exterior}) at large distance and comparing the results with the
formal expression in Eq.~(\ref{eqn:rotation_expansion}), one obtains the
quadrupole moment of the star 
\begin{equation}
    Q_{\rm r}=-\frac{J^{2}}{M}-\frac{8}{5} K M^{3}\,.
\end{equation}
Here, $Q_{\rm r}<0$ means that the star is deformed into an oblate shape. Note
that the quadrupole moments not only depend on the spin angular momentum of the
star, but also depend on the integration constant $K$, which is related to the
EoS. For black holes, the quadrupole moment is $Q_{\rm r}=-J^{2}/M$, which only
depends on the angular momentum and the mass of the black hole. This property is
guaranteed by the no hair theorem. One usually define the dimensionless
quadrupole moment,
\begin{equation}
    \bar{Q} \equiv-\frac{Q_{\rm {r }}}{M^{3} \chi^{2}}\,.
\end{equation}
Similarly, the dimensionless spin $\chi$ is defined as $\chi \equiv J/M^{2}$.

\begin{figure}
    \includegraphics[width=8cm]{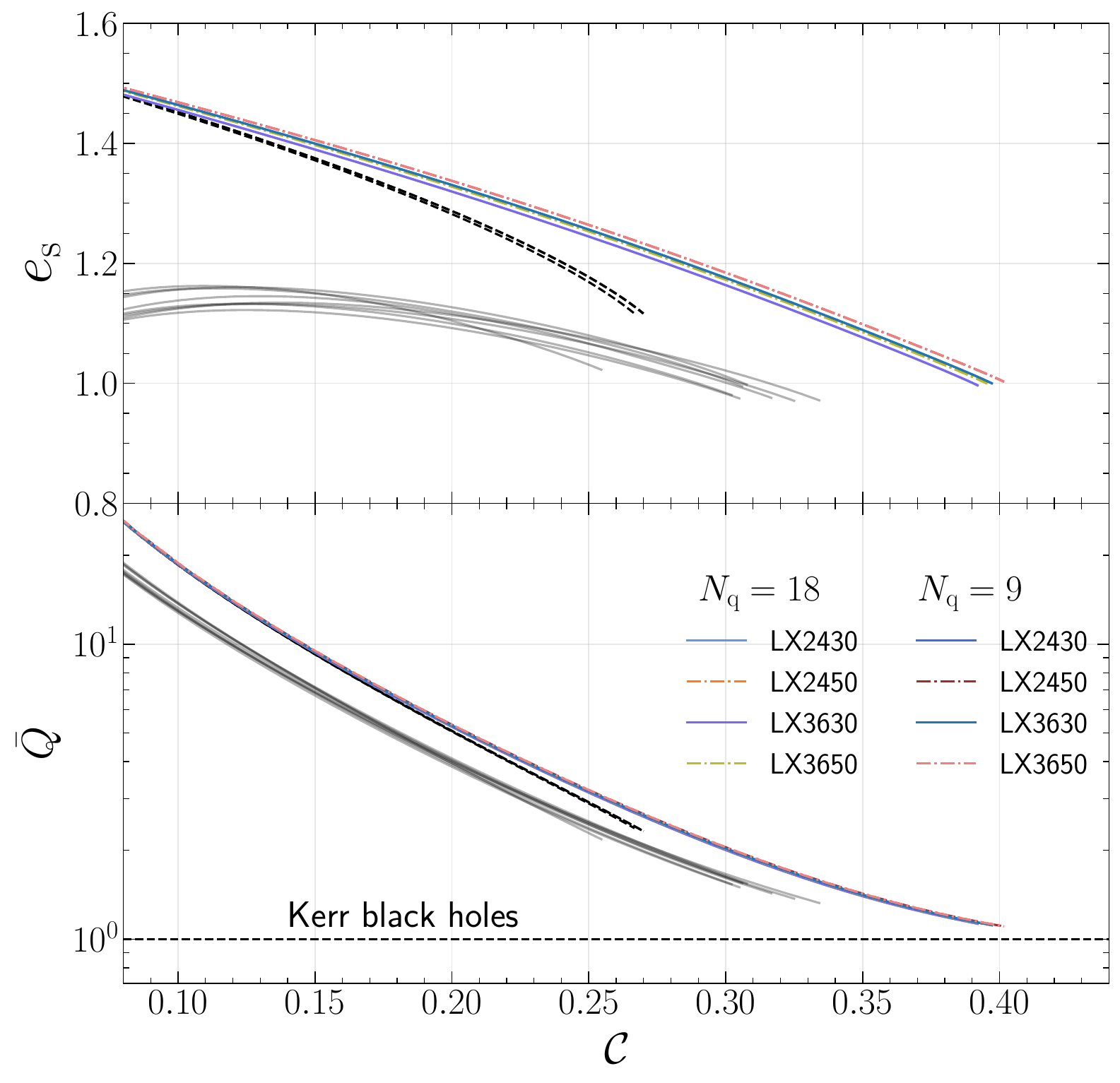}
    \caption{The upper panel shows the relation between the surface eccentricity
    and the compactness for SSs with $N_{\rm q}=18$ and $N_{\rm q}=9$, normal
    NSs (grey), and QSs (black). This panel is the same as the lower panel of
    Fig.~\ref{fig:eccentricity} but more EoSs are included.  The lower panel
    displays the relation between the dimensionless quadrupole moment and the
    compactness for SSs with $N_{\rm q}=18$ and $N_{\rm q}=9$, normal NSs
    (grey), and QSs (black).  The horizontal line represents the dimensionless
    quadrupole moment of Kerr black holes, where $\bar Q=1$.}
    \label{fig:qua_com}
\end{figure}

In Fig.~\ref{fig:qua_com}, we plot the surface eccentricity $e_{\rm s}$ and the
dimensionless quadrupole moment $\bar{Q}$ for SSs, QSs, and normal NSs. There is
a quasi-universal relation between $e_{\rm s}$ and $\mathcal C$. While the
universality of the relation between the dimensionless quadrupole moment
$\bar{Q}$ and $\mathcal C$ is tighter.  For NSs, the universal relation is quite
different from the ones for QSs and SSs.  The $\bar Q$--$\mathcal C$ universal
relation for normal NSs and QSs has been discovered by \citet{Urbanec:2013fs}. 
For SSs, the universal relation is nearly undistinguishable from that of QSs at
small compactness. But as the compactness increases, the quadrupole of SSs
becomes larger than that of QSs. This feature also appears in the universal 
relations of moment of inertia shown in Fig.~\ref{fig:I_C}. The reason is again
that SSs are much stiffer than QSs at high densities. 

The quadrupole moment $\bar Q$ and the surface eccentricity $e_{\rm s}$ both
describe the departure from spherical symmetry of the stars. Therefore, $\bar Q$
and $e_{\rm s}$ show common features: (i) for a given compactness, the
quadrupole moment and the surface eccentricity of SSs are larger than QSs, while
the quantities of QSs are larger than that of the normal NSs; (ii) the
quadrupole moment and the surface eccentricity both decrease as the compactness
increases in the range we plot.  The quadrupole moment tends to be close to the
limit of Kerr black holes. For our models of SSs, this tendency is very clear
and $\bar Q$ is very close to 1 for the stars near the maximal compactness
(corresponding to the maximal mass).  So, it is hard to distinguish Kerr black
holes and rotating SSs around maximal mass purely from these two quantities. 

\begin{figure}
    \centering
    \includegraphics[width=8cm]{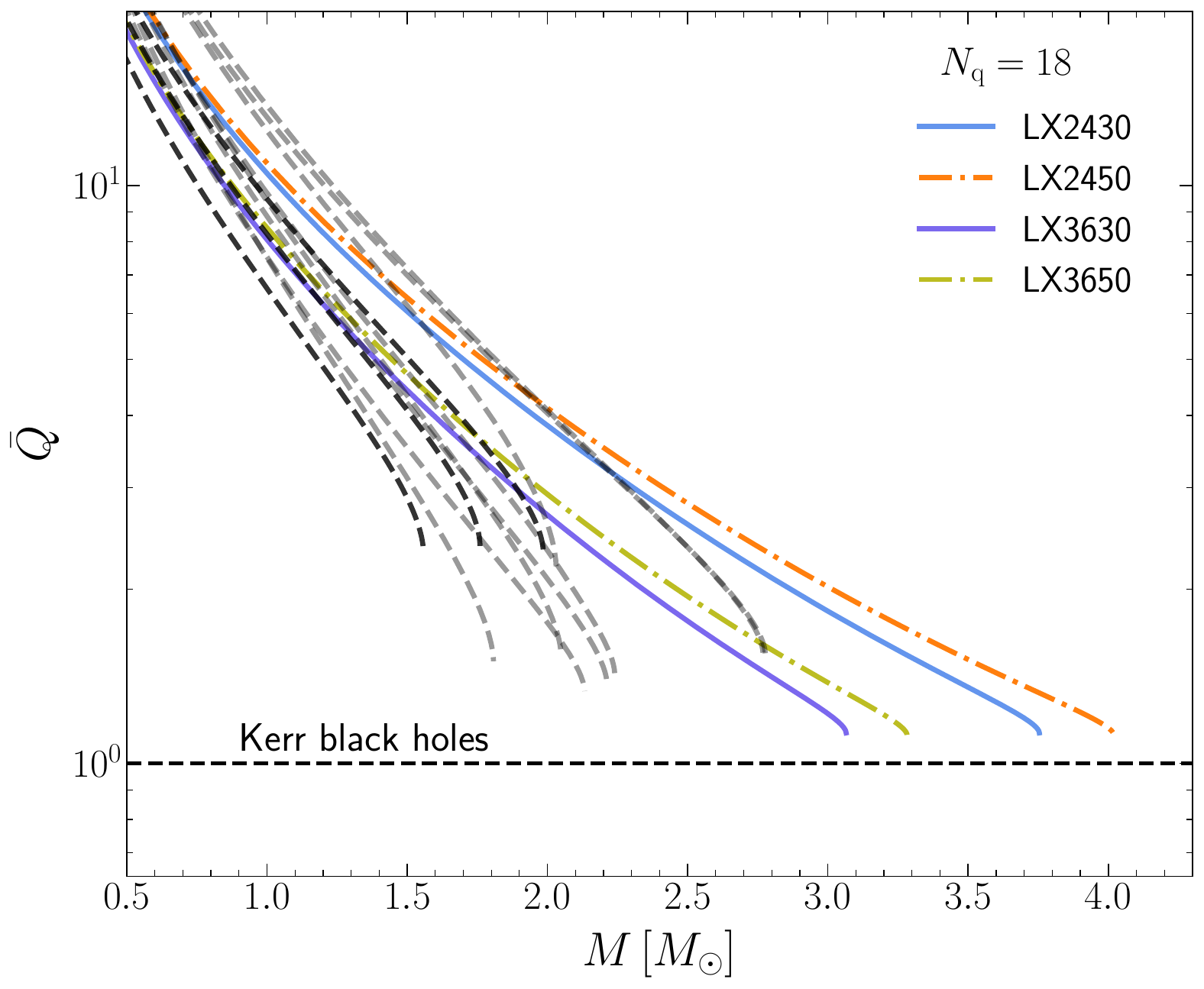}
    \caption{The dimensionless quadrupole moment versus the mass $M$ of
    non-rotating stars for SSs with $N_{\rm q}=18$. The relations for $N_{\rm
    q}=9$ have the similar trend and we ignore them for clarity of the figure.
    For comparison, the relations for normal NSs (grey) and QSs (black) are also
    plotted. The horizontal line represents the dimensionless quadrupole moment
    of Kerr black holes, where $\bar Q=1$.}
    \label{fig:nq18quadrupole}
\end{figure}

\begin{figure}
    \centering
    \includegraphics[width=8cm]{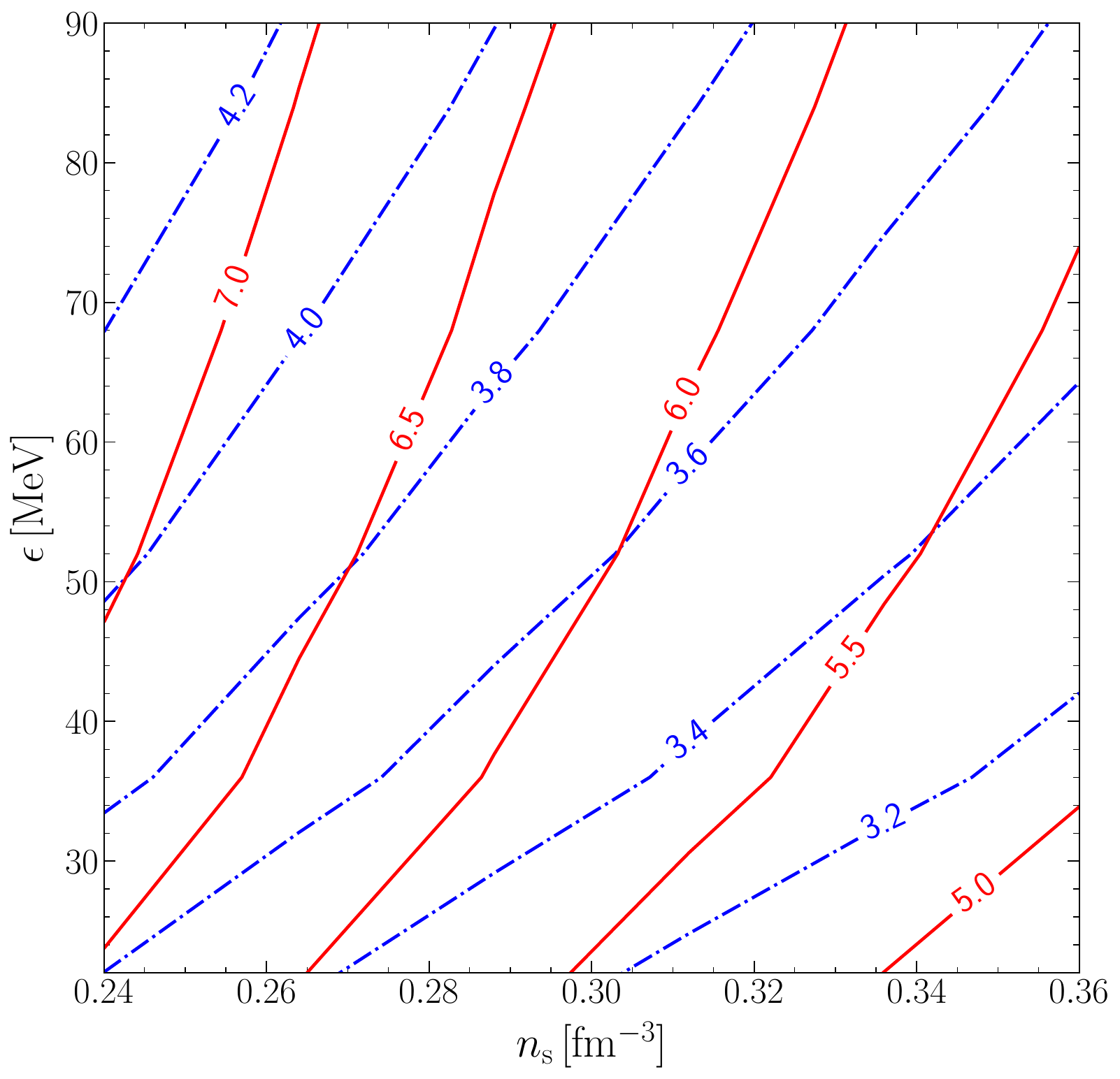}
    \caption{Contours of maximal masses $M_{\rm max}$ (blue) and dimensionless 
    quadrupole moments with $M=1.4 \, M_{\odot}$ (red) for $N_{\rm q}=18$.}
    \label{fig:contour_quadrupole_nq18}
\end{figure}

\begin{figure}
    \centering
    \includegraphics[width=8cm]{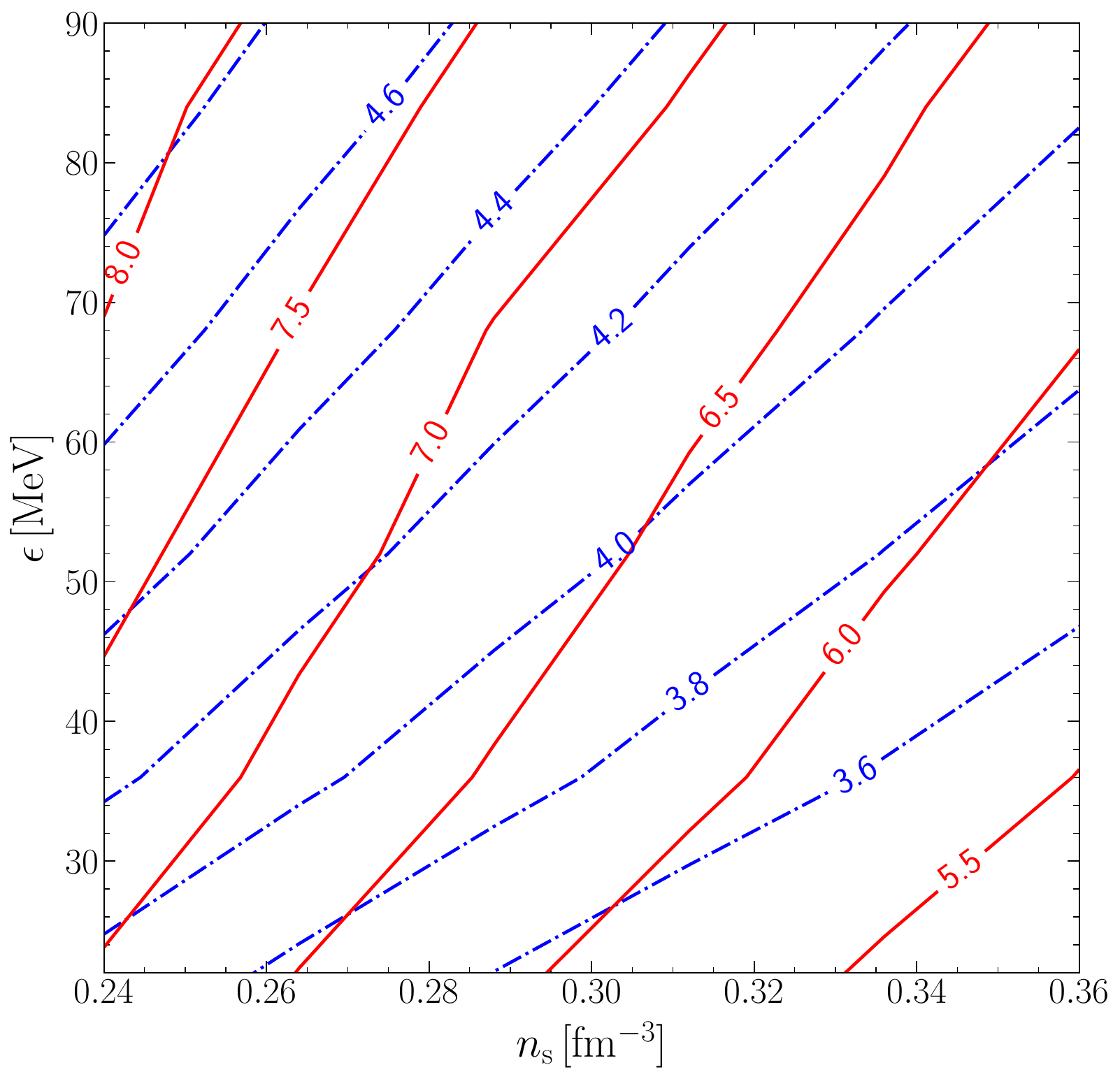}
    \caption{Same as Fig.~\ref{fig:contour_quadrupole_nq18}, but for $N_{\rm q}=9$.}
    \label{fig:contour_quadrupole_nq9}
\end{figure}

We also show the relations between the quadrupole moments and the masses of the
stars in Fig.~\ref{fig:nq18quadrupole}.  A key feature is that the quadrupole
moments decrease with the increase of the masses for all of EoS that we
consider.  The quadrupole moments also depend on the EoS strongly to some
extent. For example, the dimensionless quadrupole moment $\bar Q$ can range from
$\sim2$ to $\sim8$ at $1.4\,M_{\odot}$ for the EoSs we selected. Thus, The
measurements of the quadrupole moments can give constraints on the EoS. In
Fig.~\ref{fig:contour_quadrupole_nq18} and 
Fig.~\ref{fig:contour_quadrupole_nq9}, we display the contours of maximal mass
and quadrupole moment at $1.4 \, M_{\odot}$ for SSs with $N_{\rm q}=18$ and
$N_{\rm q}=9$. The constraints on quadrupole moments can be used to constrain
the parameter space of $\epsilon$ and $\sigma$.

The pulsed emission of X-rays originating from the surface of rotating NSs
contains the information of the strong-filed regime around the NSs, which can be
characterized by the global properties which mainly consist of mass, radius, and
spin frequency.  Detailed modelling of the emission region on the stellar
surface combined with the relativistic null geodesic of photons can be used to
construct theoretical light curves, which can then be compared with the observed
light curves to probe the masses and radii, and then constrain the EoS of the
NSs~\citep{Morsink:2007tv,Watts:2016uzu}. 

Some of the targets for X-ray observations have moderate spins
$\sim300$--$700\,\rm Hz$ \citep{Bogdanov:2008qm}.  Besides the masses and radii,
the quadrupole moments and the eccentricity of the stars also affect the light
curves of X-rays~\citep{Morsink:2007tv,Baubock:2013gna}. \citet{Morsink:2007tv}
considered the quadrupole moment and the shape of the NSs when modelling the
X-ray profiles. It is the so-called oblate-Schwarzschild approximation (OS). It
is found that the quadrupole moment and the eccentricity are important in
modelling the light curves, and for some emission geometries, the deformations
of the stars can rival the Doppler effects~\citep{Morsink:2007tv}.  The main
reason is that the oblate shape will make some certain spot locations visible
that would be invisible in the spherical cases, and vice versa.
\citet{Baubock:2012bj, Baubock:2013gna} showed that the quadrupole moments can
also induce features with narrow peaks in the X-ray flux and they also found
that the shape parameters calculated with Hartle-Thorne approximation are
consistent with the numerical results obtained by~\citet{Morsink:2007tv} to an
accuracy of $1\%$ for observed spin frequencies. 

On one hand, the universal relations between different quantities (such as $\bar
Q$ and $\mathcal C$, $e_{\rm s}$ and $\mathcal C$) can help to decrease the
dimensions of parameter space when modelling the
profiles~\citep{Baubock:2013gna}. On the other hand, the difference of the
universality for normal NSs, QSs, and SSs might be used to determine whether the
pulsars are gravitationally bound or self-bound. 

Now the NICER satellite is taking data from some X-ray pulsars and has given
certain constraints on the radii of NSs \citep{Miller:2019cac,Riley:2019yda} and
the OS approximation is commonly used in the modelling of X-ray profiles.  In
the future, the observations may also give constraints on the quadrupole moments
and the shapes of rotating stars.

For binary systems involving NSs, the quadrupole moments also contribute to GW
radiations through the quadrupole-monopole interactions
\citep{Poisson:1997ha,Yagi:2013awa,Isoyama:2018rjb,Harry:2018hke}. The leading
order effect enters into the waveform at the 2\,PN order, and the correction to
the GW phase is roughly proportional to $\sim \bar Q \chi^{2}$
\citep{Poisson:1997ha,Harry:2018hke}. Physically, it is a Newtonian effect
despite the scaling has the form of PN expansion. It may be possible
to measure or constrain the quadrupole moment with GWs. \citet{Yagi:2013awa}
performed GW data analysis for binary NSs and evaluated the possibility to
constrain quadrupole moments with the next generation ground-based GW detector
ET, and space-based detectors DECIGO/BBO. They found that although the
quadrupole moments are hard to measure due to the strong correlations with the
spins of NSs, at least one can put upper bounds on the quadrupole moments.  If
the NSs in the binary systems rotate rapidly, then the measurement of quadrupole
moments is possible \citep{Isoyama:2018rjb,Yagi:2013awa,Liu:2021dcr}.

\subsection{Third order: Corrections to the angular momentum and the moment of inertia}

Taking the integral in Eq.~(\ref{eqn:J_total}) and extending over the region
that is interior to the isodensity surface given by $R+\xi_{0}(R)+\xi_{2}(R)
P_{2}(\theta)$, one finds that only odd orders of $\Omega$ contribute to the
angular momentum. At the third order of $\Omega$,
\begin{align}
     \delta &J=-\frac{1}{6 }\left\{r^{4} j\frac{\mathrm{d} w_{1}}{\mathrm{~d} r}+r^{4} j 
     \frac{\mathrm{d} \bar{\omega}}{\mathrm{d} r}\left[h_{0}+\frac{m_{0}}{r-2 m}+\right.\right.\nonumber \\
    &\left.\left.+\frac{1}{5}\left(4 v_{2}-5h_{2}-\frac{m_{2}}{r-2 m}\right)\right]+4 r^{3} 
    \frac{\mathrm{d} j}{\mathrm{~d} r} \bar{\omega}\left(\xi_{0}-\frac{1}{5} \xi_{2}\right)\right\}{\bigg|_{r=R_{+}}}\,.
\end{align}
The moment of inertia at the second order of $\Omega$ is $\delta I=\delta
J/\Omega$.  Note that each term in the above expression is evaluated at
$r=R_{+}$ and the match conditions for $m_{0}$, $m_{2}$, and $r^{4} j\dd
w_{1}/\dd r$ at the boundary need to be considered for SSs and QSs.  The details
can be found in the Appendix~\ref{append:A}.

\begin{figure}
    \centering
    \includegraphics[width=8cm]{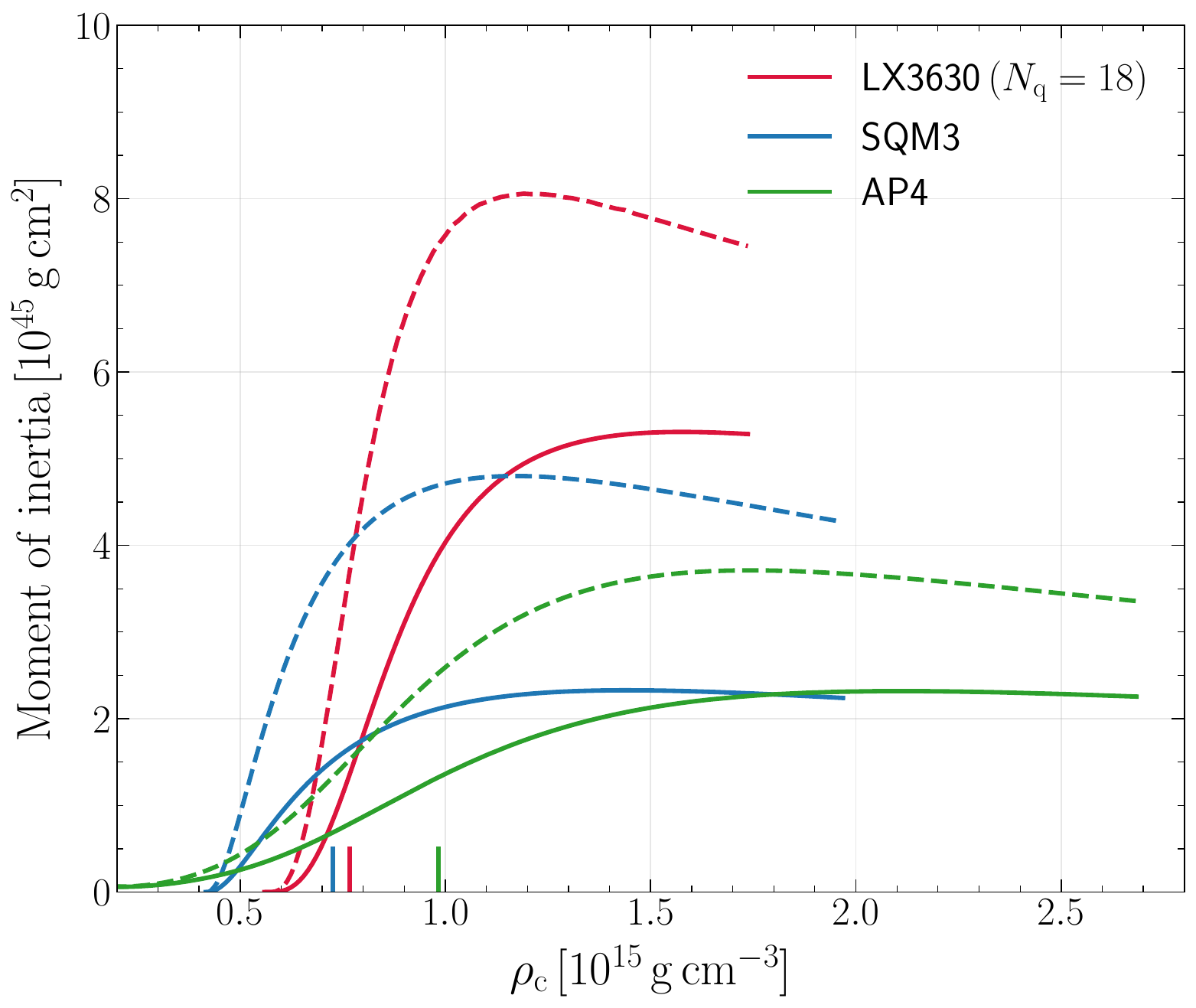}
    \caption{The moment of inertia versus central mass-energy density. The solid
    lines are the moment of inertia for non-rotating configuration. The dashed
    lines are moment of inertia with the rotation angular frequency
    $\Omega_{*}$. The moment of inertia for a smaller frequency $\Omega$, where
    the slow rotation approximation is valid, can be easily obtained by an
    upward displacement from the solid curves by the fraction
    $(\Omega/\Omega)^{2}$ of the distance to the dashed curves.  We choose EoS
    LX3630 ($\,N_{\rm q}=18$), SQM3, and AP4 in the plot. The short vertical
    lines at the bottom of the figure represent the central mass-energy density
    of non-rotating configuration with $M=1.4M_{\odot}$.}
    \label{fig:thirdmoi}
\end{figure}

In Fig.~\ref{fig:thirdmoi}, we plot the $\delta I +I$ and $I$ for representative
EoSs for SSs, QSs, and normal NSs.  Note that we take the angular frequency to
be $\Omega_{*}$. The moment of inertia for a lower frequency $\Omega$ at a
specific central density can be easily obtained by multiplying the rescaling
factor $(\Omega/\Omega_{*})^{2}$.

Under some conditions, we are interested in computing the moment of inertia as a
function of the angular velocity for a star with given baryonic mass (e.g., the
glitch processes and the spin evolution of newly-born NSs). 
\citet{Hartle:1973zza} first constructed the constant baryonic sequence based on
his early work \citep{Hartle:1967he}, and the procedures are as follows:
\begin{enumerate}
    \item Same as the first step in the constant density sequence, one obtains
    baryonic mass $\bar{M}$, gravitational mass $M$, and the radius of the star
    $R$ from the static and non-spinning configuration.
    \item The structures are calculated to the second order with the same
    central density. In order to obtain the same baryonic mass $\bar{M}$, one
    imposes that the boundary value of $p_{0}$ and $m_{0}$ deviate from zero at
    the center of the star until the corrections to baryonic mass $\delta \bar
    {M}$ is equal to zero. This means that the central density is perturbed from
    background value $\rho_{\rm c}$.
    \item Calculate the third order perturbations based on the boundary
    conditions used in the second step.
\end{enumerate}
The central idea of Hartle's approach is treating the change of central density
$\delta{\rho_{\rm c}}$ as a perturbation.  This assumption will breakdown in two
cases. First, when the star rotates sufficiently rapid, $\delta{\rho_{\rm c}}$ 
is actually not a small value and this procedure will produce large errors to
other quantities \citep{Benhar:2005gi}.  The second breakdown appears when the
mass of the star is close to the maximal mass \citep{Hartle:1973zza}. We denote 
the rotational perturbation of the baryonic mass in the constant density
sequence as $\delta \bar{M}(\rho_{\rm c},\Omega)$.  Now we want to construct a
rotating star with the same baryonic mass by perturbing the central density. The
variation of the baryonic mass is 
\begin{equation}
    \delta \bar{M}(\rho_{\rm c},\Omega)= \left.\frac{\partial \bar{M}}{\partial \Omega}\right|_
    {\rho_{\rm c}}\delta \Omega + \left.\frac{\partial \bar{M}}{\partial \rho_{\rm c}}\right|_{\Omega}\delta \rho_{\rm c}=0\,.
\end{equation}
It follows that for a given baryonic mass \citep{Hartle:1973zza,Benhar:2005gi}
\begin{equation}
   \left.\frac{ \partial  \rho_{\rm c}}{\partial \Omega}\right|_{\bar{M}} =- \bigg(\left.\frac{\partial 
   \bar{M}}{\partial \Omega}\right|_{\rho_{\rm c}} \bigg) \bigg/ \bigg( \left.\frac{\partial \bar{M}}{\partial \rho_{\rm c}}\right|_{\Omega} \bigg)\,.
\end{equation}
When the sequence is close to the maximal mass, the term $\partial
\bar{M}/\partial \rho_{\rm c}|_{\Omega}\rightarrow 0$.  Consequently, the change
of the central density $\partial  \rho_{\rm c}/\partial
\Omega|_{\bar{M}}\rightarrow \infty \,$, which violates the assumption that the
change of central density is a small correction and this approach fails.  The
solutions become unstable near the maximal masses as shown in
\citet{Hartle:1973zza}.

\citet{Benhar:2005gi} formulated another procedure to obtain the constant
baryonic mass sequence based on \citet{Hartle:1973zza}.  The procedure is as
follows:
\begin{enumerate}
    \item Same as the first step in the Hartle's constant density sequence, one
    can obtain a baryonic mass $\bar{M}$ for an assigned EoS and central density
    $\rho_{\rm c}$.
    \item Choosing an angular velocity $\Omega$ and integrating the perturbed
    equations to the third order of $\Omega$ for various values of $\rho_{\rm
    c}$, one can get a branch of solutions with the same angular velocity
    $\Omega$ but different central densities. Among these solutions, one chooses
    the one with the same baryonic mass $\bar{M}$ as the unperturbed one.
\end{enumerate}\
This approach is stable around the maximal mass. \citet{Benhar:2005gi} compared
this perturbative approach with the exact numerical solutions and found that
this algorithm is better than Hartle's in high spin frequencies. Although this
approach will also produce large errors around the maximal mass due to the fact 
$\partial \bar{M} /\left.\partial \rho_{\mathrm{c}}\right|_{\Omega} \rightarrow
0$ at the maximal mass, but the solutions are at least stable and can give more
accurate results for large spins compared to Hartle's approach.  Therefore, we
take this approach to construct the constant baryonic mass sequence in our
calculations.

\begin{figure}
    \centering
    \includegraphics[width=8cm]{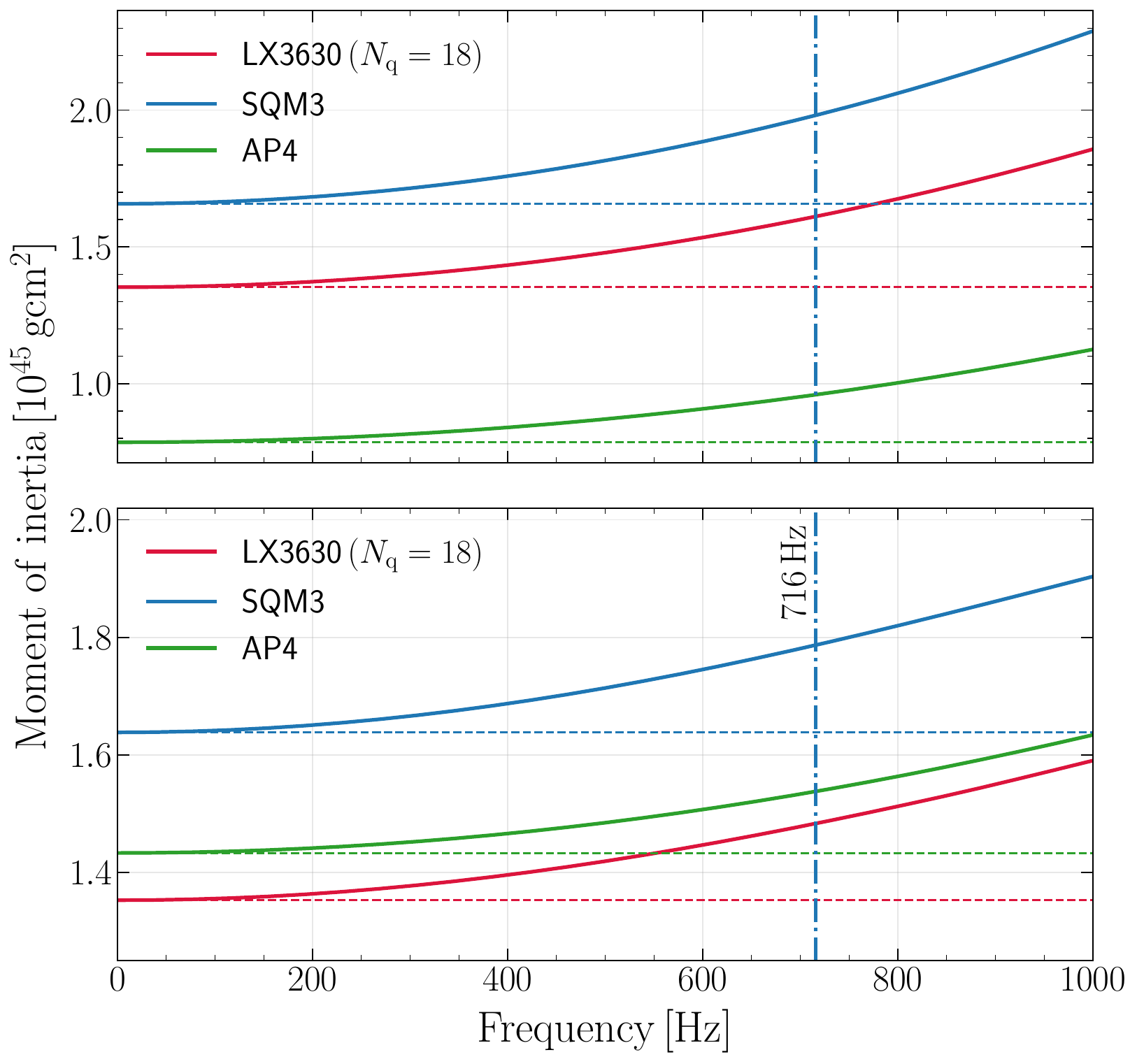}
    \caption{The upper panel shows the relation between the moment of inertia
    and spin frequency for constant central density sequence.  We take the
    central mass-energy to be $7.67\times 10^{14}\,\rm g\, cm^{-3}$, which
    corresponds to  $M=1.4M_{\odot}$ for the EoS LX3630 ($N_{\rm q}=18$). The
    solid lines are the moment of inertia for rotating stars $I+\delta I$, while
    the dashed lines represent the moment of inertia for non-rotating
    configuration $I$. The lower panel shows the moment of inertia in the
    constant baryonic mass sequence. We take the baryonic mass as
    $1.67\,M_{\odot}$, which corresponds to the gravitational mass 
    $M=1.4\,M_{\odot}$ for the EoS LX3630 ($N_{\rm q}=18$). The solid lines
    represent $I+\delta I$ while the dashed lines represent $I$. The vertical
    line at $716\,\rm Hz$ represents the observed fastest rotating pulsar,
    PSR~J1748$-$2446ad.} 
    \label{fig:third_constant}
\end{figure}

In Fig.~\ref{fig:third_constant}, we plot the moment of inertia versus the
rotating frequencies for the constant central density sequence ({\it upper
panel}) and the constant baryonic mass sequence ({\it lower panel}). The
constant central density sequence can be represented as simple quadratic
functions directly. For low spin, the correction to the moment of inertia is
very tiny. In the case of PSR~J0737$-$3039A, the rotating frequency is $\sim
45.5\,\rm Hz$ and the second order contribution can be ignored in the
discussions of Lense-Thirring precession.  But the corrections become obvious as
the star rotates sufficiently fast. For example, for PSR~J1748$-$2446ad, the
fastest spinning pulsar observed with the frequency $716\,\rm Hz$, the relative
error $\delta I/(I+\delta I)$ one makes by neglecting the contribution $\delta
I$ at order $\Omega^{2}$ are $8.8\%$ for LX3630 ($N_{\rm q}=18$), $8.3\%$ for
SQM3, and $6.8\%$ for AP4. In Table~\ref{table:deltaI}, we present the relative
errors for the constant baryonic mass sequence shown in the lower panel of
Fig.~\ref{fig:third_constant} at different rotating frequencies.  The relative
errors for LX3630 and SQM3 are larger than AP4, which results from the fact that
SSs and QSs are more compact than hadronic NSs.
 
\def\arraystretch{1.2}
\begin{table}
    \centering
    \caption{The relative error $\delta I/(I+\delta I)$ for EoSs LX3630 ($N_{\rm
    q}=18$), SQM3, and AP4 in different rotating frequencies.}
    \begin{tabular}{l | c c c}
        \hline \hline
        \multicolumn{1}{p{2.5cm}|}{ } & \multicolumn{3}{p{4cm}}{Relative errors $\delta I/(I+\delta I)\,[\%]$}\\
        \hline
        Frequency $\,[\rm Hz]$ & LX3630  & SQM3  & AP4  \\
        \hline
        100 & $0.197$ & $0.191$ & $0.149$ \\
        200 & $0.792$ & $0.755$ & $0.577$ \\
        300 & $1.77$ & $1.67$ & $1.28$ \\
        400 & $3.08$ & $2.91$ & $2.26$ \\
        500 & $4.68$ & $4.41$ & $3.47$ \\
        600 & $6.50$ & $6.13$ & $4.91$ \\
        700 & $8.48$ & $8.01$ & $6.54$ \\
        800 & $10.6$ & $10.0$ & $8.34$ \\
        900 & $12.7$ & $12.0$ & $10.3$ \\
        1000 & $15.0$ & $14.0$ & $12.3$ \\
        \hline
    \end{tabular}
    \label{table:deltaI}
\end{table}

\section{Tidal deformation and tidal Love numbers}
\label{sec:tidal}

For tidally-deformed relativistic stars, the metric $g_{tt}$ in the star's local
asymptotical rest frame can be represented
via~\citep{Flanagan:2007ix,Hinderer:2007mb}  
\begin{align}
    \label{eqn:tidal_metric}
    -\frac{\left(1+g_{t t}\right)}{2}=&-\frac{M}{r}-\frac{3 Q_{i j}}
    {2 r^{3}}\left(n^{i} n^{j}-\frac{1}{3} \delta^{i j}\right)+O\left(\frac{1}{r^{4}}\right) \nonumber\\
    &+\frac{1}{2} \mathcal{E}_{i j} x^{i} x^{j}+O\left(r^{3}\right)\,,
\end{align}
where $\mathcal{E}_{i j}$ is the tidal field generated by the companion of the
star, and $Q_{i j}$ is the quadrupole moment of the NS induced by the tidal
field. To characterize the deformations of the stars, one usually defines the
tidal deformability as 
\begin{equation}
    \lambda \equiv  -Q_{i j}/\mathcal{E}_{i j}\,,
\end{equation}
which measures the ability to be deformed by the tidal field and depends on the
EoS.  It is related to the $l=2$ Love number $k_{2}$ via $k_{2}=3\lambda
R^{-5}/2$.

To calculate the metric in Eq.~(\ref{eqn:tidal_metric}) and give the tidal
deformability $\lambda$, one introduces a $l=2$ even parity and static
perturbation on the spherical background. The metric perturbation in the
Regge-Wheeler gauge~\citep{Regge:1957td} can be represented
as~\citep{Thorne1967}
\begin{equation}
    h_{\mu\nu}^{(2m)}=Y_{2m}(\theta,\phi)
    \left[ 
      \begin{matrix}
        -e^{\nu}H_0 & H_1 & 0 & 0 \\
        H_1 & e^{\lambda} H_2  & 0 & 0 \\
        0 & 0 & \rho^2K & 0 \\
        0 & 0 & 0 & \rho^2 \sin^2\theta K
      \end{matrix}
    \right]\,,
    \label{eqn:deltag}\\
\end{equation}
where $Y_{lm}(\theta,\phi)$ are the spherical harmonics and $H_{0}$, $H_{1}$,
$H_{2}$ and $K$ are functions that only depend on $r$. Correspondingly, the
matter perturbations are 
\begin{equation}
    \delta T^{0}_{0}=-\delta \rho(r) Y_{20}(\theta, \phi) \,,\quad \delta T_{i}^{i}=\delta P(r) Y_{20}(\theta, \varphi)\,.
\end{equation}
Substituting $\delta P$ into $\delta \rho \,\dd P/\dd \rho $ and solving the
linearized Einstein equations, $\delta G^{\beta}_{\alpha}=8\pi
T^{\beta}_{\alpha}$, one obtains~\citep{Hinderer:2007mb,Damour:2009vw}
\begin{equation}
    H_{0}=-H_{2}=H\,,\quad H_{1}=0\,,\quad K^{\prime}=-H^{\prime}- H \nu^{\prime}\,,
\end{equation}
where the prime denotes the derivative respect to $r$. The ordinary differential
equations of $H(r)$ and $\alpha(r)\equiv H^{\prime}(r)$ are 
\begin{align}
    \frac{\dd H}{\dd r}=&\alpha (r)\,,\\
    \frac{\dd \alpha}{\dd r}=&-\alpha(r)\left\{\frac{2}{r}+e^{\lambda}\left[\frac{2 m(r)}{r^{2}}+4 \pi r(P-\rho)\right]\right\}\nonumber \\
    &-H\left[-\frac{6 e^{\lambda}}{r^{2}}+4 \pi e^{\lambda}\left(5 \rho+9 P+\frac{\rho+P}{\dd P / \dd \rho}\right)-\nu^{\prime 2}\right]\,.
\end{align}
One integrates the differential equations of $H(r)$ and $\alpha(r)$ to the
surface of the star $R$ with the boundary conditions $H(r)=a r^{2}$ and $\alpha
(r)=2a r$ as $r\rightarrow 0$.  Here $a$ is a constant that can be chosen
arbitrarily and will be cancelled in the calculations of the Love numbers. For
QSs and SSs, the term $\dd P/\dd \rho$ in the differential equation of $\alpha$
is not continuous across the surface of the star, just like the case in the slow
rotation. Thus, the match conditions of $H(r)$ and $\alpha(r)$ at the radius $R$
is 
\begin{equation}
    \label{eqn:H_match}
    [H]=0\,,\quad [\alpha]=[H^{\prime}]=-4\pi r^{2}H(R_{-})\rho(R_{-})/M \,.
\end{equation}
The exterior solution of $H$ can be solved
analytically~\citep{Thorne1967,Hinderer:2007mb}
\begin{equation}
    H=c_{1} Q_{2}^{2}\left(\frac{r}{M}-1\right)+c_{2} P_{2}^{2}\left(\frac{r}{M}-1\right)\,,
\end{equation}
where $c_{1}$ and $c_{2}$ are constants. The function $P_{2}^{2}$ is the
associated Legendre function of the first kind and $P_{2}^{2}(r / M-1) \sim
r^{2}$ at large $r$, while the function $Q_{2}^{2}$ is the associated Legendre
function of the second kind and $Q_{2}^{2}(r / M-1) \sim r^{-3}$ at large $r$. 
Taking the expansion of $H(r)$ at large $r$ and comparing with the multipole
moments defined in Eq.~(\ref{eqn:tidal_metric}), one obtains 
\begin{equation}
    \lambda=\frac{8M^{5}}{45}\frac{c_{1}}{c_{2}}\,,
    \quad k_{2}=\frac{3}{2}\lambda R^{-5}=\frac{4\mathcal{C}^{5}}{15}\frac{c_{1}}{c_{2}}\,.
\end{equation}
One matches the solutions of $H(r)$ and $\alpha(r)$ at the surface of the star
and gives the solutions of $c_{1}/c_{2}$ in terms of the interior solution at
$r=R$.  Then the tidal Love number $k_{2}$ can be
obtained~\citep{Hinderer:2007mb} 
\begin{align}
    \label{eqn:k2}
    k_{2}=& \frac{8 \mathcal C^{5}}{5}(1-2 \mathcal C)^{2}[2+2 {\mathcal C}(y-1)-y] \nonumber\\
    & \times \Big\{2 {\mathcal C}[6-3 y+3 {\mathcal C}(5 y-8)]\nonumber\\
    &+4 {\mathcal C}^{3}\left[13-11 y+{\mathcal C}(3 y-2)+2 \mathcal C^{2}(1+y)\right] \nonumber\\
    &+3(1-2 \mathcal C)^{2}[2-y+2 \mathcal C(y-1)] \ln (1-2 \mathcal C)\Big\}^{-1}\,,
\end{align}
where $y\equiv R\alpha(R)/H(R)$. For QSs or SSs, one needs to take into account
the match conditions in  Eq.~(\ref{eqn:H_match}), and $y(R)$ can be represented
as  
\begin{equation}
    y(R)=\frac{R\alpha(R_{-})}{H(R_{-})}-\frac{4\pi R^{3}\rho({R_{-})}}{M}\,.
\end{equation}
The second term contributes crucially to the tidal Love numbers of QSs or SSs
and cannot be ignored. The tidal deformability can be calculated with the
relation $\lambda=2k_{2}R^{5}/3$. For later analysis of GW constraints, we will
concentrate on the dimensionless tidal deformability 
\begin{equation}
    \Lambda=\frac{2k_{2}}{3 \mathcal C^{5}}\,.
\end{equation}

\begin{figure}
    \includegraphics[width=8cm]{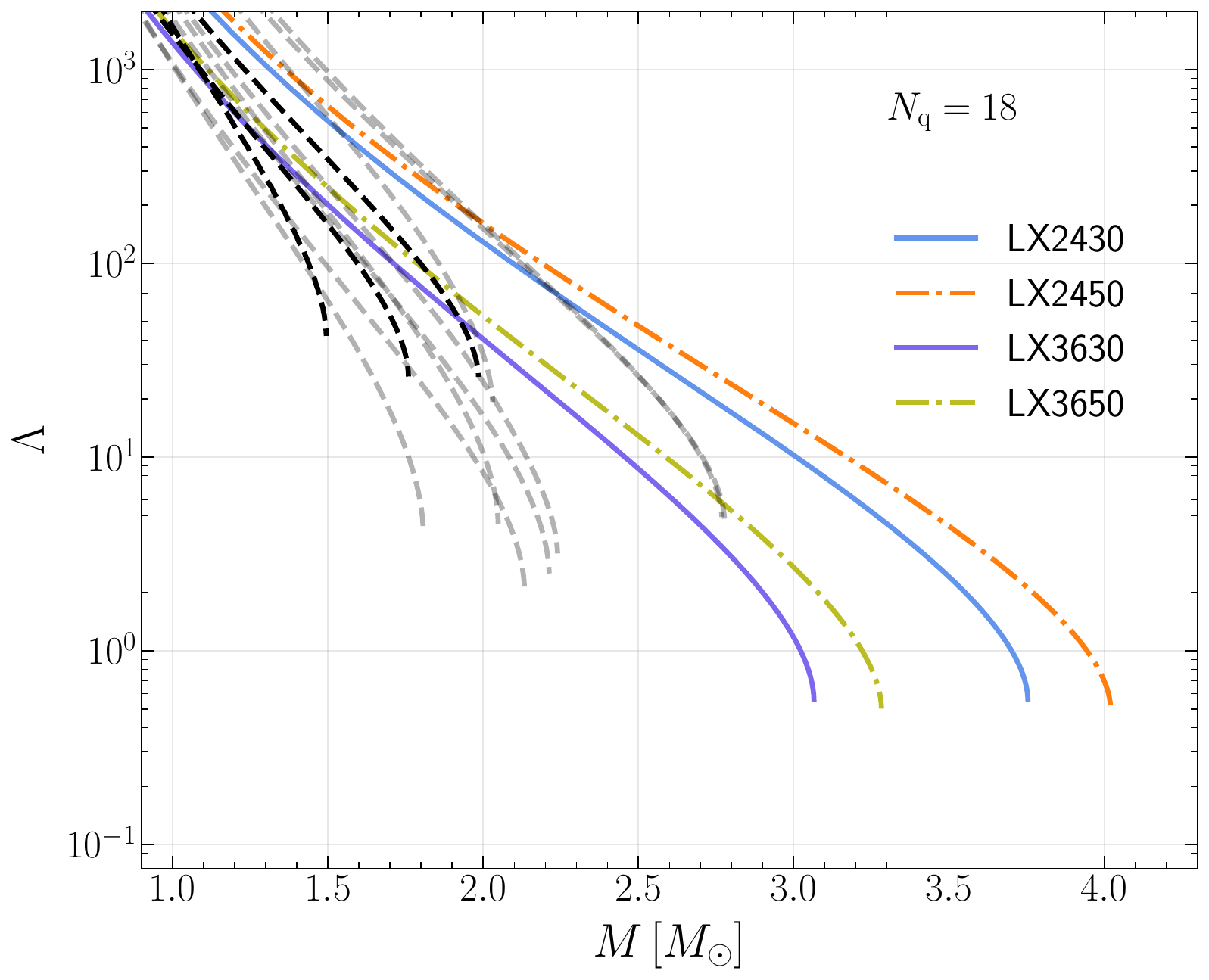}
    \caption{Dimensionless tidal deformability $\Lambda$ as a function of the
    mass $M$ for SSs ($N_{\rm q}=18$), QSs (black dashed lines), and normal NSs
    (grey dashed lines). For SSs with $N_{\rm q}=9$, the trend of $\Lambda$ is
    basically the same as $N_{\rm q}=18$, and we do not show them here for
    clarity.}
    \label{fig:Lambda_M}
\end{figure}

In Fig.~\ref{fig:Lambda_M}, we display the relation between the dimensionless
tidal deformabilities and the masses for SSs, QSs, and normal NSs. For the mass
range we plot, a common feature is that $\Lambda$ decreases with the increase of
the mass because the star becomes more and more compact and harder to be
deformed. For SSs, as the potential depth $\epsilon$ increases and the surface
baryonic density $n_{\rm s}$ decreases, the EoS becomes stiffer, which leads to
larger maximal masses and tidal deformabilities. Compared to normal NSs and QSs,
SSs are very compact near the maximal masses and the dimensionless tidal
deformabilities can extend to the value smaller than one. For Schwarzschild
black holes, the tidal deformabilities are zero since $1-2\mathcal C$ in
Eq.~(\ref{eqn:k2}) becomes zero. This feature is guaranteed by the no hair
theorem~\citep{Damour:2009vw}. 

One can notice that tidal deformability is proportional to the fifth power of
the radius $R$. Therefore, constraining or measuring tidal deformability of NSs
can provide important information on the EoS of NSs. Actually, the tidal 
deformations of NSs have imprints on the GWs from binary NSs. At the early stage
of inspiral, the dynamical motion can be treated as point particles. But once
the binary system evolves to the late stage of inspiral, the finite size effects
induced by the tidal interactions will affect the motions of binary system and
contribute to the GW  emission~\citep{Flanagan:2007ix}.  The tidal contributions
to the evolution of GW phases first enter at 5\,PN. It is actually a Newtonian
term in spite of scaling with PN order. Since the energy goes to deform the star
and the induced quadrupole moments will contribute to the GW radiation, the 
phase evolution will be faster than non-spinning point particles with the same
mass~\citep{Dietrich:2020eud}. The phase corrections depend on a parameter
$\tilde \Lambda$, which is a mass-weighted linear combination of the
dimensionless tidal deformabilities of two stars~\citep{Flanagan:2007ix}
\begin{equation}
    \tilde{\Lambda}=\frac{16}{13} \frac{\left(m_{1}+12 m_{2}\right) m_{1}^{4} 
    \Lambda_{1}+\left(m_{2}+12 m_{1}\right) m_{2}^{4} \Lambda_{2}}{\left(m_{1}+m_{2}\right)^{5}}\,,
\end{equation}
where $m_{1,2}$ and $\Lambda_{1,2}$ represent the masses and the tidal
deformabilities of the binary components respectively.

\begin{figure}
    \centering
    \includegraphics[width=8cm]{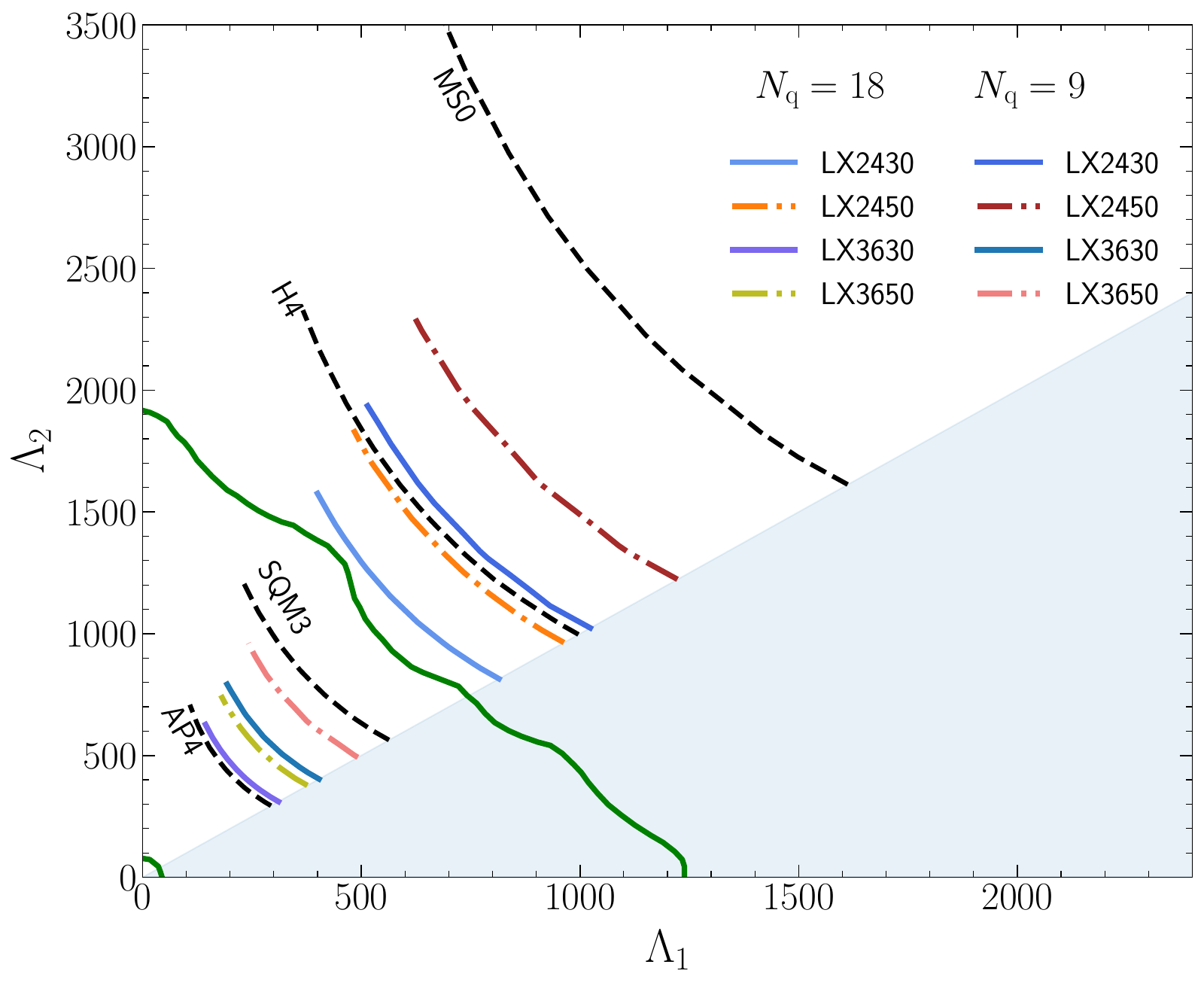}
    \caption{Posterior of tidal deformabilities (green lines) for the binary NS
    in GW170817 with restricted low spins.  The data of the posterior is taken
    from~\citet{LIGOScientific:2018hze}. The green line indicates the enclosure
    of the $90\%$ credible regions of $\tilde{\Lambda}$, $300_{-230}^{+420}$.
    The primary mass $m_{1} \in(1.36,1.60) \,{M}_{\odot}$ and the secondary mass
    $m_{2} \in (1.16,1.36) \,{M}_{\odot}$. The tidal deformability for SSs with 
    $N_{\rm q}=18$ and $N_{\rm q}=9$ are displayed. For comparison, we also plot
    the tidal deformabilities for AP4, H4, MS0, and SQM3. The shaded region is
    $\Lambda_{1}>\Lambda_{2}$.}
    \label{fig:tidal_posterior}
\end{figure}

The GWs from the binary NS inspiral, GW170817, give the constraints on the tidal
deformabilities for the first time
~\citep{LIGOScientific:2017vwq,LIGOScientific:2018cki,LIGOScientific:2018hze}.
In the discovery paper, \citet{LIGOScientific:2017vwq} placed a $90\%$ upper
limit of $\tilde{\Lambda} \leq 800$ for low spin prior. With a linear expansion
of $\Lambda(m)$ at fiducial mass $1.4 \, M_{\odot}$, they also gave
$\Lambda(1.4M_{\odot})\leq 800$. In a following paper,
\citet{LIGOScientific:2018hze} extended the range of the GW frequencies from
$30\,\rm Hz$ in the initial analysis~\citep{LIGOScientific:2017vwq} down to
$23\,\rm Hz$.  Besides, several sophisticated and more accurate waveform models
augmented with other physical effects (such as spins) are used to do data
analysis. Under minimal assumptions about the nature of the compact objects,
\citet{LIGOScientific:2018hze} constrained the tidal deformability $\tilde
\Lambda$ in the range $\tilde \Lambda \in (0,630)$ for a high spin prior and
$\tilde \Lambda \in (70,720)$ for a low spin prior.
\citet{LIGOScientific:2018cki} complemented the study
of~\citet{LIGOScientific:2018hze} with the assumptions that GW170817 comes from
the inspiral of a binary NS whose masses and spins are consistent with the
galactic binaries. They concluded that the tidal deformability for a $1.4 \,
M_{\odot}$ NS is in the range $\tilde \Lambda \in (70,580)$ at a $90\%$
incredible level. For QSs, \citet{Miao:2021nuq} used GW170817 data and gave a
systematic study with Bayesian inference.

In Fig.~\ref{fig:tidal_posterior}, we take the posterior
in~\citet{LIGOScientific:2018hze} and plot the $90\%$ credible lines for the low
spin case. This posterior uses minimal assumptions on the nature of the compact
objects. The tidal deformabilities for several SSs with $N_{\rm q}=18$ and
$N_{\rm q}=9$ are presented. For comparison, we also show some selected models
of normal NSs and QSs. The constraints rule out several stiff normal EoSs (MS0,
H4) and models of SSs with very low surface baryonic densities (LX2430, LX2450)
at a $90\%$ credible level. Recall that the surface baryonic density is
inversely proportional to the cubic of $\sigma$. Thus the constraints indicate
that the repulsive core cannot extend too large. 

\begin{figure}
    \centering
    \includegraphics[width=8cm]{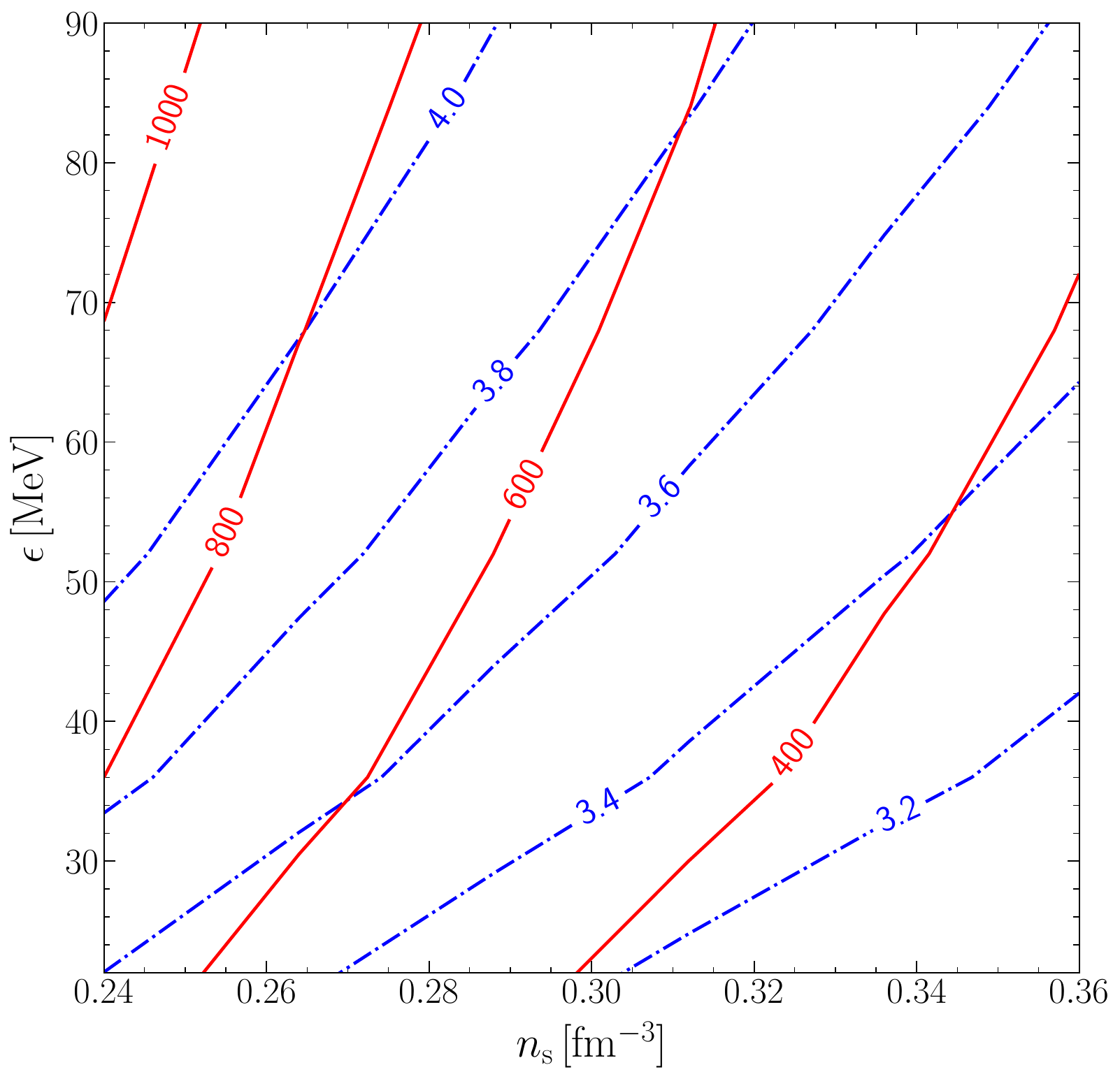}
    \caption{Contour lines of $\Lambda(1.4M_{\odot})$ (red) and maximal mass
    (blue) for $N_{\rm q}=18$.}
    \label{fig:tidal_nq18_contour}
\end{figure}

\begin{figure}
    \centering
    \includegraphics[width=8cm]{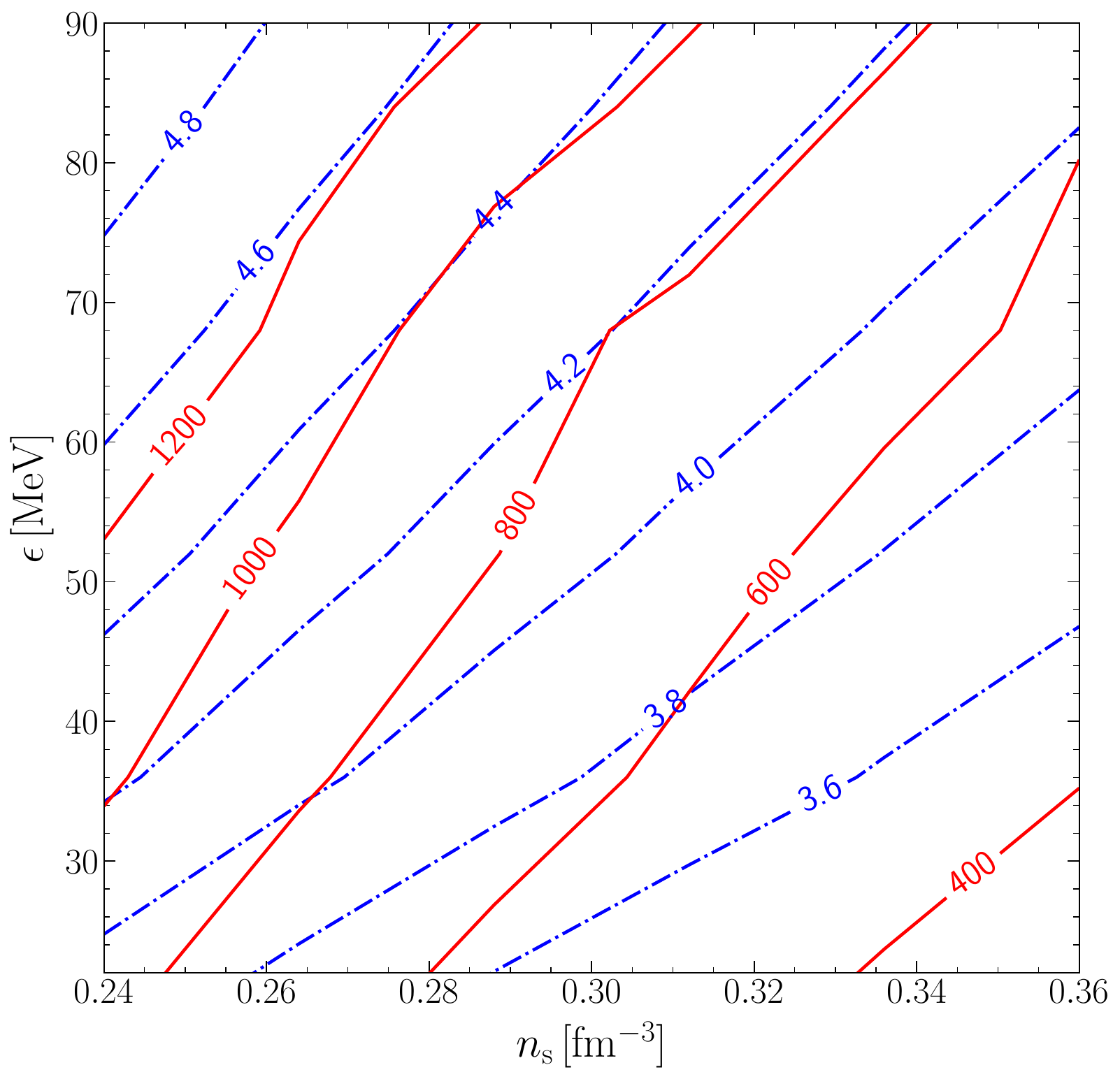}
    \caption{Same as Fig.~\ref{fig:tidal_nq18_contour}, but for $N_{\rm q}=9$.}
    \label{fig:tidal_nq9_contour}
\end{figure}

Using the posterior of tidal deformabilities in~\citet{LIGOScientific:2017vwq},
\citet{LZX2019} constrained the parameter space of the potential depth
$\epsilon$ and surface baryonic density $n_{\rm s}$ for SSs with $N_{\rm q}=18$.
Based on their work, we plot the contours for tidal deformabilities and maximal
masses for $N_{\rm q}=18$ and $N_{\rm q}=9$ in Fig.~\ref{fig:tidal_nq18_contour}
and Fig.~\ref{fig:tidal_nq9_contour} respectively. If we take the conservative
constraint $\Lambda(1.4M_{\odot})\leq800$ in the initial work
\citep{LIGOScientific:2017vwq}, the maximal mass at least should be less than
$4.2 \, M_{\odot}$ in the parameter space we choose.  

\section{I-Love-Q universal relations}
\label{sec:universal}

\citet{Yagi:2013awa,Yagi:2013bca} found remarkable EoS-insensitive universal
relations between the dimensionless moment of inertia $\bar{I}=I/M^{3}$,
quadrupole moment $\bar Q$, and the tidal deformability $\Lambda$, which is the
so-called I-Love-Q universal relations. The relative errors of the analytical
fits \citep{Yagi:2013awa,Yagi:2016bkt} connecting any of two quantities in the
I-Love-Q relations hold to $1\%$ for a variety of EoSs, including models for
normal NSs and QSs.
 
The I-Love-Q relations are useful in many aspects. For example, if one obtains
the moment of inertia from the aforementioned pulsar timing technique, then an
accurate estimation of the rotational quadrupole moment $\bar{Q}$ can be made.
The eccentricity, quadrupole moment, and moment of inertia affect the modelling
of the X-ray profiles of pulsars 
\citep{Morsink:2007tv,Baubock:2012bj,Baubock:2013gna,Gao:2020zcd}. The finite
size effects from rotation and tidal interactions will contribute to the
continuous GW emission
\citep{Hinderer:2007mb,Yagi:2013awa,Harry:2018hke,Dietrich:2020eud}.  Therefore,
the I-Love-Q relations can be used to break degeneracies between some parameters
in the modelling of X-ray profiles and waveform of GWs
\citep{Yagi:2013awa,Yagi:2016bkt,Baubock:2013gna,Silva:2017uov}. In this way,
the parameter space can be reduced and other parameters in the modelling can be
obtained more accurately \citep{Yagi:2016bkt,Baubock:2013gna}.

\begin{figure*}
    \centering
    \includegraphics[width=17cm]{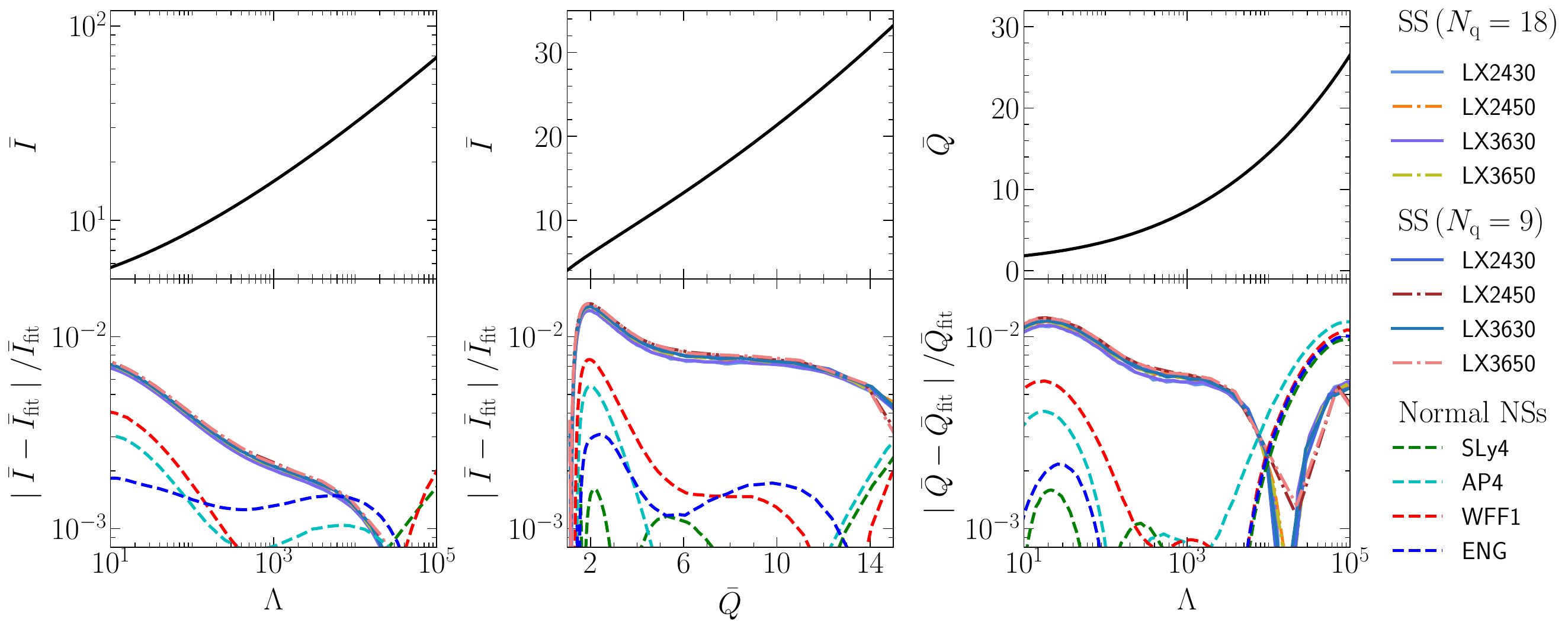}
    \caption{The EoS-insensitive universal relations between dimensionless
    quantities $\bar{I}$, $\bar{Q}$, and $\Lambda$.  The quantities are
    normalized via $\bar I\equiv I/M^{3}$, $\bar Q \equiv -Q_{\mathrm{r}}/(M^{3}
    \chi^{2})$, and $\Lambda \equiv \lambda/M^{5}$. Here $M$ is the mass of the
    non-rotating configuration, $\chi=J/M^{2}$ is the dimensionless spin, and
    $J$ is the angular momentum in the first order of $\Omega$. In the first
    row, we show the fitting curves of I-Love-Q relations taken from
    \citet{Yagi:2016bkt}. In the second row, we show the relative errors between
    the fitting values and the numerical data for SSs and normal NSs.}
    \label{fig:I_Love_Q}
\end{figure*}

So far, we have calculated static, slowly rotating, and tidally deformed SSs. 
Compared to other kinds of EoSs, SSs in the Lennard-Jones model admits $\dd
P/\dd \rho>1$ with the causality still being satisfied.  Therefore, it is
worthwhile to see whether the I-Love-Q universal relations still hold for SSs.
\citet{Yagi:2013awa,Yagi:2016bkt} showed that the I-Love-Q relations can be
fitted by fifth-order polynomials in the form 
\begin{equation}
    \ln y_{i}=\sum_{k=0}^{4} c_{k}\left(\ln x_{i}\right)^{k}\,,
\end{equation}
with great accuracy. Here $y_{i}$ and $x_{i}$ are any two quantities in
$\bar{I}$, $\bar Q$, and $\Lambda$, and $c_{k}$ are fitting coefficients.  In
the first row of Fig.~\ref{fig:I_Love_Q}, we show the fitting curves with the
fitting constants $c_{k}$ given in \citet{Yagi:2016bkt}. In the second row, we
show the relative errors between the fitting values and the numerical data. For
SS, the relative errors are smaller than $1\%$ for SSs in the most regions of
the parameter space. Some departures occur in the $\bar I$-$\bar Q$ and $\bar
Q$-$\Lambda$ relations, but the relative errors are still in the order of $1\%$.
We come to the conclusion that the I-Love-Q relations still hold despite $\dd
P/\dd \rho>1$.

\begin{figure}
    \centering
    \includegraphics[width=8cm]{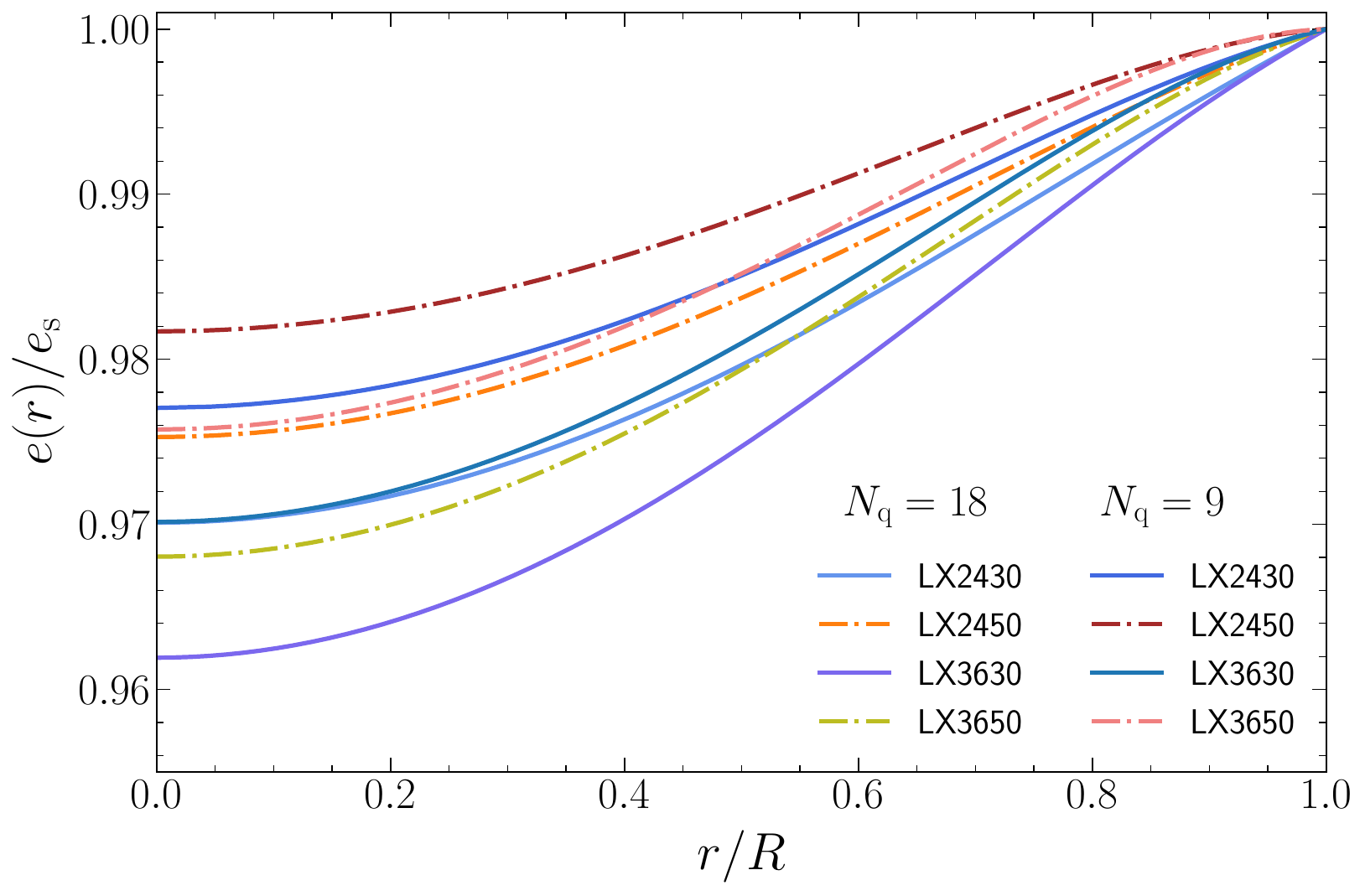}
    \caption{The eccentricity $e(r)$ at a given isodensity surface inside of the
    SSs divided by the eccentricity at the star's surface $e_{\rm s}$. Here we
    choose $M=1.4 \, M_{\odot}$ and relations for $N_{\rm q}=18$ and $N_{\rm
    q}=9$ are displayed.}
    \label{fig:isodensity_eccentricity}
\end{figure}

Another interesting feature is that the relative errors of SSs are nearly the
same for the EoS models that we choose. An instinctive reason is that the EoS
models come from the same form of mathematical equations (the Lennard Jones
model). We can further understand this feature (at least for the $\bar I$--$\bar
Q$ relation) with the eccentricities of those stars. After the discovery of the
I-Love-Q relations, \citet{Yagi:2014qua} explored the reason why this
universality occurs. They suggested that the universality results from an
emergent approximate symmetry.  With the increase of the compactness, the
variation of the eccentricity defined in Eq.~(\ref{eqn:eccentricity}) inside
stars decreases, which leads to a self-similarity of the isodensity surface
inside of the stars. This self-similarity indicates some common matter
distributions, therefore the universal behavior occurs for the exterior
multipole moments. Motivated by this argument, In
Fig.~\ref{fig:isodensity_eccentricity}, we plot the eccentricities inside of the
stars $e(r)$ divided by the surface eccentricity $e_{\rm s}$ for different
choices of parameters. We can notice that the variations of eccentricities at
given radial coordinate are very narrow for the selected models, about
$1.8$--$3.8\%$. The values and the range of the variations of eccentricities
are much smaller than that of different normal NSs used in the analysis of
\citet{Yagi:2014qua}. Therefore, if we use this scenario of how the universality
occurs, the features with nearly the same relative errors for SSs can be
understood, at least for the $\bar I$--$\bar Q$ relation.

In above discussions, we only consider slow rotation. For 
rapidly rotating NSs, \citet{Doneva:2013rha} extended the computation of the 
$\bar I$--$\bar Q$ relation to the mass shedding limit and found that the universality 
is lost. \citet{Chakrabarti:2013tca} discovered that it is still universal 
among various EoSs if one uses some dimensionless parameters to characterize the magnitude of rotation. 
More interestingly, one of the parameters involves the radius of NSs and a new 
universal relation expressing the radius with mass and spin parameter was found, which 
can be used as a powerful tool for radius measurement. \citet{Pappas:2013naa} discovered that 
the first four multipoles of NSs are related in a way that is independent of the EoS of NSs, 
which let us describe the geometry around NSs with only a few parameters quite accurately.
Because of the ultra-compact nature, the universal relations and spacetime geometry of rapidly 
rotating SSs may give us more valuable information. We leave them to future study.

\section{Discussions and Conclusions}
\label{sec:discussion}

Bulk strong matter at several times of nuclear densities may restore the
three-light-flavor symmetry~\citep{Xu:2003xe,Xu:2018zdo}.  At this energy level,
the quarks may not be deconfined and form quark clusters, which we call
strangeons. The residual strong interactions can trap the strangeons in the
potential well and the whole star is in a solid state. Therefore, we conjecture
that the pulsar-like compact objects could be actually SSs rather than NSs or
QSs. We use the Lennard-Jones model, which is parametrized by the potential
depth $\epsilon$ and the surface baryonic density $n_{\rm s}$, to describe the
EoS of SSs.  Though simple, it provides a powerful physically motivated
framework to study strangeon stars, complementing the parametrization usually
seen for NSs \citep{Read:2008iy}. In the Lennard-Jones model, the EoS is very
stiff due to the non-relativistic nature of the particles and the strong
repulsive force between the strangeons in the short distance. The astronomical
observations may give certain constraints on the parameters and even verify or
falsify the existence of SSs. Thus, we calculate the static, slowly rotating, and
tidally deformed SSs in details and briefly discuss some existing and possible
future observations that can constrain the EoS of SSs. The results in this work
are ready to be used in various scenarios.

In the calculations of the static and non-spinning background, all the parameter
space in our model can produce maximal mass $M_{\rm max}$ larger than
$2.5\,M_{\odot}$. In our model, ultra-compact stars near the maximal mass can 
invade into the region $\dd P/\dd \rho>1$ with the causality limit still being
satisfied. We also compared the structures of SSs with the analytical Tolman IV
and Tolman VII solutions discussed in~\citet{Tolman:1939jz},
\citet{Lattimer:2004sa}, and \citet{Lattimer:2012nd}. We found that the SSs can
possess maximal mass larger than the ones given by Tolman VII solution but is
still lower than that of Tolman IV solution. If pulsar-like compact stars are
SSs, a much long-lived star will form in the remnant of the GW170817
event~\citep{LZX2019}.  A stiffer EoS predicts smaller central mass-energy
density at maximal mass. If future observations of the shapiro time delay with 
pulsar timing technique and post-merger signals from binary mergers support the
existence of massive pulsar-like compact objects with mass larger than
$2.5\,M_{\odot}$, the phase transition from hadrons to free quarks may not
happen and the existence of SSs is favored.  

For slowly rotating SSs, we calculated the structures to the third order of
angular frequency $\Omega$. In the first order, the star remains to be spherical
and the local inertial frame is dragged. We calculated the angular momentum $J$
and the moment of inertia $I$ of SSs and some representative models of normal
NSs and QSs. Based on the work of~\citet{Lattimer:2000nx} and
\citet{Bejger:2002ty}, we studied the universal relations between the moment of
inertia and the compactness of the star. At low compactness, this universal
relation for SSs is basically the same as QSs. But as the compactness becomes
larger, EoS of QSs becomes soft while SSs are still very stiff. The universal 
relation will deviate from QSs and the moment of inertia is always close to the
limit set by the incompressible fluid. The frame dragging effect will induce
Lense-Thirring precession in the double pulsar system PSR~J0737$-$3039A/B. The
periastron advance due to the spin-orbit coupling may be detectable in the
upcoming years with the SKA~\citep{Hu:2020ubl}. A $10\%$ precision of the
moment-of-inertia measurement can give informative constraints on the parameter
space of $\epsilon$ and $n_{\rm s}$ in our model.

To the order $\Omega^{2}$, the star is deformed. On the spherical part, we
calculated the spherical stretching and the change in gravitational mass and the
baryonic mass. Differently from NSs, SSs have a hard surface with finite
density. The match conditions at the surface of star cannot be ignored and the
corrections must be crucially considered.  On the quadrupole part, we calculated
the quadrupole moments, eccentricities of isodensity surface, and investigated
the universal relation between the dimensionless quadrupole moments and
compactness discussed by~\citet{Urbanec:2013fs}. We found similar features shown
in the relations between the moment of inertia and the compactness. At large
compactness, the relations for SSs deviate from QSs.  We also find
quasi-universal relations between the surface eccentricity and the compactness
for SSs, QSs, and normal NSs.  The relations basically have the same features as
the relations between the quadrupole moments and the compactness. We also found
that the eccentricity and the compactness for SSs and QSs satisfy
quasi-universal relations, which are distinct from the relation for normal NSs
shown in \citet{Baubock:2013gna}.

To the third order of $\Omega$, we studied the corrections to the angular
momentum and the moment of inertia for the constant central density sequence and
the constant baryonic mass sequence. We found that for moderate spins, the
corrections of moment of inertia $\delta I$ are very small compared to the
zeroth-order contribution $I$. But for rapidly rotating stars, the corrections
can be up to $\sim 10\%$ for the EoSs we considered. For rapidly rotating NSs, 
\citet{Benhar:2005gi} found that the relative errors of the perturbative
approach compared to the results calculated from numerical relativity can be
reduced largely if the third order contributions are considered. This conclusion
could also be used for SSs.  Our calculations may be useful to study the spin
evolutions of newly-born SSs or glitch processes in pulsars.

For the tidally deformed SSs, we calculated the tidal deformabilities with the
appropriate match conditions at the surface of the stars. We used the posterior
of GW170817 to give a constraint on the parameter space of $\epsilon$ and
$n_{\rm s}$.  If we take the constraint of $\Lambda(1.4\,M_{\odot})\leq
800$~\citep{LIGOScientific:2017vwq}, we then come to the conclusion that the
maximal mass cannot be larger than $\sim 4.2\,M_{\odot}$ both for $N_{\rm q}=18$
and $N_{\rm q}=9$. In the future, smaller values of $\Lambda(1.4{M_{\odot}})$
are expected.

Based on the calculations of slow rotation and tidal deformation, we studied the
I-Love-Q universal relations \citep{Yagi:2013awa,Yagi:2013bca}. The universal
relations still hold although $\dd P/\dd \rho >1$. We also discussed the nearly
the same relative errors compared to the fitting formula given
in~\citet{Yagi:2014qua} especially for the $\bar I$-$\bar Q $ relation.  The
I-Love-Q relations and other universal relations such as the $I$-$\mathcal C$
and $\bar Q$-$\mathcal C$ relations can be used to study the GWs from the
binaries and the modelling of X-ray profiles.

A main concern of our calculations is that we take the perfect fluid assumption
to calculate the perturbations but SSs are actually in a solid state. The key
parameter for the calculations of the perturbations for solid components is the
shear modulus $\mu$. For NSs, the structure can be roughly divided into a
superdense fluid core and a solid crust. The interactions in the crust are
dominated by electromagnetic force and the mean shear modulus in the crust is
about $\mu \simeq 4\times 10^{29}\,\rm
erg/cm^{3}$~\citep{Ushomirsky:2000ax,Owen:2005fn}. \citet{Penner2011} developed
a framework to study the tidal response of NSs with solid crusts. They found
that the elasticity of the solid crust provides a small correction to the tidal
deformability. Recently, \citet{Gittins:2020mll} presented detailed formalism
that describes the static perturbations on the relativistic NSs with solid crust
and corrected some inconsistencies in~\citet{Penner2011}.  The results shows
that the inclusion of the solid crust has a negligible effect on the tidal
deformability of a NS, in the range of $\sim 10^{-8}$--$10^{-7}$. 

However, in our model, the interactions between strangeons are dominated by
strong force. The detailed calculations of the shear modulus is still not
performed, but it should be much larger than that of the NS's crust.  If the
burst oscillation frequencies observed in low-mass X-ray binaries correspond to
the first few torsional modes of SSs, the shear modulus should be about one
thousand times of the NS's crust, say $\mu \simeq 4\times 10^{32}\,\rm
erg/cm^{3}$ \citep{Xu:2003xe,Owen:2005fn}. We do not know how large the tidal
deformability will deviate from the fluid case for SSs, but we can obtain some
key insights of this problem from the calculations of tidal deformabilities for
crystalline color superconducting phase~\citep{Lau:2017qtz,Lau:2018mae}. QSs
composed of crystalline color superconducting phase are rigid with extremely
high shear modulus~\citep{Alford:2000ze,Mannarelli:2007bs}. The shear modulus is
approximately given by \citep{Mannarelli:2007bs}
\begin{equation}
    \mu=2.47 \, \mathrm{MeV} / \mathrm{fm}^{3}\left(\frac{\Delta}
    {10 \, \mathrm{MeV}}\right)^{2}\left(\frac{\mu_{\rm q}}{400 \, \mathrm{MeV}}\right)^{2}\,,
\end{equation}
where $\mu_{\rm q}$ is the average quark chemical potential and $\Delta$ is the
gap parameter which is in the range of $5$--$25\,\rm MeV$.  If the shear modulus
is taken to be $\mu=4\times 10^{32}\,\rm erg/cm^{3}$ ($\Delta \simeq 5\,\rm
MeV$), according to the results in \citet{Lau:2017qtz,Lau:2018mae}, the
deviations of the tidal deformabilities from the fluid case are in the order of
$1\%$. For larger gap parameters, the deviations can be very large and more
interestingly, they found that the universal relations between the moment of
inertia and tidal deformability will deviate from the ones for fluid stars. If
we take the shear modulus to be about $\mu \simeq 4\times 10^{32}\,\rm
erg/cm^{3}$, the corrections may not be very large and the results discussed in
our work is a good approximation. But if the shear modulus is actually larger,
then we need to consider the corrections due to the elasticity.

For the slow rotation approximation, \citet{Carter1975} developed Hartle-Thorne
approximation and gave the method to construct an elastic rotational solid star~
based on the elastic perturbation theory in general
relativity~\citep{carter1973elastic,Andersson:2020phh}. To the first order of
$\Omega$, the star remains to be spherical and~\citet{Quintana1976} found that
the moment of inertia is the same as that of fluid case. For higher order
perturbations, the star will be deformed and one needs to consider the effects
of solid state. The deviations to the fluid cases are also determined by the
value of shear modulus, which may have the same order of effects as the tidal.
Anyway, we ignore the solid aspects, namely the shear modulus, of the stars in
this study. In future, we will present the calculations of the shear modulus and
use the elastic perturbations in general relativity to study the structures of
SSs.

The nature of the supra-nuclear matter in compact stars closely relates to the
non-perturbative QCD at low energy levels, which still remains to be an unsolved
important problem in physics. The proposed SSs perceive pulsar-like compact
objects as solid stars composed of strangeons. We give a study on the structures
of rotating and tidally-deformed SSs within  a perturbative approach.  So far,
the SSs have passed many tests. The future observations of X-rays, radio
signals, and GWs will give us more information on the SSs and the bulk of strong
matter at low energy scales. 

\section*{Acknowledgements}

We thank the anonymous referee for comments.
This work was supported by the National SKA Program of China (2020SKA0120300),
the National Key R\&D Program of China (2017YFA0402602), the National Natural
Science Foundation of China (11975027, 11991053, U1831104, 11721303), the Young
Elite Scientists Sponsorship Program by the China Association for Science and
Technology (2018QNRC001), the Max Planck Partner Group Program funded by the Max
Planck Society, the Strategic Priority Research Program of the Chinese Academy
of Sciences (XDB23010200), and the High-Performance Computing Platform of Peking
University.

\section*{Data Availability}


The data underlying this paper will be shared on reasonable request to the
corresponding authors.



\bibliographystyle{mnras}
\bibliography{refs} 




\appendix

\section{Ordinary differential equations related to slowly rotating relativistic stars}
\label{append:A}

The metric for slowly rotating relativistic star to the third order of $\Omega$
can be represented as 
\begin{align}
    {\rm d} s^{2}=&\left\{-e^{\nu(r)}\left[1+2\big(h_{0}+h_{2}P_{2}(\cos\theta)\big)\right] +r^{2}\omega^{2}\sin^{2}\theta  \right\}\dd t^{2} \nonumber\\
    &+e^{\lambda(r)}\left[1+2 \big(m_{0}+m_{2}P_{2}(\cos\theta)\big)/\big(r-2 m \big)\right] \dd r^{2} \nonumber\\
    &+r^{2}\left[1+2 (v_{2}-h_{2})P_{2}(\cos\theta)\right]\left(\dd \theta^{2}+\sin ^{2} \theta \dd \phi^{2}\right) \nonumber \\
    &-2r^{2}\sin^{2}\theta \left[\omega+w_{1}-\frac{1}{\sin\theta}\frac{\dd P_{3}(\cos\theta)}{\dd \theta }w_{3}\right]\dd \phi \dd t\nonumber\\
    &-4r^{2}\sin^{2}\theta P_{2}(\cos\theta)\big(v_{2}-h_{2}\big)\omega \dd \phi \dd t\,.
\end{align}
Combining this metric with the stress-energy tensor in
Eq.~(\ref{eqn:stress_energy}) and applying the linearized Einstein equations, a
systematic ordinary differential equations which determine the stars' structures
can then be obtained. 

In the following, we will give these differential equations, the boundary
conditions at the center of the star, and the match conditions between the
interior and the exterior solutions order by order. Most of the equations can be
found in
\citet{Hartle:1967he,Hartle:1968si,Hartle:1973zza,Chandra1974,Benhar:2005gi,Urbanec:2013fs}.
Note that some typos in the original work \citep{Hartle:1967he} have been
corrected.

\subsection{The first order}

To the first order of $\Omega$, the star remains to be spherical and only the
spacetime is dragged due to the rotation of the star. The fluid motion is not
determined by the angular velocity $\Omega$ relative to a distant observer but
by the angular velocity $\bar \omega$ with respect to the local inertial frame,
which is defined as
\begin{equation}
    \bar \omega =\Omega -\omega \,.
\end{equation}
This quantity satisfies the differential equation 
\begin{equation}
    \frac{1}{r^{4}} \frac{\dd}{\dd r}\left(r^{4} j \frac{\dd \bar \omega}{\dd r}\right)+
    \frac{4}{r} \frac{\dd j}{\dd r} \bar{\omega}=0\,,
\end{equation}
where $j=e^{-(\nu+\lambda)/2}$. In practice, we follow \citet{Sumiyoshi1999} and
\citet{Benhar:2005gi} and introduce two variables  
\begin{equation}
    \xi=j\bar \omega \,, \quad u=r^{4}j\frac{\dd \bar \omega}{\dd r}\,,
\end{equation}
which satisfy 
\begin{equation}
    \label{eqn:uxi}
    \frac{\dd \xi}{\dd r}=\frac{u}{r^{4}}-\frac{4 \pi r^{2}(\rho+P)\xi}{r-2 m}\,,\quad
    \frac{\dd u}{\dd r}=\frac{16 \pi r^{5} (\rho+P)\xi}{r-2 m}\,,
\end{equation}
in the interior of the star. The boundary conditions of $\xi$ and $u$ at the
center of the star are 
\begin{equation}
    \label{eqn:uxicenter}
    \xi(0)=j_{\rm c}\bar \omega_{\rm c}\,,\quad u(0)= 0\,,
\end{equation}
where $j_{\rm c}$ and $\bar \omega_{\rm c}$ are the values of $j$ and $\bar \omega$ at the
center of the star. Integrating Eq.~(\ref{eqn:uxi}) with the boundary conditions
at the center of the star, one obtains the interior solutions. Outside of the
star, the stress-energy tensor $T_{\alpha\beta}$ vanishes and the solutions of
Eq.~(\ref{eqn:uxi}) are 
\begin{equation}
    \xi(r)=\Omega-\frac{2J}{r^{3}}\,,\quad u(r)=6J\,,
\end{equation}
where $J$ is the angular momentum of the star. Taking the matching conditions at
the surface of the star
\begin{equation}
    [\xi]=0\,,\quad [u]=0\,,
\end{equation}
the angular momentum $J$, angular velocity $\Omega$, and the moment of inertia
$I$ can be represented as 
\begin{equation}
    J=u(R)/6\,, \quad\Omega=\xi(R)+\frac{2J}{R^{3}}\,, \quad I\equiv J/\Omega \,.
\end{equation}

\subsection{The second order}

To the order $\Omega^{2}$, the stars will be deformed and the perturbations can
be divided into $l=0$ and $l=2$ parts in the spherical harmonics expansions,
which correspond to spherical stretching and quadrupole deformations
respectively. The isodensity surface at radial coordinate $r$ in the
non-rotating configuration is displaced to 
\begin{equation}
    r+\xi_{0}(r)+\xi_{2}(r)P_{2}(\cos\theta)\,,
\end{equation}
in the rotating configuration.
The displacement functions $\xi_{0}$ and $\xi_{2}$ can be represented as 
\begin{equation}
    \xi_{0}(r)=-p_{0}(\rho+P) /(\mathrm{d} P / \mathrm{d} r)\,,\quad
    \xi_{2}(r)=-p_{2}(\rho+P) /(\mathrm{d} P / \mathrm{d} r)\,,
\end{equation}
where $p_{0}$ and $p_{2}$ are related to the metric perturbation via
\begin{align}
    \label{eqn:p0}
    p_{0}=&-h_{0}+\frac{r^{3}\xi^{2}}{3(r-2m)}+h_{0\rm c}\,,\\
    \label{eqn:p2}
    p_{2}=&-h_{2}-\frac{r^{3}\xi^{2}}{3(r-2m)} \,,
\end{align}
where $h_{0\rm c}$ is a constant. The above two equations can also be treated as
the definitions of $p_{0}$ and $p_{2}$.

\subsubsection{Spherical part}

The spherical part of the perturbations can be obtained by integrating the
differential equations of the ``mass perturbation factor'' $m_{0}$ and the
``pressure perturbation factor'' $p_{0}$
\begin{align}
    \frac{\dd m_{0}}{\dd r} =&\frac{u^{2}}{12 r^{4}}+\frac{8 \pi r^{5}(\rho+P) \xi^{2}}{3(r-2 m)}+4 \pi r^{2}(\rho+P) \frac{\dd \rho}{\dd P} p_{0}\,,\\
    \frac{\dd p_{0}}{\dd r}=& \frac{u^{2}}{12r^{4}(r-2m)} +\frac{2r^{2}\xi}{3(r-2m)}\left[\frac{u}{r^{3}}+\frac{r-3m-4\pi r^{3}P}{r-2m}\xi\right] \nonumber\\
   &-\frac{(1+8\pi r^{2}P)m_{0}}{(r-2m)^{2}}-\frac{4\pi r^{2}(\rho+P)p_{0}}{r-2m}\,.
\end{align}
In practice, one can integrate the equations with the boundary conditions at
$r\rightarrow 0$ 
\begin{align}
    m_{0}(r)= &\frac{4\pi}{15}\big(\rho_{\rm c}+P_{\rm c}\big)\left[\frac{\dd \rho}{\dd P}\mid_{r=0} +2\right]\xi(0)^{2}r^{5}\,,\\
    p_{0}(r)=& \frac{1}{3}\xi(0)^{2}r^{2}\,.
\end{align}
where $\rho_{\rm c}$ and $P_{\rm c}$ are the energy density and the pressure at
the center of the star.  Outside of the star, the analytical solution of $m_{0}$
is 
\begin{equation}
    m_{0}(r)=\delta M-\frac{J^{2}}{r^{3}}\,.
\end{equation}
The constant $\delta M$ is the correction of the gravitational mass at order
$\Omega^{2}$. The match condition of $m_{0}$ at surface of the star is 
\begin{equation}
    [m_{0}]=\frac{4 \pi R^{3} \rho\left(R_{-}\right)(R-2 M) p_{0}(R)}{M}\,.
\end{equation}
Then the total gravitational mass of the rotating star can be represented as 
\begin{equation}
    M+\delta M=M+m_{0}(R)+\frac{J^{2}}{R^{3}}+\frac{4 \pi R^{3} \rho\left(R_{-}\right)(R-2 M) p_{0}(R)}{M}\,.
\end{equation}
Once $p_{0}$, $\delta M$ and $J$ are calculated, the interior solution of
$h_{0}$ can be obtained from Eq.~(\ref{eqn:p0}), while the exterior solution of
$h_{0}$ is 
\begin{equation}
    h_{0}(r)=-\frac{\delta M}{r-2 M}+\frac{J^{2}}{r^{3}(r-2 M)}\,.
\end{equation}
The spherical stretching of isodensity surface at the radial coordinate $r$ can
be represented as 
\begin{equation}
    \xi_{0}(r)=p_{0} r(r-2 m) \Big/\left(m+4 \pi r^{3} P\right)\,,
\end{equation}
and the variation of the stellar radius $\delta R$ is then 
\begin{equation}
    \delta R = \xi_{0}(R)=\frac{R(R-2M)p_{0}(R)}{M}\,.
\end{equation}

\subsubsection{Quadrupole part}
We obtain the quadrupole deformation of the star by integrating the differential equations 
of $h_{2}$ and $v_{2}$
\begin{align}
    \label{eqn:v2}
    \frac{\dd v_{2}}{\dd r}=& \frac{1}{r(r-2 m)}\bigg\{-2 h_{2}\left(m+4 \pi r^{3} P\right)+\big(r-m+ 4 \pi r^{3} P\big)  \nonumber\\
    & \times \left[\frac{8 \pi r^{5}(\rho+P) \xi^{2}}{3(r-2 m)}+\frac{u^{2}}{6 r^{4}}\right]\bigg\} \,,\\
    \label{eqn:h2}
    \frac{\dd h_{2}}{\dd r}=& h_{2}\Bigg\{\frac{r^{2}}{2\left(m+4 \pi r^{3} P\right)}\left[8 \pi(\rho+P)-\frac{4 m}{r^{3}}\right]
    -\frac{2\left(m+4 \pi r^{3} P\right)}{r(r-2 m)}\Bigg\} \nonumber\\
    &-\frac{2 v_{2}}{m+4 \pi r^{3} P} +\frac{u^{2}}{6 r^{5}}\left[\frac{m+4 \pi r^{3} P}{r-2 m}-\frac{r}{2\left(m+4 \pi r^{3} P\right)}\right]\nonumber \\
    &+\frac{8 \pi r^{5}(\rho+P) \xi^{2}}{3(r-2 m)}\left[\frac{m+4 \pi r^{3} P}{r(r-2 m)}+\frac{1}{2\left(m+4 \pi r^{3} P\right)}\right]\,.
\end{align}
The general interior solutions of $h_{2}$ and $v_{2}$ can be obtained by
combinations of homogeneous solutions and particular solutions, 
\begin{equation}
    v_{2}=v_{2}^{\rm p}+C\,v_{2}^{\rm h} \,, \quad h_{2}=h_{2}^{\rm p}+C\,h_{2}^{\rm h}\,,
\end{equation}
where $C$ is a constant to be determined. The particular solutions $v_{2}^{\rm
p}$ and $h_{2}^{\rm p}$ can be calculated by integrating
Eqs.~(\ref{eqn:v2}--\ref{eqn:h2}) and taking the boundary conditions near the
center of the star
\begin{equation}
    v_{2}^{\rm p}=B r^{4}\,,\quad h_{2}^{\rm p} =A r^{2}\,.\\
\end{equation}
The constants $A$ and $B$ satisfy 
\begin{equation}
    B+2\pi\big(\frac{1}{3}\rho_{\rm c}+P_{\rm c}\big) A=\frac{2 \pi}{3}\left(\rho_{\rm{c}}+P_{\rm{c}}\right)\xi(0)^{2}\,.
\end{equation}
The homogeneous solutions $v_{2}^{\rm h}$ and $h_{2}^{\rm h}$ can be obtained by
integrating 
\begin{align}
    \frac{\dd v_{2}^{\rm h}}{\dd r}=& \frac{1}{r(r-2 m)}\left\{-2 h_{2}^{\rm h}\left(m+4 \pi r^{3} P\right)\right\} \,,\\
    \frac{\dd h_{2}^{\rm h}}{\dd r}=& h_{2}^{\rm h}\Bigg\{\frac{r^{2}}{2\left(m+4 \pi r^{3} P\right)}\left[8 \pi(\rho+P)-\frac{4 m}{r^{3}}\right]
    -\frac{2\left(m+4 \pi r^{3} P\right)}{r(r-2 m)}\Bigg\} \nonumber\\
    &-\frac{2 v_{2}^{\rm h}}{m+4 \pi r^{3} P} \,,
\end{align}
with the boundary conditions at $r\rightarrow 0$
\begin{equation}
    v_{2}^{\rm h}=B_{1} r^{4}\,,\quad h_{2}^{\rm h} =A_{1} r^{2}\,.
\end{equation}
The constants $A_{1}$ and $B_{1}$ satisfy the relation 
\begin{equation}
    B_{1}+2\pi\Big(\frac{1}{3}\rho_{\rm c}+P_{\rm c}\Big) A_{1}=0\,.
\end{equation}
The exterior solutions of $v_{2}$ and $h_{2}$ can be calculated analytically 
\begin{align}
    h_{2}=&J^{2}\left(\frac{1}{M r^{3}}+\frac{1}{r^{4}}\right)+K Q_{2}^{2}\left(\frac{r}{M}-1\right)\,,\\
    v_{2}=&-\frac{J^{2}}{r^{4}}+K \frac{2 M}{[r(r-2 M)]^{1 / 2}} Q_{2}^{1}\left(\frac{r}{M}-1\right)\,,
\end{align}
where $Q_{2}^{1}$ and $Q_{2}^{2}$ are Legendre functions of the second kind and
$K$ is a constant to be determined together with $C$ by matching the interior
and exterior solutions at the boundary of the star. The match conditions are  
\begin{equation}
    [v_{2}]=0\,,\quad [h_{2}]=0\,.
\end{equation}
The quadrupole moment can then be read off from the Newtonian potential far from
the star 
\begin{equation}
    Q=-\frac{8 K M^{3}}{5}-\frac{J^{2}}{M}\,.
\end{equation}
The function $m_{2}$ can be obtained from an algebraic relation 
\begin{equation}
    m_{2}=(r-2 m)\left[-h_{2}+\frac{8 \pi r^{5}(\epsilon+P) \xi^{2}}{3(r-2 m)}+\frac{u^{2}}{6 r^{4}}\right]\,,
\end{equation}
and the match condition at the surface of the star is 
\begin{equation}
    [m_{2}]=-8 \pi R^{5}\rho(R_{-}) \xi(R)^{2}/3
\end{equation}
Once we know the solutions of $v_{2}$, $h_{2}$ and $m_{2}$, the function $p_{2}$ can be 
calculated from the relation in Eq.~(\ref{eqn:p2}). The eccentricity of the isodensity 
surface can be defined as \citep{Hartle:1968si}
\begin{equation}
    e(r)=\left[-3\left(v_{2}-h_{2}+\xi_{2} / r\right)\right]^{1 / 2}\,,
\end{equation}
and the surface eccentricity $e_{\rm s}$ can be obtained by taking $r=R$.

\subsection{The third order}

In the third order, the $l=1$ function $w_{1}$ and the $l=3$ function $w_{3}$
will enter.  As we mentioned in the main text, the correction to the angular
momentum and moment of inertia do not depend on the function $w_{3}$. Therefore,
we only concentrate on the calculation of $w_{1}$. The differential equation is
\citep{Hartle:1973zza,Benhar:2005gi}
\begin{equation}
    \label{eqn:w1}
     \frac{\dd}{\dd r}\left(r^{4} j \frac{\dd w_{1}}{\dd r}\right)+4r^{3} \frac{\dd j}{\dd r} w_{1}=r^{4}D_{0}-\frac{1}{5} r^{4} D_{2}\,,
\end{equation}
where 
\begin{align}
    r^{4}D_{0}=& -u \frac{\dd}{\dd r}\left[\frac{m_{0}}{r-2m}+h_{0}\right]-\frac{16\pi r^{5}\xi (\rho+P)}{r-2m}\nonumber\\
    & \times \left[\frac{2m_{0}}{r-2m}+(\frac{\dd \rho}{\dd P}+1)p_{0}+\frac{2r^{3}\xi^{2}}{3(r-2m)}\right]\,,\\
    \frac{r^{4}D_{2}}{5}=& \frac{u}{5}\frac{\dd}{\dd r}\left[4v_{2}-5h_{2}-\frac{m_{2}}{r-2m}\right]-\frac{16\pi r^{5}\xi (\rho+P)}{r-2m}\nonumber\\
    & \times \left[\frac{2m_{2}}{r-2m}+(\frac{\dd \rho}{\dd P}+1)p_{2}-\frac{2r^{3}\xi^{2}}{3(r-2m)}\right]\,.
\end{align}
Similarly with the first order, we introduce two variables 
\begin{equation}
    \xi_{1}=jw_{1}\,,\quad u_{1}=r^{4}j\frac{\dd w_{1}}{\dd r}\,,
\end{equation}
and Eq.~(\ref{eqn:w1}) can then be written as 
\begin{align}
    \frac{\dd \xi_{1}}{\dd r}=&\frac{u_{1}}{r^{4}}-\frac{4 \pi r^{2}(\rho+P) \xi_{1}}{r-2 m} \,,\\
    \frac{\dd u_{1}}{\dd r}=&\frac{16 \pi r^{5}(\rho+P) \xi_{1}}{r-2 m}+r^{4} D_{0}-\frac{r^{4}}{5} D_{2}\,.
\end{align}
Inside of the star, the general solutions are combinations of particular
solutions and homogeneous solutions 
\begin{equation}
    \xi_{1}=\xi_{1}^{\rm p}+C_{1}\,\xi_{1}^{\rm h} \,, \quad u_{1}=u_{1}^{\rm p}+C_{1}\,u_{1}^{\rm h}\,,
\end{equation}
where $C_{1}$ is a constant. The particular solutions can be obtained with the
boundary conditions at the center of the star 
\begin{equation}
    \xi_{1}^{\rm p}(0)=0\,,\quad u_{1}^{\rm p}(0)=0\,.
\end{equation}
Setting $D_{0}=D_{2}=0$, one obtains the homogeneous equations. The solutions of
$\xi_{1}^{\rm h}$ and $u_{1}^{\rm h}$ can be obtained by integrating the
homogeneous equations with the asymptotical conditions at $r\rightarrow 0$ 
\begin{align}
    \xi_{1}^{\rm h}(r)=&A_{2}\left[1-\frac{2\pi}{5}\big(\rho_{\rm c}+P_{\rm c}\big)r^{2}\right]\,,\\
     u_{1}^{\rm h}(r)=&A_{2}\left[\frac{16\pi}{5}\big(\rho_{\rm c}+P_{\rm c}\big)r^{5}\right]\,,
\end{align}
where $A_{2}$ is a constant that can be chosen arbitrarily. Outside of the star,
$\xi_{1}$ and $u_{1}$ can be solved analytically 
\begin{equation}
    \xi_{1}(r)=\frac{2\delta J}{r^{3}}+G(r)\,,\quad u_{1}(r)=-6\delta J +Y(r)\,.
\end{equation}
The functions $G(r)$ and $Y(r)$ can be represented as \citep{Benhar:2005gi}
\begin{align}
    G(r)=&-\frac{12 J^{3}}{5 r^{7}}-\frac{4 J^{3}}{5 M r^{6}}+\frac{J K}{40 M^{3} r^{4}}\bigg[108 r^{4} \ln \left(\frac{r}{r-2 M}\right) \nonumber\\
    &-288 r^{3} M \ln \left(\frac{r}{r-2 M}\right)+33 r^{4}-240 r^{3} M \nonumber\\
    &+336 r^{2} M^{2}+256 M^{3} r-96 M^{4} \nonumber\\
    &+192 r M^{3} \ln \left(\frac{r}{r-2 M}\right)+12 r^{4} \ln \left(\frac{r}{r-2 M}\right)\bigg]-\frac{33 J K}{40 M^{3}}\,,\\
    Y(r)=&\frac{84J^{3}}{5r^{4}}+\frac{24J^{3}}{5Mr^{3}} +\frac{24KJQ_{2}^{2}\big(\frac{r}{M}-1\big)}{5}-\frac{48KMJQ^{1}_{2}\big(\frac{r}{M}-1\big)}{5\big(r(r-2M)\big)^{1/2}}\,.
\end{align}
The constant $\delta J$ is the angular momentum at order $\Omega^{3}$, which can
be obtained together with $C_{1}$ by matching the solutions of $\xi_{1}$ and
$u_{1}$ at the boundary of the star.  Note that the match conditions of
$\xi_{1}$ and $u_{1}$ are 
\begin{align}
    [\xi_{1}]=&0\,,\\
    [u_{1}]=&-\frac{16 \pi R^{6} \rho\left(R_{-}\right)\xi(R)}{M}\left(p_{0}(R)-\frac{1}{5} p_{2}(R)\right)\nonumber\\
    &-\frac{4\pi R^{3}\rho(R_{-})u(R)p_{0}(R)}{M}-\frac{8\pi R^{5}\rho(R_{-})u(R)\xi(R)^{2}}{15(R-2M)}\,.
\end{align} 
The contribution of the moment of inertia at the second order of $\Omega$ is
$\delta I=\delta J/\Omega$.

\bsp	
\label{lastpage}
\end{document}